\documentclass[a4paper,12pt]{article}
\usepackage[dvipdfmx]{graphicx}
\usepackage{amssymb}
\usepackage{amsmath}
\usepackage{latexsym}
\usepackage{supertabular}
\usepackage{slashed}
\usepackage{booktabs}
\usepackage{cases}
\usepackage{bigints}
\usepackage{mysty}
\usepackage[abs]{overpic}
\usepackage{amsfonts}
\usepackage{comment-k}
\usepackage{color}
\usepackage[large]{subfigure}

\newcommand{\nn}{\nonumber}
\newcommand{\simlt}{\lower.5ex\hbox{$\; \buildrel < \over \sim \;$}}
\newcommand{\simgt}{\lower.8ex\hbox{$\; \buildrel > \over \sim \;$}}

\def\dfrac{\displaystyle\frac}
\def\sss{\scriptscriptstyle}

\def\gsim{~{\rlap{\lower 3.5pt\hbox{$\mathchar\sim$}}\raise 1pt\hbox{$>$}}\,}
\def\lsim{~{\rlap{\lower 3.5pt\hbox{$\mathchar\sim$}}\raise 1pt\hbox{$<$}}\,}
\def\numt#1#2{#1 \times 10^{#2}}

\def\ra{\rightarrow}
\def\cerenkov{$\check{\rm C}$erenkov}

\def\dmns{\delta_{\sss{\rm CP}}}

\def\dm#1#2{\delta m^2_{#1#2}}

\def\dCP{\delta_{\rm CP}}
\def\Erec{E_{\rm rec}}
\def\effe{\varepsilon_e}
\def\effm{\varepsilon_\mu}
\def\ssq#1#2#3{\sin^2 #1\theta_{#2#3}}
\def\s#1#2#3{\sin #1\theta_{#2#3}}

\def\dchisq{\Delta\chi^2}
\def\dchisqmin{(\Delta\chi^2)_{\rm min}}
\def\dchisqmh{\Delta\chi^2_{\rm MH}}
\def\odchisqmh{\overline{\dchisqmh}}
\def\T2KKO{T2KK/T2KO}
\def\T2K122{{\rm T2K}_{122}}
\def\nonqe{non-CCQE}
\def\nuance{{\tt Nuance}}
\def\nuancev{{\tt Nuance v3.504}}
\def\polfit{POLfit }
\def\nubar{\bar{\nu}}
\def\matter{matter} 

\incomment{com}

\preprint{KEK-TH-1868, UT-16-21}
\title{Revisiting T2KK and T2KO physics potential and $\nu_\mu$ - $\bar{\nu}_\mu$ beam ratio}

\renewcommand{\thefootnote}{\fnsymbol{footnote}}
\author[a,b]{Kaoru Hagiwara\footnote{kaoru.hagiwara@kek.jp}}
\author[c]{Pyungwon Ko\footnote{pko@kias.re.kr}}
\author[d]{Naotoshi Okamura\footnote{nokamura@iuhw.ac.jp}}
\author[e]{and Yoshitaro Takaesu\footnote{takaesu@hep-th.phys.s.u-tokyo.ac.jp}}

\affiliation[a]{Theory Center, KEK, 1-1 Oho, Tuskuba, Ibaraki 305-0801, Japan}
\affiliation[b]{Department of Accelerator Science, Sokendai, 1-1 Oho,
Tsukuba, Ibaraki 305-0801, Japan}
\affiliation[c]{School of Physics, KIAS, 85 Hoegi-ro, Dongdaemun-gu,
Seoul 130-722, Korea}
\affiliation[d]{Department of Radiological Sciences, International
University of Health and Welfare, 2600-1 Kitakanemaru, Ohtawara, Tochigi
324-8501, Japan}
\affiliation[e]{Department of Physics, University of Tokyo, Tokyo
113-0033, Japan}

\abstract{
We revisit the sensitivity study of the Tokai-to-Kamioka-and-Korea (T2KK)
and Tokai-to-Kamioka-and-Oki (T2KO) proposals where a water \cerenkov\,
detector with the 100 kton fiducial volume is placed in Korea ($L = 1000$ km) and Oki island ($L = 653$ km) in Japan, respectively, in addition to the
 Super-Kamiokande for determination of the neutrino mass
hierarchy
and leptonic CP phase ($\dmns$). We systematically study the running
ratio of the $\nu_\mu$ and
$\bar{\nu}_\mu$ focusing beams with dedicated background estimation for the $\nu_e$ appearance and $\nu_\mu$ disappearance
signals, especially improving treatment of the neutral current
$\pi^0$ backgrounds.
Using a $\nu_\mu$ - $\bar{\nu}_\mu$ beam ratio between 3\,:\,2 and
2.5\,:\,2.5 (in unit of $10^{21}$POT with the proton energy of 40 GeV),
the mass hierarchy determination with 
the median sensitivity of 3\,-\,5 $\sigma$ by the T2KK
and 
1\,-\,4 $\sigma$ by the T2KO experiment are expected when $\ssq{}23 = 0.5$, depending on the mass hierarchy pattern
and CP phase. These sensitivities are enhanced (reduced) by $30\%$ -
$40\%$ in $\dchisq$ when
$\ssq{}23 = 0.6\, (0.4)$.
The CP phase is measured with the uncertainty of $20^\circ$ - $50^\circ$ by
the T2KK and T2KO using the $\nu_\mu$ - $\bar{\nu}_\mu$ focusing beam ratio between 3.5\,:\,1.5 and 1.5\,:\,3.5.
These findings indicate that 
inclusion of the $\bar{\nu}_\mu$ focusing beam improves the
sensitivities of the T2KK and T2KO experiments to both the mass
hierarchy determination and leptonic CP phase measurement
simultaneously with the preferred beam ratio being between 3\,:\,2 - 2.5\,:\,2.5 ($\times 10^{21}$POT).
 }

\begin{document}

\maketitle

\renewcommand{\thefootnote}{\arabic{footnote}}
\setcounter{footnote}{0}
\renewcommand{\include}[1]{}
\renewcommand\documentclass[2][]{}

\section{Introduction}
\label{intro}
After the accurate measurements of $\ssq213$ by DayaBay~\cite{An:2012eh,An:2013uza,An:2013zwz,An:2014ehw,An:2015rpe,An:2016bvr},
Reno~\cite{Ahn:2012nd,RENO:2015ksa} and Double
Chooz~\cite{Abe:2012tg,Abe:2013sxa,Abe:2014lus,Abe:2014bwa,Abe:2015rcp} experiments,
determination of the neutrino mass hierarchy and CP violating phase in the
Maki-Nakagawa-Sakata (MNS) mixing matrix~\cite{Maki:1962mu} has been
the next targets in the neutrino physics. 

Ideas of extending the Tokai-to-Kamioka (T2K) experiment with additional
water \cerenkov\, detectors placed in
Korea (Tokai-to-Kamioka-and-Korea, T2KK, experiment~\cite{Hagiwara:2004iq,Ishitsuka:2005qi,Hagiwara:2005pe,Hagiwara:2006vn,Kajita:2006bt,Hagiwara:2006nn,Huber:2007em,Hagiwara:2009bb,Dufour:2010vr,Hagiwara:2011kw,Hagiwara:2012mg,Dufour:2012zr})
or in Oki island (Tokai-to-Kamioka-and-Oki, T2KO,
experiment~\cite{Badertscher:2008bp,Hagiwara:2012mg}) 
has been proposed to address those questions~\footnote{For the CP phase
measurement, there are also proposals to utilize
neutrinos from muon decays at rest~\cite{Ciuffoli:2014ika,Evslin:2015pya,Ge:2016xya}.}.
It has been shown that the T2KK experiment with a 100 kton
fiducial-volume detector in Korea in addition to the SK detector is an
appealing proposal 
if we can use the J-PARC neutrino beam with 0.64 MW beam power and the
$2.5^\circ$ - $3.0^\circ$
off-axis angle at the
SK~\cite{Hagiwara:2005pe,Hagiwara:2006vn,Hagiwara:2006nn,Hagiwara:2009bb,Hagiwara:2011kw}.
The authors of Ref.~\cite{Hagiwara:2005pe,Hagiwara:2006vn,Hagiwara:2006nn} investigated the
sensitivities to the mass hierarchy and CP phase with the $\nu_\mu$
focusing beam in a simple manner, ignoring the effects of neutral current (NC) $\pi^0$ backgrounds,
miss-identification of a muon as an electron, and smearing of
reconstructed neutrino energy.
Authors of Ref.~\cite{Hagiwara:2009bb} then re-evaluated the physics
potential of the same T2KK setup with careful consideration on those effects.

Inclusion of $\bar{\nu}_\mu$ focusing beams may improve the sensitivity
of long-baseline oscillation experiments to the mass
hierarchy since the matter effects, which enhance the mass-hierarchy difference
in neutrino oscillation patterns, appear in the opposite way in $\nu_\mu$ and
$\bar{\nu}_\mu$ oscillations.
The impacts of including the
$\bar{\nu}_\mu$ focusing beam in the T2KK experiment was studied in Ref.~\cite{Hagiwara:2011kw}.
The authors considered the running ratio of the $\nu_\mu$ and $\bar{\nu}_\mu$
focusing beams of 5\,:\,0 and 2.5\,:\,2.5 in the unit of protons on
target (POT) and argued that including
the $\bar{\nu}_\mu$ focusing beam improves the sensitivity to the mass hierarchy
determination significantly.
 The impacts of anti-neutrino beams was also
studied in Ref.~\cite{Ishitsuka:2005qi} for a different
 T2KK setup; two 270 kton detectors are each placed at Kamioka
 and Korea, receiving $2.5^\circ$ off-axis beams with the beam power of
 4 MW and the total running time of eight years.
Physics potential of the T2KO experiment was also investigated~\cite{Hagiwara:2012mg} with a
similar analysis and conclusion as in Ref.~\cite{Hagiwara:2011kw}. However, those
studies again did not consider the effects of the NC $\pi^0$
backgrounds, miss-identified muon, and events from other
neutrino-nucleus interactions than the charged-current quasi-elastic
(CCQE) one. Therefore, it is not very clear whether
the $\nu_\mu$ - $\bar{\nu}_\mu$ focusing beam ratio of 1\,:\,1 is the best
for the mass hierarchy
determination and CP phase measurement.

In this paper, we revisit the sensitivity study of the T2KK~\cite{Hagiwara:2009bb,Hagiwara:2011kw} and T2KO~\cite{Hagiwara:2012mg} experiments for the
neutrino mass hierarchy and CP phase, studying the 
dependence of the sensitivities on the $\nu_\mu$ - $\bar{\nu}_\mu$
focusing beam ratio systematically with dedicated estimation of
backgrounds. Especially, the treatment of the NC
$\pi^0$ backgrounds is improved in this analysis. The NC $\pi^0$
backgrounds is estimated using a realistic $\pi^0$ rejection probability
based on the \polfit (Pattern Of Light fitter)
algorithm~\cite{Barszczak:2005sf}, and the contribution form the coherent $\pi^0$ production
process is taken into
account, which is neglected in the previous
analysis~\cite{Hagiwara:2009bb}. The uncertainty of the NC $\pi^0$ backgrounds is also reconsidered including the uncertainty
from the axial masses in the models of neutrino-nucleus scattering cross sections \cite{Rein:1980wg,Rein:1982pf}.

The remaining part of this paper is organized as follows. After describing the T2KK and T2KO
experimental
setups in Section~\ref{sec:setup}, our analysis details are discussed in
Section~\ref{sec:chi2}. Results for the sensitivity of the T2KK and T2KO
experiments to the mass hierarchy determination and CP phase
measurements are presented in Sections~\ref{sec:MH} and \ref{sec:CP}, and our main
conclusions are summarized in Section~\ref{sec:conclusion}.

\section{Simulation details of T2KK and T2KO experiments}
\label{sec:setup}
In this section, we fix our notation and introduce useful approximated formulae for
the $\nu_\mu \rightarrow \nu_\mu$ and $\nu_\mu
\rightarrow \nu_e$
oscillation probabilities. We then describe the experimental
setups and discuss the simulation
details of the expected signal event number in those experiments, taking
into account of smearing of
reconstructed neutrino energy due to the Fermi motions of target nuclei,
detector resolution and contamination of events from non-CCQE
neutrino-nucleus interactions.
Simulation of the background events are also discussed: the NC
single-$\pi^0$ background and its uncertainty, the secondary neutrino beam
backgrounds, and miss-identified muon/electron backgrounds.

\subsection{Neutrino oscillations in \matter}
\label{sec:oscillation}
We briefly review the neutrino oscillation probabilities in
\matter, presenting analytic approximations for the $\nu_\mu \rightarrow \nu_\mu$
($\nu_\mu$ disappearance) and $\nu_\mu
\rightarrow \nu_e$ ($\nu_e$ appearance) oscillation modes,
which are useful for understanding the physics potential of
the T2KK and T2KO experiments qualitatively. 

We work in the three neutrino flavor scheme, where the neutrino flavor eigenstate 
$\left|\nu_{\alpha} \right\rangle$ ($\alpha=e,\mu,\tau$) 
are mixtures of the three mass eigenstates 
$\left|\nu_{i} \right\rangle$
 with their masses $m_i$ ($i=1,2,3$) as
\begin{equation}
 \left|\nu_{\alpha} \right\rangle=
 \sum^{3}_{i=1} U_{\alpha i}
\left|\nu_{i} \right\rangle.
\end{equation}
Here $U$ is the Maki-Nakagawa-Sakata (MNS)~\cite{Maki:1962mu} matrix,
which can be parameterized with the three mixing angles, $\theta_{12},
\theta_{13}, \theta_{23}$, and three phases, $\dmns$, $\phi_1$, $\phi_2$
\cite{Beringer:1900zz}. Among them, two phases can be eliminated in
lepton number conserving processes, remaining one relevant
phase, $\dmns$, to neutrino oscillation experiments.
The definition regions of the four parameters are
chosen as $0\leq \theta_{12},\theta_{13},\theta_{23}\leq \pi/2$ and
$-\pi \leq \dmns \leq \pi$.

The probability 
that an initial flavor eigenstate $\left|\nu_\alpha\right\rangle$ 
with energy $E$ is observed as a flavor eigenstate 
$\left|\nu_\beta\right\rangle$
after traveling a distance $L$ in the matter of density 
$\rho(x)$ $(0<x<L)$ is given by
\begin{eqnarray}
 P_{\nu_\alpha \to \nu_\beta}
 &=&
\left|
\left\langle \nu_\beta \right|
\exp\left({-i\int_0^LH(x)dx}\right)
\left|\nu_{\alpha} \right\rangle
\right|^2 \,,
\label{eq:PPP}
\end{eqnarray}
where the Hamiltonian inside \matter\, is
\begin{eqnarray}
 H(x)&=&
\dfrac{1}{2E}U
\left(
\begin{array}{ccc}
 0 & 0 & 0 \\
 0 & \dm21 & 0 \\
 0 & 0 & \dm31
\end{array}
\right)
U^\dagger
+
\dfrac{a(x)}{2E}
\left(
\begin{array}{ccc}
 1 & 0 & 0 \\
 0 & 0 & 0 \\
 0 & 0 & 0
\end{array}
\right)\,\nn\\
&=&\dfrac{1}{2E}
{\tilde U(x)}
\left(
\begin{array}{ccc}
 \lambda_1(x) & 0 & 0 \\
 0 & \lambda_2(x) & 0 \\
 0 & 0 & \lambda_3(x)
\end{array}
\right)
{\tilde U}^\dagger(x)\,
\label{eq:Hm}
\end{eqnarray}
with
$\delta m_{ij}\equiv m^2_i -m^2_j$.
$a(x)/2E$ is the effective potential due to electrons in \matter\, as
\begin{equation}
 a(x) = 2\sqrt{2} G_F E^{} n_e(x)
  \simeq \numt{7.56}{-5} \mbox{{[eV$^2$]}} 
  \left(\dfrac{\rho(x)}{\mbox{{g/cm$^3$}}}\right)
  \left(\dfrac{E}{\mbox{{GeV}}}\right),
\label{eq:matt_a}
\end{equation}
where $G_F$ is the Fermi constant and 
$n_e(x)$ is the electron number density. In the translation from $n_e(x)$ to $\rho(x)$,
we assume that the number of the neutron in \matter\, is same as that of proton. $\lambda_i(x)/2E$
and $\tilde{U}(x)$ are
the eigenvalues and the corresponding unitary matrix of the Hamiltonian
at the distance $x$, respectively.
To a good approximation~\cite{Itow:2001ee,Arafune:1997hd,Hagiwara:2011kw},
the matter density along the T2K, T2KO and T2KK baselines
can be replaced by the averaged one, $\rho(x) \simeq \bar{\rho}$, and so as
 $a(x)$ in Eq.~\eqref{eq:Hm}, $a(x) \simeq \bar{a}$.
Then the oscillation probability, $ P_{\nu_\alpha \to \nu_\beta}$, can be expressed compactly
by using $x$-independent eigenvalues, $\lambda_i$, and a unitary matrix, $\tilde
 U$, as
\begin{subequations}
\begin{eqnarray}
 P_{\nu_\alpha \to \nu_\beta}
&=&
{\delta}_{\alpha\beta}
-4\sum_{i>j}
 {\rm Re}({\tilde U}^{\ast}_{\alpha i}{\tilde U}_{\beta i}^{}
     {\tilde U}_{\alpha j}^{}{\tilde U}^{\ast}_{\beta j})
 \sin^2\dfrac{{\tilde \Delta}_{ij}}{2} 
-2\sum_{i>j}
 {\rm Im}({\tilde U}^{\ast}_{\alpha i}{\tilde U}_{\beta i}^{}
     {\tilde U}_{\alpha j}^{}{\tilde U}^{\ast}_{\beta j})
 \sin{\tilde \Delta}_{ij}\,,~~~~~~~
\label{eq:PPP2}\\
 \tilde{\Delta}_{ij}
&\equiv&
 \dfrac{\lambda_i -\lambda_j}{2E}L\,.
 \label{eq:P}
\end{eqnarray}
\end{subequations}
Our numerical results are based on the above solution,
Eq.~(\ref{eq:PPP2}), and the main results are not affected significantly 
by the matter density profile
as long as the mean matter density is chosen appropriately~\cite{Hagiwara:2011kw,Hagiwara:2012mg}.

In this study we are mainly interested in the $\nu_\mu
\rightarrow \nu_e$ and $\nu_\mu \rightarrow \nu_\mu$ oscillation modes and their charge
conjugated ones. It is useful to write them down
in the approximated analytic forms as \cite{Hagiwara:2005pe} 
\begin{subequations}
 \begin{align}
  P_{\nu_\mu \rightarrow \nu_\mu} \simeq&\,\, 1 -\sin^22\theta_{\rm
  atm}\left\{ \left( 1+A^\mu \right)\sin^2\Delta_{31}
  +B^\mu\sin(2\Delta_{31}) \right\} +C^\mu, \label{eq:Pmumu}\\
  P_{\nu_\mu \rightarrow \nu_e} \simeq&\,\,
 4\sin^2\theta_{13}\sin^2\theta_{23}\left\{ \left( 1 +A^e
  \right)\sin^2\Delta_{31} +B^e\sin(2\Delta_{31}) \right\} +C^e,
\label{eq:Pmue}
 \end{align}
\end{subequations}
where $\sin\theta_{\rm atm} \equiv \sin\theta_{23}\cos\theta_{13}$ and
$\Delta_{ij} \equiv \delta m^2_{ij}L/4E$. Here
$A^\alpha, B^\alpha$ and $C^\alpha (\alpha = \mu, e)$ are corrections
due to the matter effect and smaller
mass difference $\delta m^2_{21}$:
\begin{subequations}
 \begin{align}
 A^\mu \simeq&\, 0, \\
B^\mu \simeq&\, \Delta_{21}\cos^2\theta_{12}, \\
C^\mu \simeq&\, 0, \\
 A^e \simeq&\, \frac{aL}{2\Delta_{31}E}
  -\Delta_{21}\frac{\s212}{\tan\theta_{23}\s{}13}\sin\dmns, \\
B^e \simeq&\, -\frac{aL}{4E}
  +\frac{\Delta_{21}}{2}\frac{\s212}{\tan\theta_{23}\s{}13}\left(\cos\dmns -2\ssq{}12\right), \\
C^e \simeq&\, \Delta^2_{21}\ssq212\cos^2\theta_{23}.
 \end{align}
\label{eq:ABC}
\end{subequations}
In these expressions we retain up to the sub-leading terms of $\Delta_{21}, \ssq{}13$ and
$aL/4E$. The corresponding probabilities for anti-neutrino
oscillations can be obtained from the above expressions by reversing the sign of the matter effect
term ($a \rightarrow -a$) and the CP phase ($\dmns \rightarrow -\dmns$). These
expressions are valid as long as those three parameters are negligibly
smaller than unity; this is the case for T2K, T2KO and T2KK experiments, where typically
$L/E \sim {\cal O}(10^2-10^3) [1/{\rm eV}^2]$. 

The $\nu_e$ appearance mode plays more important role in
determining the mass hierarchy (i.e., the sign of $\Delta_{31}$) than
the $\nu_\mu$ disappearance mode. This is because the appearance mode may
have sensitivity to the mass hierarchy around oscillation peaks through
the $A^e$ parameter, while the disappearance mode is lack of sensitivity
around oscillation peaks since $A^\mu \simeq 0$.
On the other hand, the disappearance mode is important in constraining
the $\theta_{\rm 23}$ mixing angle, which still has large uncertainty~\cite{Beringer:1900zz}.
The $\nu_e$ appearance mode also has sensitivity to
the CP phase. It is sensitive to the sine of $\dmns$ around the
oscillation peaks, mainly through the $A^e$ parameter; on the other
hand, it is
sensitive to the the cosine of $\dmns$
between
oscillation maxima and minima, mainly through the $B^e$
parameter. Therefore, if we try to obtain the full information of the
$\dmns$, it is not enough to observe just around the first oscillation peak, as we will see later.  

\subsection{Experimental setups}
\label{sec:exp}
We use the $\nu_{\mu}$ and $\bar{\nu}_{\mu}$ focusing beam fluxes
 from the J-PARC with the proton energy of 40 GeV \cite{T2Kflux}. In
 Fig.~\ref{fig:flux1}, we show the
fluxes corresponding to $10^{21}$ POT (protons on target) at the SK.
\begin{figure}[t]
 \centering
\resizebox{1.0\textwidth}{!}{
\includegraphics[width=1.0\textwidth]{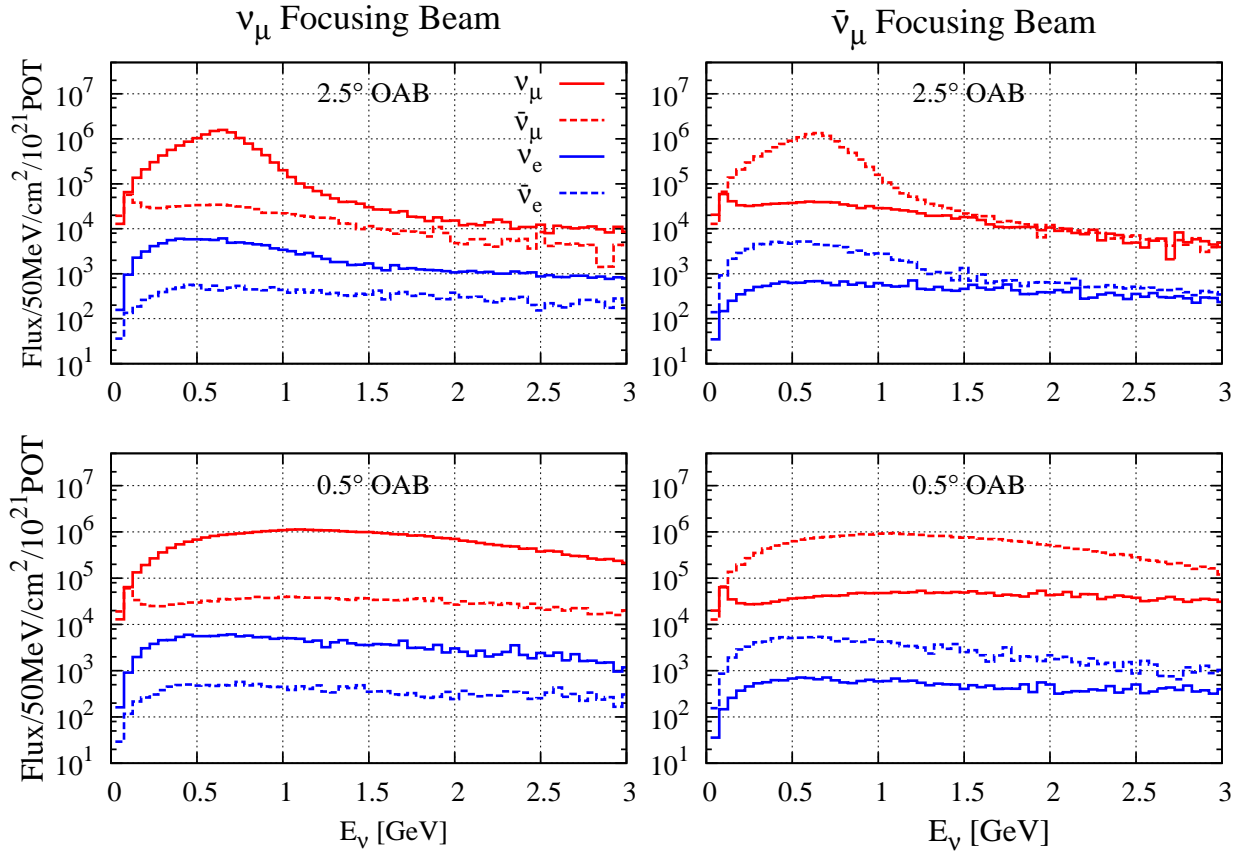}
}
  \caption{The neutrino fluxes in $\nu_{\mu}$ and $\bar{\nu}_{\mu}$ focusing
 beams at the SK as functions
 of neutrino energy. The left and right plots are for the
 $\nu_{\mu}$ and $\bar{\nu}_{\mu}$ focusing beams, while the upper
 and lower plots are for the $2.5^{\circ}$ and
 $0.5^{\circ}$ off-axis beams (OAB), respectively. In each plot, the fluxes of $\nu_{\mu}$ (solid red),
 $\bar{\nu}_{\mu}$ (dashed red), $\nu_e$ (solid blue) and ${\bar{\nu}_e}$
 (dashed blue) are shown. The fluxes are normalized to
  $10^{21}$ POT with 40 GeV proton energy.
}
\label{fig:flux1}
\end{figure}
The $\nu_{\mu} \,(\bar{\nu}_{\mu})$ focusing beams
include the primary, $\nu_{\mu} \,(\bar{\nu}_{\mu})$, and secondary,
$\bar{\nu}_{\mu} \,(\nu_{\mu})$, $\nu_e$, $\bar{\nu}_e$, components, and
we take them into account in our analyses.  

The baseline length from the J-PARC to the SK and Oki detectors are taken
  to be 295 km \cite{Abe:2011ks} and 653 km \cite{Hagiwara:2012mg},
  respectively. The baseline length to a detector in Korea (Kr detector) can be taken from
 1000 km to 1300 km in South Korea~\cite{Ishitsuka:2005qi,Hagiwara:2005pe}.
In this study, we place a Kr detector at the shortest baseline length,
$L = 1000$ km, to receive the J-PARC neutrino beams with the smallest
off-axis angle \cite{Hagiwara:2005pe}, which is preferred in terms of the
  sensitivity to the mass hierarchy determination~\cite{Hagiwara:2005pe,Hagiwara:2006vn,Hagiwara:2012mg,Hagiwara:2009bb}.
For the nominal 2.5$^\circ$ off-axis angle at the SK, a Kr detector
  receives the $\sim 1^\circ$ off-axis beam (OAB); the case of
  $3.0^\circ$ OAB at the SK is also investigated,
  corresponding to the $0.5^\circ$ OAB at a Kr
  detector~\cite{Hagiwara:2012mg}.   
On the other hand, variation of the off-axis
 angle does not affect sensitivities of the T2KO experiment 
 to the mass hierarchy and CP phase measurements significantly~\cite{Hagiwara:2012mg}, and we only
 consider the $2.5^\circ$ off-axis angle at the SK for
 the T2KO experiment, corresponding to $0.9^\circ$
  OAB at the Oki detector~\cite{Hagiwara:2012mg}.

The averaged matter densities, $\bar{\rho}$, along the baseline between J-PARC and SK,
 Oki or Kr detectors  have been evaluated in Refs.~\cite{Hagiwara:2011kw,Hagiwara:2012mg} and
 taken as in Table~\ref{tb:detector}
in this study. 
It is sufficient to use
 those averaged densities in those long-baseline neutrino experiments~\cite{Hagiwara:2011kw}, and
 we neglect small effects from the variation of the matter density along
 those baselines. 
\begin{table}[t]
\begin{center}
\begin{tabular}{cccccc}
\hline\hline\addlinespace[2 pt]
Detector & $L$ [km] & FV [kton] & $\bar{\rho}\, [{\rm g}/{\rm cm}^3]$ & OA [deg.] \\[2 pt]
\hline\addlinespace[2 pt]
SK & 295 & 22.5 & 2.60 \cite{Hagiwara:2012mg} & 2.5/3.0 \\
Oki & 653 & 100 & 2.75 \cite{Hagiwara:2012mg} & 0.9/- \cite{Hagiwara:2012mg}\\
Kr & 1000 & 100 & 2.90 \cite{Hagiwara:2012mg} & 1.0/0.5 \cite{Hagiwara:2012mg}\\
\hline\hline
\end{tabular}
\end{center}
\caption{Summary of the parameters related to detectors at
  Kamioka (SK), Oki island (Oki) and Korea (Kr). $L$ is the baseline
  length between the J-PARC and a detector, FV is the fiducial volume of a
  detector, $\bar{\rho}$ is the average matter density along a baseline,
  and OA is the off-axis angle of the J-PARC neutrino (anti-neutrino) beams at a
  detector. The first and second OA angles at the Oki and Kr detectors are related to
  the corresponding OA angles at the SK. These parameter values are used as
 default
  in our simulation unless otherwise mentioned.}
\label{tb:detector}
\end{table}

\subsection{Signal events}
\label{sec:signal}
In this subsection we describe how to estimate the signal
event numbers at the SK, Oki and Korea detectors.
We consider charged-current quasi-elastic (CCQE) events, $\nu_l\,
n\rightarrow l\,p$ or $\bar{\nu}_l\,p\rightarrow \bar{l}\,n\, (l = \mu$
or $e$),
 from the $\nu_\mu\rightarrow \nu_\mu$ and $\nu_\mu\rightarrow
 \nu_e$ oscillation modes and their charge conjugated modes 
as signal events. The CCQE events are identified as
events with only one \cerenkov\, ring from an electron or muon where
the visible energy of the ring is required to be larger than 200 MeV.
Since neutrino beam direction at a far detector is understood well in long-baseline
 experiments, we can reconstruct incoming neutrino energy for the CCQE events as~\cite{Abe:2013hdq}
\begin{equation}
 \Erec = 
\dfrac{m_p^2 -(m_n -E_n^b)^2 - m_\ell^2 +2(m_n -E_n^b)E_\ell}
{2(m_n -E_n^b -E_\ell+ p_\ell \cos\theta_\ell)},
\label{eq:erec1}
\end{equation}
assuming that target
nucleons are at rest.
Here $E_\ell$, $p_\ell$ and $\theta_\ell$ are
the charged lepton's energy,
magnitude of the three-momentum and polar angle about the neutrino beam
direction; $m_p, m_n$ and $m_l$ are the mass of
a proton, neutron and charged-lepton, respectively, and $E_n^b$ is the
neutron binding energy in the target nucleus. 
For the anti-neutrino events,
$m_p$ and $m_n$ should be exchanged and $E_n^b$ should be replaced with
the proton binding energy $E_p^b$ in Eq.~(\ref{eq:erec1}).
It should be noted that
in reality the reconstructed energy may be
different from the incoming neutrino energy due to the Fermi motion of target
nucleons inside nuclei and finite detector resolutions for
lepton momenta and scattering angles.

The number of the signal events in the $i_{\rm th}$ energy bin,
$E^i_{\rm rec} < E < E^{i+1}_{\rm rec}$, from the $\nu_{\alpha}
\rightarrow \nu_{\beta}$ oscillation mode 
at the water \cerenkov\, detector $D\, (=$ SK, Oki, Kr) via the $X$-type
neutrino-nucleus interaction ($X=$ CCQE, non-CCQE) are calculated as
\begin{align}
N^{i,X}_D&(\nu_\alpha \rightarrow \nu_\beta) = \nn\\
& 
\int^{E^{i+1}_{\rm rec}}_{E^i_{\rm rec}} dE_{\rm rec} \int_0^{\infty} dE_\nu\,
 \Phi_{\nu_\alpha}^{D}(E_\nu)~
 P_{\nu_\alpha\to\nu_\beta}(E_\nu, \bar{\rho}^D)
 \sum_{Z=H,O} N_Z \,\hat{\sigma}_{\nu_\beta Z}^{X}(E_\nu)~ 
 S^{X}_{\nu_{\beta} Z}(E_\nu, E_{\rm rec}),
\label{eq:N}
\end{align}
where $E^i_{\rm rec} = 0.05 {\rm GeV} \times i$,
$E_{\nu}$ is an incoming neutrino energy, 
$\Phi_{\nu_\alpha}^D$ is the flux of $\nu_{\alpha}$ at the detector $D$,
$P_{\nu_{\alpha}\to \nu_{\beta}}$
is the neutrino oscillation probability including the matter effects with
the mean matter density $\bar{\rho}^D$, and $N_Z$ is the number of the nucleus $Z$ (hydrogen ($H$) or oxygen ($O$)) in
the detector.
$\hat{\sigma}^{X}_{\nu_\beta Z}$
is the
cross section of the $X$-type $\nu_\beta$--$Z$ interaction after
imposing a CCQE selection
cuts.
The smearing function $S^X_{\nu_\beta Z}(E_\nu, E_{\rm rec})$ returns the
probability that the energy $E_{\rm rec}$ is reconstructed from an
event induced by an incoming neutrino with the energy $E_\nu$, taking into account the Fermi motion of the target
nucleons and detector resolutions.
The detection
 efficiency of \cerenkov\, rings and
 the electron/muon identification efficiencies will be discussed in
 Sections~\ref{sec:bg} and \ref{sec:chi2}.

In order to estimate the cross sections of the CCQE signal,
 $\hat{\sigma}^X_{\nu_\beta Z}$, we generate events induced by the neutrino and anti-neutrino
charged-current interactions with the Monte-Carlo event generator
\nuancev~\cite{Casper:2002sd}, imposing the CCQE selection criteria:
\begin{subequations} 
\label{eq:criteriaCC}
\begin{eqnarray}
&&
\mbox{{Only one charged lepton}}~(\ell=\mu^\pm \mbox{ or } e^\pm)~
\mbox{{with}}~|p_\ell|>200 \mbox{{~MeV}},
\label{eq:CC1} \\
&&
\mbox{{No high energy}}~\pi^\pm~(|p_{\pi^\pm}|>200\mbox{{~MeV}})\,,
\label{eq:CC2}\\
&&
\mbox{{No high energy}}~\gamma~(|p_{\gamma}|>30\mbox{{~MeV)}}\,,
\label{eq:CC3}\\
&&
\mbox{{No}}~\pi^0,~K_{S},~K_{L}~\mbox{{and}}~K^\pm\,.
\label{eq:CC4}
\end{eqnarray}  
\end{subequations}
The lower limit of the lepton momentum in the first criterion,
(\ref{eq:CC1}), is from the threshold of the
water \cerenkov\, detector for muons.
$\pi^\pm$ with $|p|>200$~MeV and
$\gamma$ with $|p|>30$~MeV as well as
$\pi^0, K_{S}, K_{L}$ and $K^\pm$ (which are assumed to
 decay inside a detector)
give rise to additional rings.
Events with such additional 
rings are not selected as the CCQE events and are removed.
 The survived events after imposing the selection cuts consist of the
  genuine CCQE
  events and the other charged-current events (non-CCQE events).
Some of the non-CCQE events arise from single soft $\pi^{\pm}$ emission via the $\Delta$ resonance~\cite{Hagiwara:2009bb}.
We parameterize the CCQE and non-CCQE cross sections for target nuclei after imposing
the selection criteria \eqref{eq:criteriaCC} and summarize them in
Appendix~\ref{sec:appA0}.  

The smearing effects due to the Fermi motion and
detector resolution shown in Table~\ref{tab:resolution} are taken into account by smearing
functions.
\begin{table}[t]
\centering
  \begin{tabular}{lcc}
   \hline\hline\addlinespace[2 pt]   
   & ${\delta p}/{p}~~(\%)^{}_{}$& $\delta \theta$ (degree)\\
   \hline
   $\mu$ & $1.7 +0.7/{\sqrt{p\mbox{[GeV]}}}$ & $1.8^\circ$ \\
   $e$ & $0.6+2.6/{\sqrt{p\mbox{[GeV]}}}$ & $3.0^\circ$ \\
   \hline\hline
  \end{tabular}
 \caption{The momentum and angular resolutions 
for muons and electrons at the SK detector~\cite{Ashie:2005ik}.}
 \label{tab:resolution}
\end{table}
We made fitting formulae of the anti-neutrino smearing functions for numerical
simulations and show them in Figs.~\ref{fig:smearfunc_a} for incoming
anti-neutrino energy of 1 and 2 GeV.
The explicit expressions of the anti-neutrino smearing functions are found in 
 Appendix~\ref{sec:appA}. Those for neutrinos are in Ref.~\cite{Hagiwara:2009bb}.
\begin{figure}[t]
 \centering
\resizebox{1.0\textwidth}{!}{
\includegraphics[width=0.5\textwidth]{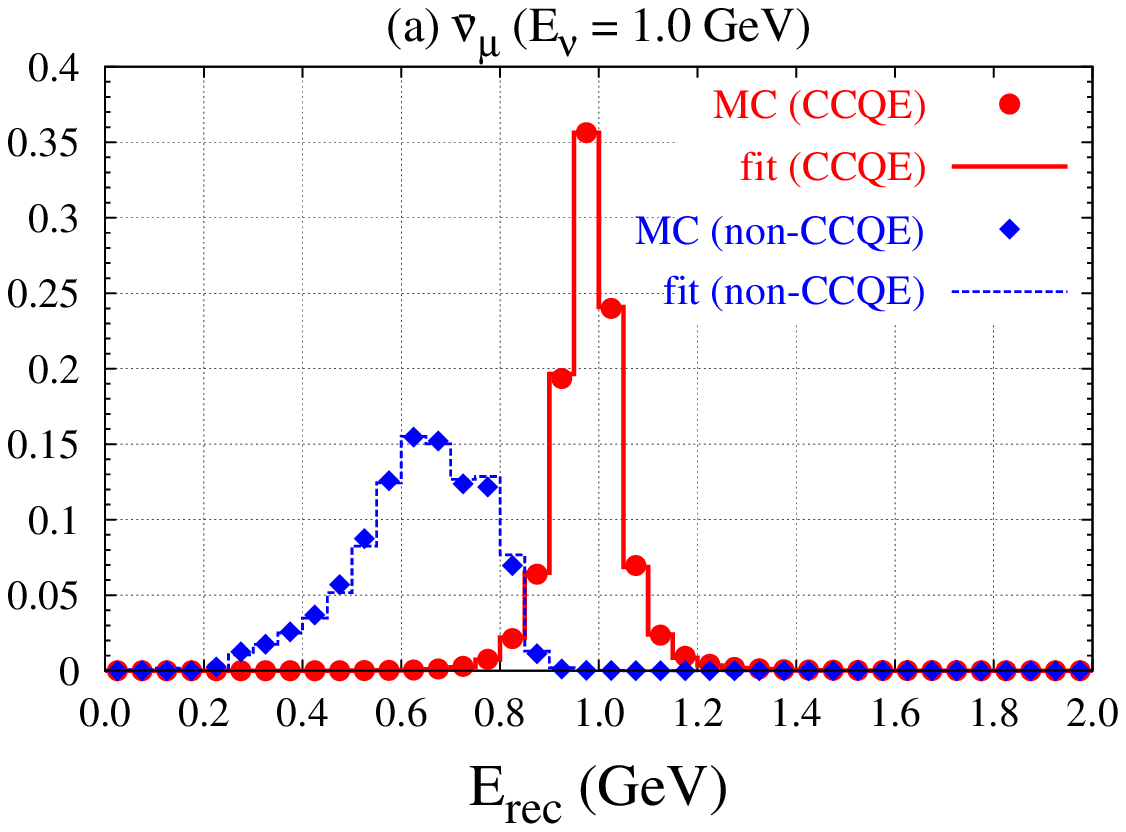} 
\includegraphics[width=0.5\textwidth]{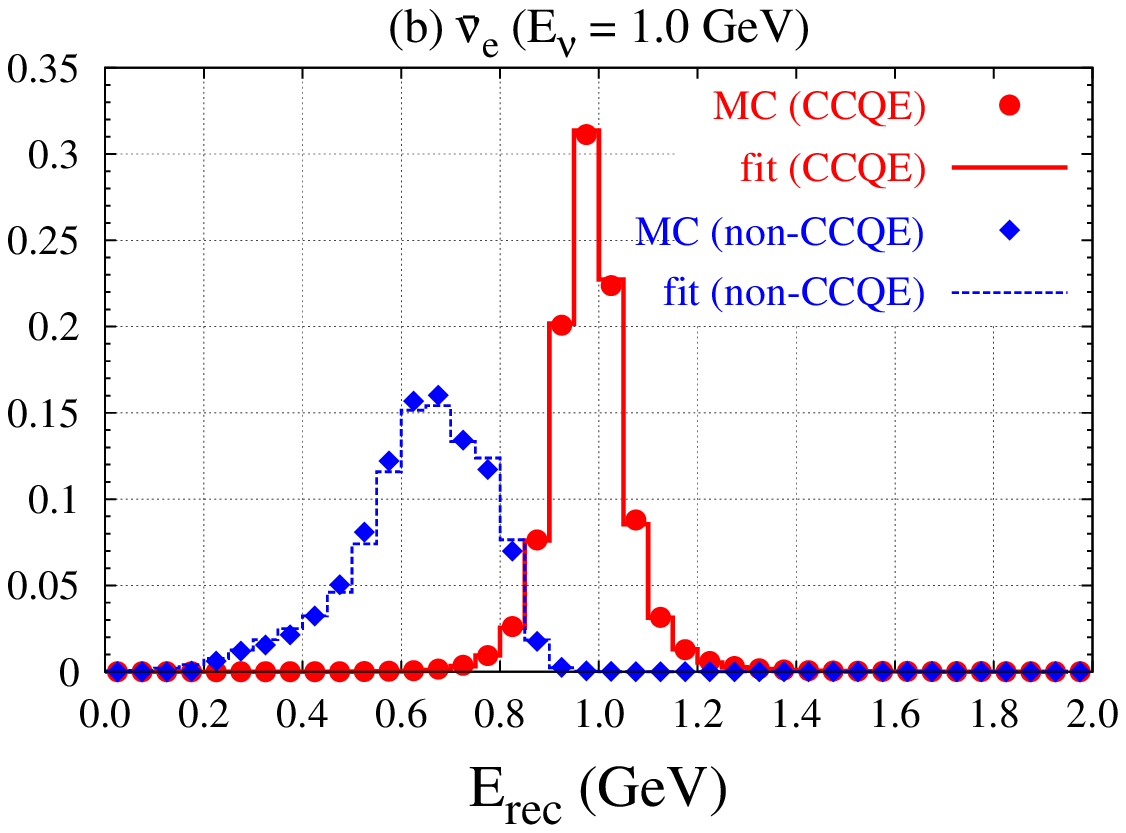} 
}
\resizebox{1.0\textwidth}{!}{
\includegraphics[width=0.5\textwidth]{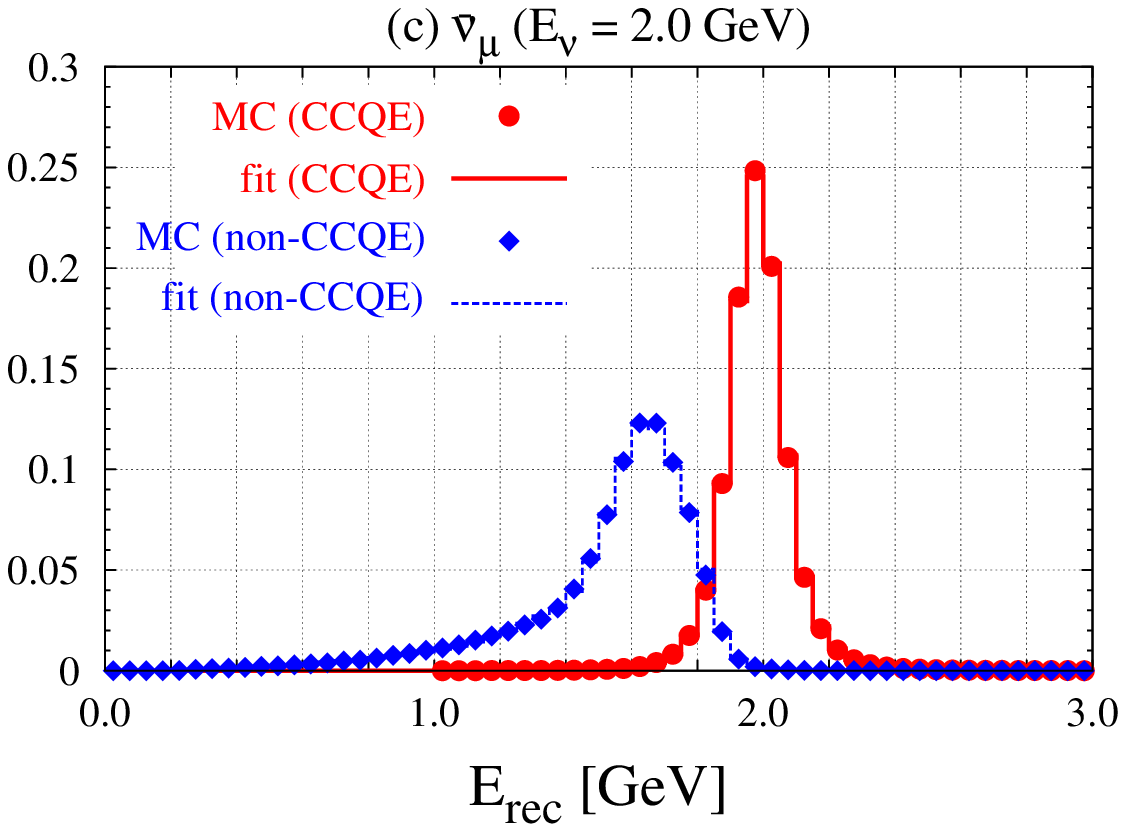} 
\includegraphics[width=0.5\textwidth]{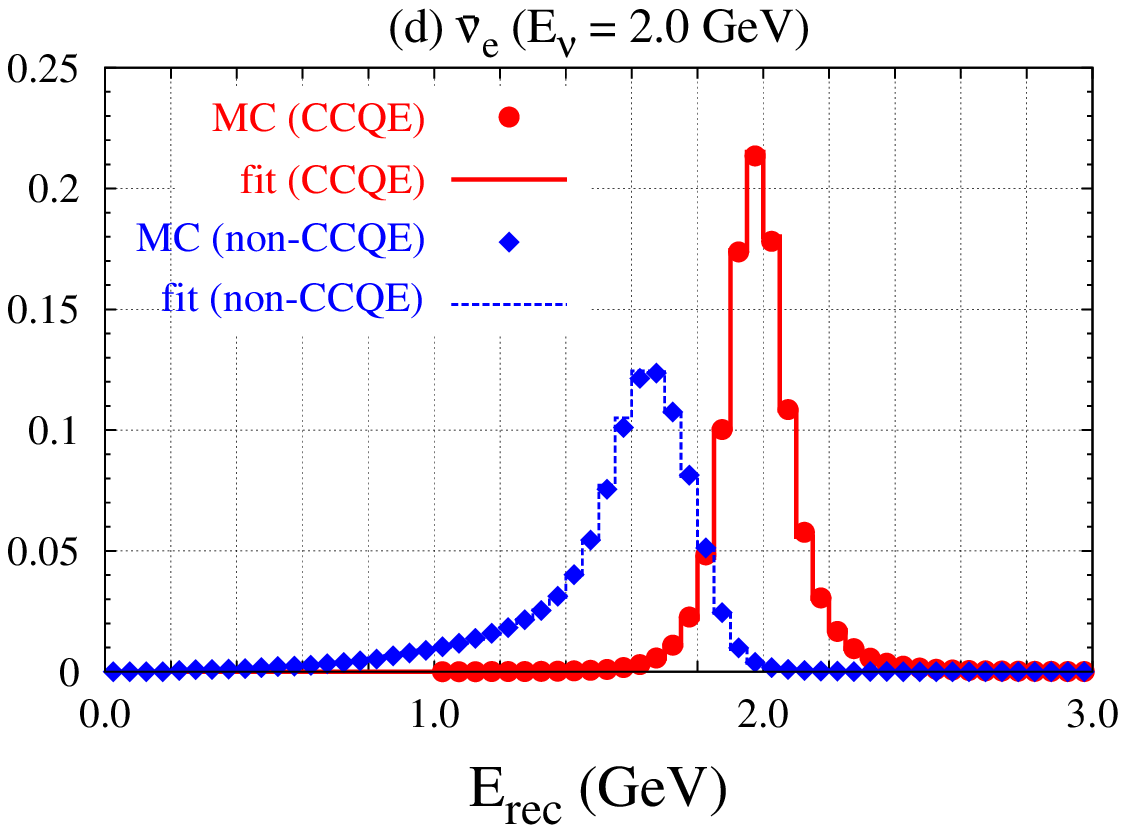} 
}
\caption{The normalized reconstructed energy distributions of CCQE (solid
   circles) and non-CCQE (solid diamonds) events initiated by
  monocromatic anti-neutrinos: (a) $\bar{\nu}_\mu$ and (b) $\bar{\nu}_e$ with $E_\nu = 1$ GeV; (c) $\bar{\nu}_\mu$ and (d)
  $\bar{\nu}_e$ with $E_\nu = 2$ GeV. Those events are generated by
  \nuancev~\cite{Casper:2002sd}, imposing the CCQE
   selection cuts~\eqref{eq:criteriaCC} and applying the detector
   resolutions in Table~\ref{tab:resolution}
  to the produced muons and electrons. The solid-red and dotted-blue histograms show the
   fitting formulae of the CCQE and non-CCQE smearing functions, respectively.}
\label{fig:smearfunc_a}
\end{figure}
The red circles show 
simulated distributions for genuine CCQE interactions,
while the red
histograms show the distributions based on the fitting formulae.
The blue diamonds and
histograms are for non-CCQE interactions. We see that the fitting
formulae describe the simulated distributions well. 
Anti-neutrinos can interact with protons in hydrogens, in addition to
Oxygens.
Thus, the reconstructed energy distibutions
show sharper peak than those
for neutrinos since protons in hydrogens do not have the Fermi motion and are almost
at rest for anti-neutrinos with ${\cal O}(1)$ GeV energy. (The smearing
functions for neutrinos are shown in Fig.3 in
Ref.~\cite{Hagiwara:2009bb}.) 

\subsection{Background events}
\label{sec:bg}
In this section we discuss the sources of background events taken into account in
this study: neutral-current (NC) single-$\pi^0$ events,
secondary neutrinos in the $\nu_\mu$ and $\bar{\nu}_\mu$ focusing beams and
misidentified muon and electron events. 

NC single-$\pi^0$ events can be a substantial background source for
the $\nu_e$ and $\bar{\nu}_e$ appearance modes, where one of
the photons from a $\pi^0$ decay is lost, or the produced $\pi^0$ is so
energetic that it decays to unresolved photons, mimicking an electron ring.
NC neutrino-nucleus scatterings occurs through the quasi-elastic (NCQE),
resonant $\pi^0$ production (NCRes), coherent $\pi^0$ production (NCCoh) or deep inelastic (NCDI)
scatterings.
The number of the NC single-$\pi^0$ events in the $i_{\rm th}$ energy bin,
$E^i_{\rm rec} < E < E^{i+1}_{\rm rec}$, at the water \cerenkov\,
detector $D$ (= SK, Oki, Kr) via the $Y$-type
neutrino-nucleus interaction ($Y=$ NCQE, NCRes, NCCoh and NCDI) induced by the $\nu_\alpha$ component
of the $\nu_\mu$ or $\nubar_\mu$ focusing beam are calculated as
\begin{align}
N^{i,Y}_{\pi^0, D}&(\nu_\alpha) = 
\int^{E^{i+1}_{\rm rec}}_{E^i_{\rm rec}} dE_{\rm rec} \int_0^{\infty} dE_\nu \,
 \Phi_{\nu_\alpha}^{D}(E_\nu)
 \sum_{Z=H,O} N_Z \,\hat{\sigma}_Z^Y(E_\nu)~ 
 S^Y_Z(E_\nu, E_{\rm rec}),
\label{eq:Npi0}
\end{align}
where $E^i_{\rm rec} = 0.05\,{\rm GeV} \times i$, and $\hat{\sigma}^Y_Z$ and $S^Y_Z$ are the
cross section and smearing function for the $Y$-type $\nu$--$Z$ interaction after
imposing the NC single-$\pi^0$ selection
criteria:
\begin{subequations} 
\label{eq:criteriaNC}
\begin{eqnarray}
&&
\mbox{{No charged leptons,}}
\label{eq:NC1}\\
&&
\mbox{{Only one $\pi^0$,}}\\
&&
\mbox{{No high energy}}~\pi^\pm~(|p_{\pi^\pm}|>200\mbox{{~MeV}}),
\label{eq:NC2}\\
&&
\mbox{{No high energy}}~\gamma~(|p_{\gamma}|>30\mbox{{~MeV)}},
\label{eq:NC3}\\
&&
\mbox{{No}}~K_{S},~K_{L}~\mbox{{and}}~K^\pm\,.
\label{eq:NC4}
\end{eqnarray}  
\end{subequations}
The first condition, Eq.~(\ref{eq:NC1}), selects NC events, while the
other conditions eliminate multi-ring events.
The $\pi^0$ momentum distributions from the $\nu_\mu$ focusing beams
after imposing the above criteria are shown in Fig.~\ref{fig:pi0mom} for various
off-axis beam angles. 
The NC single-$\pi^0$ events are
produced more with smaller off-axis beam angle because the fluxes of such
neutrino beams are distributed in higher energy region as shown in
Fig.~\ref{fig:flux1}.
This is a disadvantage of using neutrino beams with smaller off-axis
angles at a far detector, and a low $\pi^0$ misidentification probability
is needed especially for the CP phase measurements. 
\begin{figure}[t]
\centering
\resizebox{1.0\textwidth}{!}{
 \includegraphics[width=0.5\textwidth]{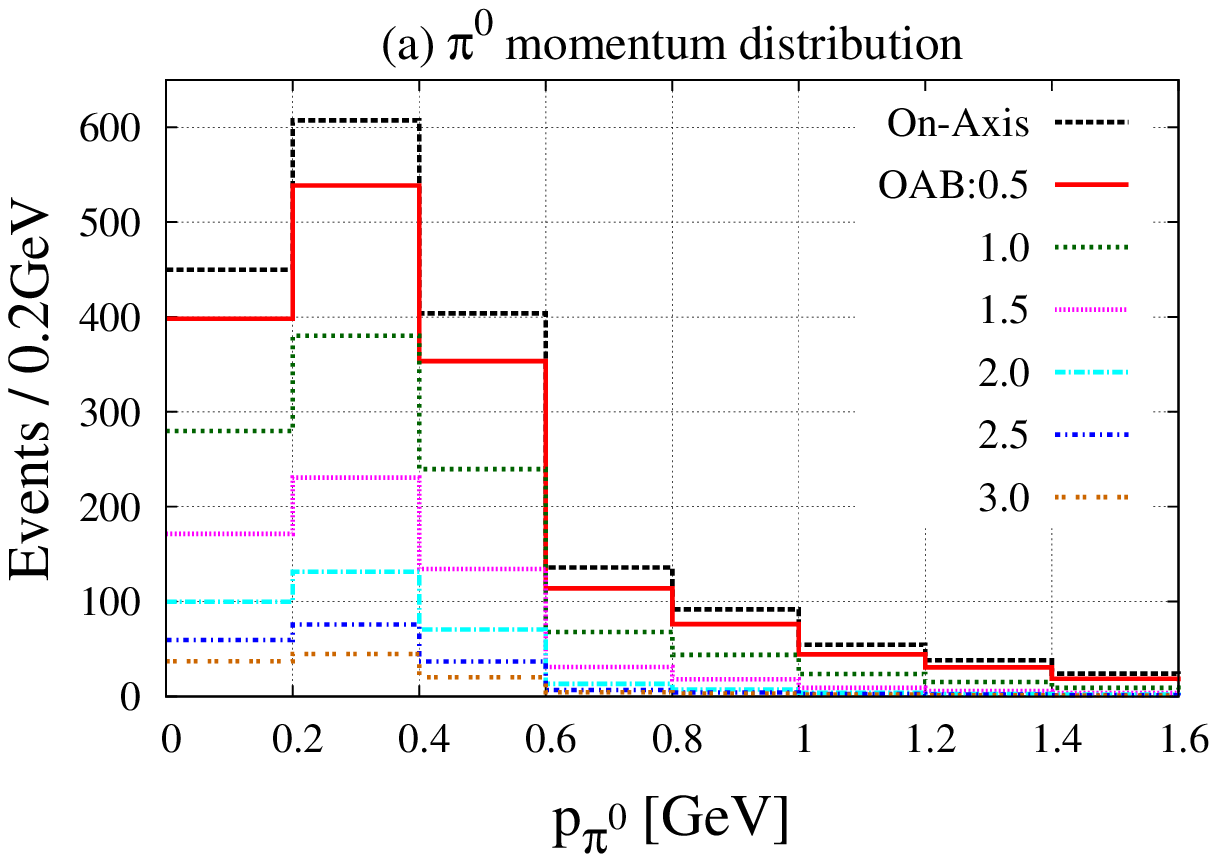} 
  \includegraphics[width=0.5\textwidth]{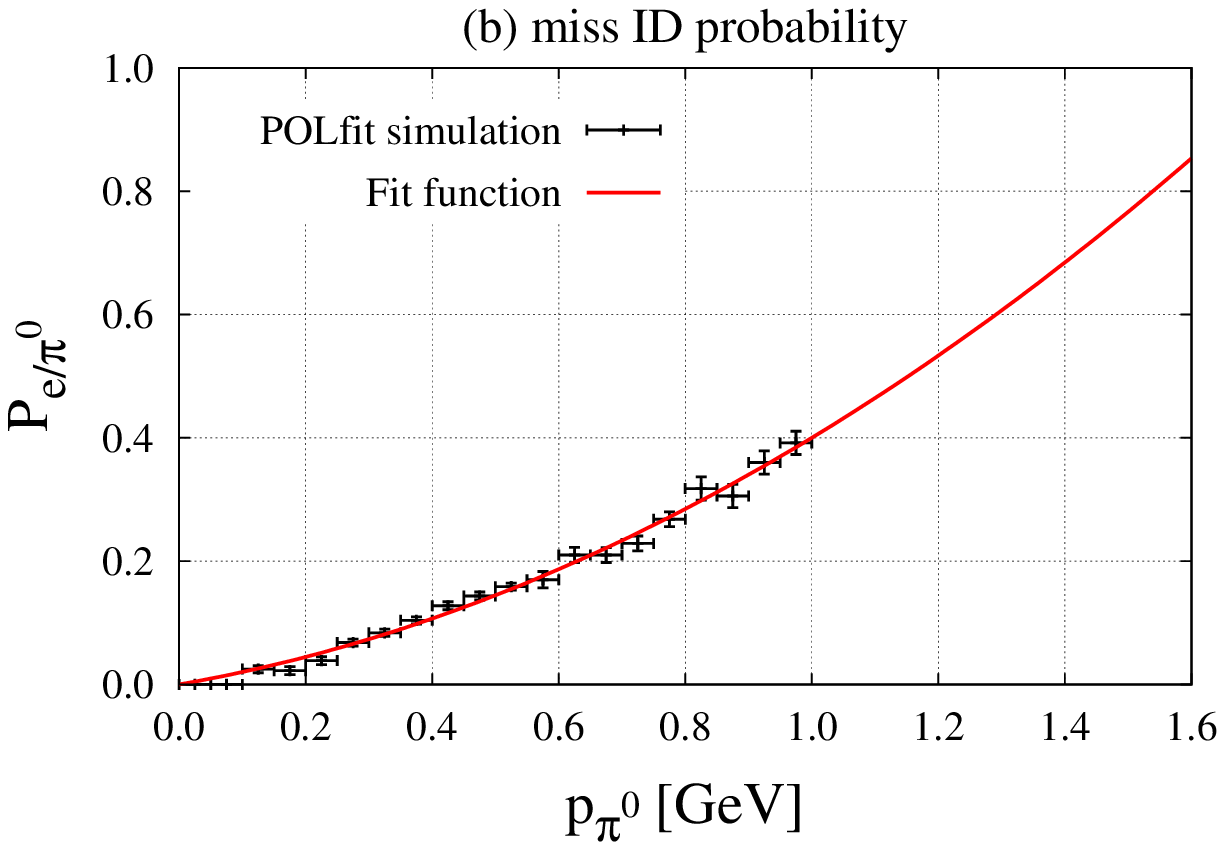} 
}
\caption{(a) The $\pi^0$ momentum distributions of NC single-$\pi^0$ events
 selected by the criteria (\ref{eq:criteriaNC}) for various off-axis
 angles at a far detector. The event numbers are obtained with a 100 kton water target at
 $1000$ km away from the J-PARC and the $\nu_\mu$ focusing beam flux corresponding to
 $5\times 10^{21}$ POT with the proton energy of 40 GeV. 
(b) The probability of misidentifying a $\pi^0$ to an $e^{\pm}$ as a
 function of $\pi^0$ momentum, based on the \polfit algorithm
 \cite{Barszczak:2005sf}. The black points show simulated
 data by the T2K collaboration~\cite{Plfit_okumura}, and the red curve
 show the fitted function to the data, Eq.~\eqref{eq:polfitfunc}.}
\label{fig:pi0mom}
\end{figure}
We parameterize the $\pi^0$ misidentification probability, $P_{e/\pi^0}$,
used in our analysis as
a function of $\pi^0$ momentum $x$ [GeV] as
\begin{align}
P_{e/\pi^0} (x) =& \,a x (x+b), \label{eq:polfitfunc}\\
a =& \,0.222 \,\,[1/{\rm GeV}^2],\nn\\
b =& \,0.802 \,\,[{\rm GeV}],
\end{align}
 based on the simulation
\cite{Plfit_okumura} of the \polfit $\pi^0$-rejection algorithm\footnote{Recently,
 more efficient $\pi^0$ rejection algorithm has been developed by the T2K collaboration~\cite{Abe:2015awa}, and our NC $\pi^0$ background
 estimation may be regarded as a conservative one.}
\cite{Barszczak:2005sf}.
The reference data and the fitted
function are shown in Fig.~\ref{fig:pi0mom}(b).
The misidentification probability is kept less than 0.2 for $p_{\pi^0} < 0.6$ GeV, where
the $\pi^0$ backgrounds mostly distributes. 
The background events are then selected from the
simulated NC single-$\pi^0$ events according to the misidentification
probability, and the reconstructed energy of each background event is calculated with Eq.~(\ref{eq:erec1}), assuming the misidentified $\pi^0$ as an electron. 

The NC single-$\pi^0$ backgrounds significantly affect the sensitivity to
the mass hierarchy and CP phase
\cite{Hagiwara:2009bb}, and it is important to include their uncertainty properly in
our analyses. 
One of the
major uncertainty sources of the NC single-$\pi^0$ backgrounds is
modeling of 
  neutrino-nucleus interactions. 
In Fig.~\ref{fig:mAunc} we show the reconstructed
 energy distributions of the NC
 single-$\pi^0$ backgrounds calculated with \nuance\, for different
 neutrino-nucleus interactions. We see that the $\pi^0$ backgrounds
 mainly distribute in low energy region, where the contributions from
 resonant and coherent single-$\pi^0$ production processes
 dominate. These processes are implemented in \nuance\, based on the
 Rein-Sehgal's calculations \cite{Rein:1980wg,Rein:1982pf}.
 Among the modeling parameters of the NC neutrino-nucleus interactions, axial
 form-factor masses ($m_A$) have
 not been measured accurately.
 Therefore, we vary the axial masses of
 the resonant and coherent single-pion production processes within their
 uncertainties: $m_A^{\rm Res} = 1.1
 \pm 0.11$ GeV \cite{A.KabothfortheT2K:2013fva} and $m_A^{\rm Coh} = 1.03 \pm 0.28$ GeV
 \cite{AguilarArevalo:2010zc}.
As shown in Fig.~\ref{fig:mAunc}, those uncertainties can be well approximated by
13\% and 15\% normalization uncertainties for the NC resonant and coherent
single-$\pi^0$ backgrounds, respectively.

\begin{figure}[t]
\centering
\resizebox{1.0\textwidth}{!}{
  \includegraphics[width=0.5\textwidth]{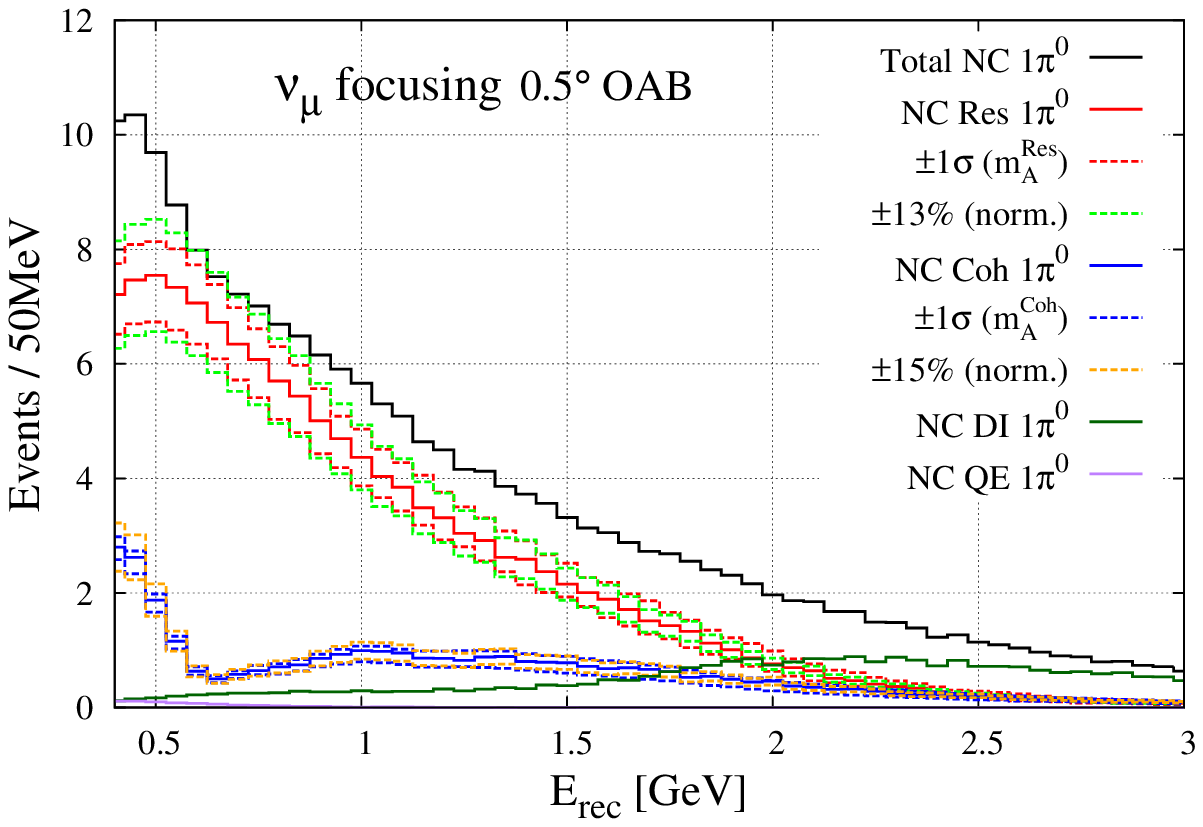} 
  \includegraphics[width=0.5\textwidth]{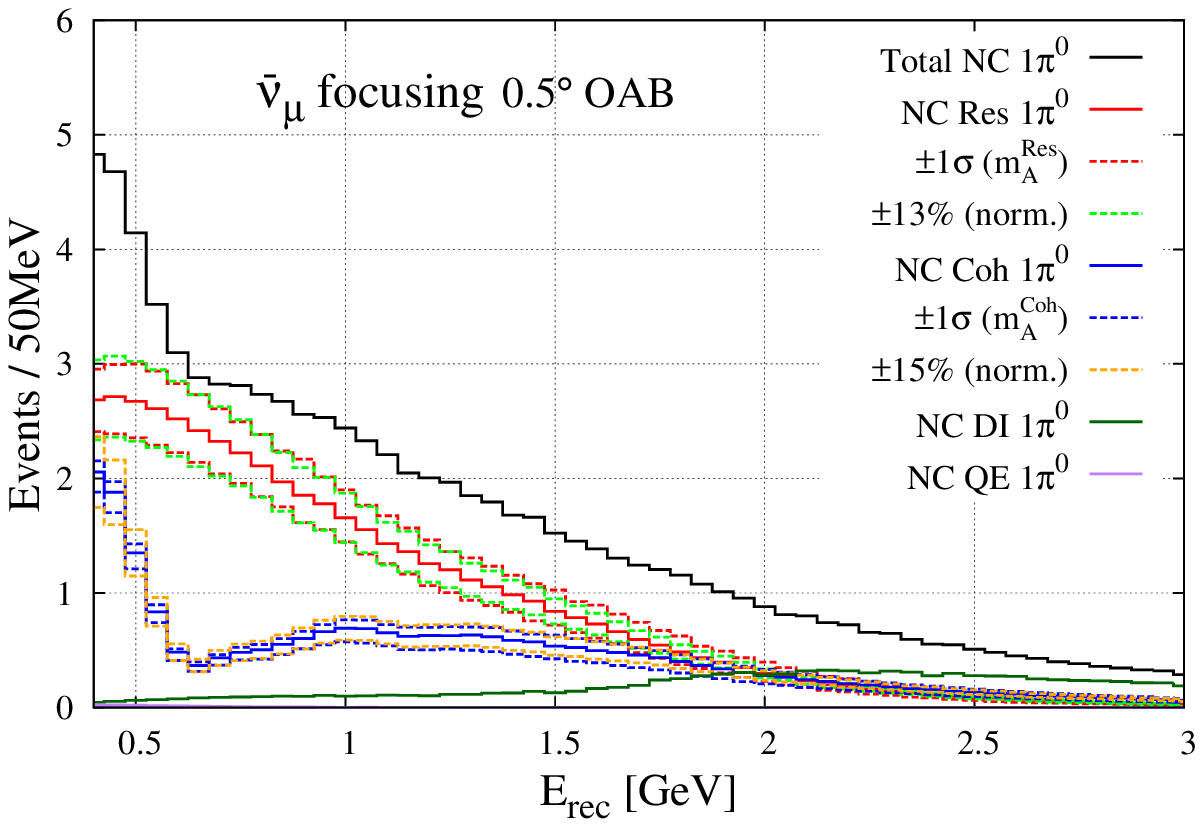} 
  }
\caption{The reconstructed energy distributions of the NC single-$\pi^0$
 backgrounds for the $\nu_\mu$ (left) and $\bar{\nu}_\mu$ (right)
 focusing beams. The solid-red and solid-blue histograms show the
 NC resonant and coherent single-$\pi^0$ components calculated with the
 axial masses ($m_A^{\rm Res}$ and $m_A^{\rm Coh}$) of 1.1 and 1.03 GeV,
 respectively. The dashed-red and dashed-blue histograms show the
  1$\sigma$ uncertainty ranges of the resonant
 and coherent axial masses, respectively.
The solid-green and purple
  histograms show the NC deep-inelastic (DI) and quasi-elastic (QE)
 single-$\pi^0$ components, and the black histogram is for the total
  NC single-$\pi^0$ backgrounds. 
Event numbers are calculated for a 100 kton detector using the
 $0.5^\circ$ off-axis beam with the $\nu_\mu\, (\bar{\nu}_\mu)$ flux
 corresponding to $5\times 10^{21}$ POT with the proton energy of 40 GeV.
  } 
\label{fig:mAunc}
\end{figure}

Another major source of
 uncertainty of the NC single-$\pi^0$ backgrounds arises from 
the $\pi^0$ misidentification probability,
Eq.~(\ref{eq:polfitfunc}). The T2K collaboration estimated 10.8\% uncertainty in the NC-$\pi^0$ background estimation due to the \polfit algorithm \cite{Plfit_okumura}.
 Since our modeling of the $\pi^0$ misidentification
probability is based on the \polfit algorithm, we assign $11\%$
uncertainty to the normalization of the NC single-$\pi^0$ backgrounds due to the $\pi^0$ misidentification. 

All in all, we include the 13\% and 15\% normalization
uncertainties for the NC resonant and coherent single-$\pi^0$
backgrounds, respectively, and 11\% normalization uncertainty for the total NC
single-$\pi^0$ backgrounds. This treatment allows independent
normalization corrections for the resonant and coherent
NC single-$\pi^0$ backgrounds.
 
The $\nu_\mu\,(\bar{\nu}_\mu)$ focusing beams contain not only $\nu_{\mu}\,(\bar{\nu}_{\mu})$ but also
other neutrino flavors, $\nu_e$, $\bar{\nu}_e$ and $\bar{\nu}_\mu\,
(\nu_\mu)$, secondary neutrino beams. Especially, for the $\nu_\mu
\rightarrow \nu_e$ and $\bar{\nu}_\mu \rightarrow \bar{\nu}_e$ oscillation modes,
the $\nu_e$ and $\bar{\nu}_e$ secondary beams become major background
sources. We simulate these secondary-neutrino events in the same way as the signal
events described in Sec.~\ref{sec:signal}.    

There is also some probability of misidentifying a muon (electron) \cerenkov\, ring as an electron
(muon),
$P_{e/\mu}$
($P_{\mu/e}$). Although these probabilities depend on the detector design and
performance, we assume the same probabilities for the SK and a far detector in Oki and
Korea as
\begin{equation}
P_{e/\mu} = P_{\mu/e} = 1\pm 1\%.
\end{equation}     

\section{$\chi^2$ analysis}
\label{sec:chi2}
Using the simulated signal and background events, we estimate the sensitivity of the T2KK and T2KO experiments to the
mass hierarchy and CP phase ($\delta_{\rm CP}$), performing the
$\chi^2$ analysis. The $\chi^2$ function used in this study can be
written as
\begin{eqnarray}
\chi^2 \equiv
  \chi^2_{\rm SK} 
+ \chi^2_{\rm Oki/Kr}
+ \chi^2_{\rm sys}
+ \chi^2_{\rm para}.
\label{eq:def_chi^2}
\end{eqnarray}
The first two terms
measure deviations of data from the theoretical predictions
at the SK and a far detector in Oki or Korea,
\begin{align}
 \chi^2_{D}
 =& \sum_{i} \left\{
\left(
\dfrac
{(N_{\mu,D}^{i})^{\rm fit} - (N_{\mu,D}^{i})^{\rm input}}
{\sqrt{(N^i_{\mu,D})^{\rm fit}}}
\right)^2
+
\left(
\dfrac
{(\overline{N}_{\mu,D}^{i})^{\rm fit} - (\overline{N}_{\mu,D}^{i})^{\rm input}}
{\sqrt{(\overline{N}^i_{\mu,D})^{\rm fit}}}
\right)^2 \right.\nn\\
&\left.\hspace{1.5em}+
\left(
\dfrac
{(N_{e,D}^{i})^{\rm fit} - (N_{e,D}^{i})^{\rm input}}
{\sqrt{(N^i_{e,D})^{\rm fit}}}
\right)^2
+
\left(
\dfrac
{(\overline{N}_{e,D}^{i})^{\rm fit} - (\overline{N}_{e,D}^{i})^{\rm input}}
{\sqrt{(\overline{N}^i_{e,D})^{\rm fit}}}
\right)^2
\right\},
\label{eq:chi_N}
\end{align}
 where 
$(N_{\mu,D}^{i})^{\rm input}$ and 
$(N_{e,D}^{i})^{\rm input}$
denote the $\mu$- and $e$-like event numbers, respectively, in the $i$-th bin of
the $\Erec$ distributions measured at a detector $D\,(=$ SK, Oki, Kr)
from the $\nu_{\mu}$ focusing beam, and $(\overline{N}_{\mu,D}^{i})^{\rm input}$ and 
$(\overline{N}_{e,D}^{i})^{\rm input}$ are those from the
$\bar{\nu}_{\mu}$ focusing
beam. 
The summation runs over all the $E_{\rm rec}$ bins
from $0.4$~GeV to $5.0$~GeV at both the SK and a far (Oki or Kr) detectors.

The $\mu$- and $e$-like event numbers are calculated using the CC
signal and NC single-$\pi^0$ background events ($N^{i,X}_D$ and
$N^{i,Y}_{\pi^0,D}$ defined by Eqs.~\eqref{eq:N} and \eqref{eq:Npi0}) as
\begin{subequations} 
 \begin{align}
\left(N_{\mu,D}^{i}\right)^{\rm input}
=&\,
\left(
1-P_{e/\mu}^D
\right)\effm^D
  \sum_{X,\nu_\alpha}
  \left\{ N_D^{i,X}(\nu_\alpha\!\rightarrow\!\nu_\mu) +N_D^{i,X}(\nu_\alpha\!\rightarrow\!\nubar_\mu)\right\} \nn\\
&\hspace{4.3em}+
P_{\mu/e}^D \,\effe^D
  \sum_{X,\nu_\alpha}
  \left\{ N_D^{i,X}(\nu_\alpha\!\rightarrow\!\nu_e)
+N_D^{i,X}(\nu_\alpha\!\rightarrow\!\nubar_e) \right\},
\label{eq:Nm}\\
\left(N_{e,D}^{i}\right)^{\rm input}
=&\,
P_{e/\mu}^D \,\effm^D
  \sum_{X,\nu_\alpha}
    \left\{ N_D^{i,X}(\nu_\alpha\!\rightarrow\!\nu_\mu) 
  +N_D^{i,X}(\nu_\alpha\!\rightarrow\!\nubar_\mu)\right\} \nn\\
&+
(1-P_{\mu/e}^D) \,\effe^D
  \sum_{X,\nu_\alpha}
   \left\{ N_D^{i,X}(\nu_\alpha\!\rightarrow\!\nu_e)
+N_D^{i,X}(\nu_\alpha\!\rightarrow\!\nubar_e) \right\} \nn\\
&+\sum_{Y,\nu_\alpha} N^{i,Y}_{\pi^0,D}(\nu_\alpha),
\label{eq:Ne}
 \end{align}
\label{eq:Nme}
\end{subequations}
where $\effm^D\,(\effe^D)$ is
the efficiency of detecting muon (electron) \cerenkov\, rings, and $P_{e/\mu}^D$ ($P_{\mu/e}^D$) is the probability of misidentifying
the detected muon (electron) \cerenkov\, ring as an electron (muon).
Detected neutrino flavors are already summed in $N^{i,Y}_{\pi^0,D}$.
The anti-neutrino event numbers $(\overline{N}^i_{\mu,D})^{\rm input}$ and 
$(\overline{N}^i_{e,D})^{\rm input}$ are calculated with similar
expressions as Eqs.~(\ref{eq:Nm}) and (\ref{eq:Ne}).
It should be noted that we neglect statistical
fluctuations in the input event numbers, and those event numbers should be considered as
averaged ones. The reconstructed energy distributions for the
$\mu$- and $e$-like events are shown in Figs.~\ref{fig:erec_sk} - \ref{fig:erec_t2kk}, which are calculated using the input parameter
values in Table~\ref{tb:parameters}.
\begin{table}[t]
\begin{center}
\begin{tabular}{lcc}
\hline\hline\addlinespace[2 pt]
Systematic parameters ($S$)  & Input value ($S_{\rm input}$) & Uncertainty
	 ($\delta S$)\\[2 pt]
\hline\addlinespace[2 pt]
Fiducial volume of detectors ($f_V^D$) & 1.00 & 0.03 \cite{Hagiwara:2009bb}\\
Neutrino flux at a detector ($f_{\nu_{\alpha}}^D)$ & 1.00 & 0.03 \cite{Hagiwara:2009bb}\\
CCQE cross sections ($f^{\rm CCQE}_{\nu_\beta}$) & 1.00 & 0.03 \cite{Hagiwara:2009bb}\\
Non-CCQE cross sections ($f^{\rm nonCCQE}_{\nu_\beta}$) & 1.00 & 0.20 \cite{Hagiwara:2009bb}\\
Misidentified NC $\pi^0$ events
 ($f_{\pi^0}^{\rm NC}$) &
     1.00 & 0.11 \\
Misidentified NC resonant $\pi^0$ events
 ($f^{\rm NCRes}_{\pi^0}$) & 1.00 & 0.13 \\
Misidentified NC coherent $\pi^0$ events ($f^{\rm NCCoh}_{\pi^0}$) & 1.00 & 0.15 \\
Detection efficiency of electron \cerenkov\, rings ($\epsilon_{e}^D)$ & 0.90 & 0.05 \cite{Hagiwara:2009bb}\\
Detection efficiency of muon \cerenkov\, rings ($\epsilon_{\mu}^D)$ & 1.00 & 0.01 \cite{Hagiwara:2009bb}\\
$\mu$-to-$e$ miss-ID probability ($P_{e/\mu}^D$) & 0.01 & 0.01 \cite{Hagiwara:2009bb}\\
$e$-to-$\mu$ miss-ID probability ($P_{\mu/e}^D$) & 0.01 & 0.01 \cite{Hagiwara:2009bb}\\
\hline\addlinespace[2 pt]
\hline\addlinespace[2 pt]
Physical parameters ($P$)  & Input value ($P_{\rm input}$) & Uncertainty
	 ($\delta P$)\\[2 pt]
\hline\addlinespace[2 pt]
$\ssq212$ & 0.875 & 0.024 \cite{Beringer:1900zz}\\
$\ssq213$ & 0.095 \cite{Beringer:1900zz} & 0.005 \cite{Zhang:2015fya}\\
$\ssq{}23$ & 0.5 & 0.1 \cite{Beringer:1900zz}\\
$\dm21\hspace{1em}{\rm [eV]}^2$ & $7.50\times 10^{-5}$ & $0.20\times 10^{-5}$ \cite{Beringer:1900zz}\\
$|\dm32|\hspace{1em}{\rm [eV]}^2$ & $2.32\times 10^{-3}$ & $0.10\times 10^{-3}$ \cite{Beringer:1900zz}\\
$\dmns$ & $0^\circ$ & - \\
$\bar{\rho}^{\rm SK}\hspace{1em}[{\rm g}/{\rm cm}^3]$ & $2.60$ & 6\% \cite{Hagiwara:2012mg}\\
$\bar{\rho}^{\rm Oki}\hspace{0.8em}[{\rm g}/{\rm cm}^3]$ & $2.75$ & 6\% \cite{Hagiwara:2012mg}\\
$\bar{\rho}^{\rm Kr}\hspace{1.1em}[{\rm g}/{\rm cm}^3]$ & $2.9$ & 6\% \cite{Hagiwara:2012mg}\\
\hline\hline
\end{tabular}
\end{center}
\caption{The systematic and physical parameters in the $\chi^2$ function,
 Eq.~(\ref{eq:def_chi^2}), where $D$ stands for the detector site (SK,
 Oki and Kr), and $\nu_\alpha$ or $\nu_\beta$ denotes neutrino species ($\nu_\mu,\bar{\nu}_\mu,\nu_e$ and $\bar{\nu}_e$). These input values and
 uncertainties are used in the sensitivity study otherwise
 mentioned. 
}
\label{tb:parameters}
\end{table}
\begin{figure}[t]
 \centering
\resizebox{0.8\textwidth}{!}{
  \includegraphics[width=0.4\textwidth]{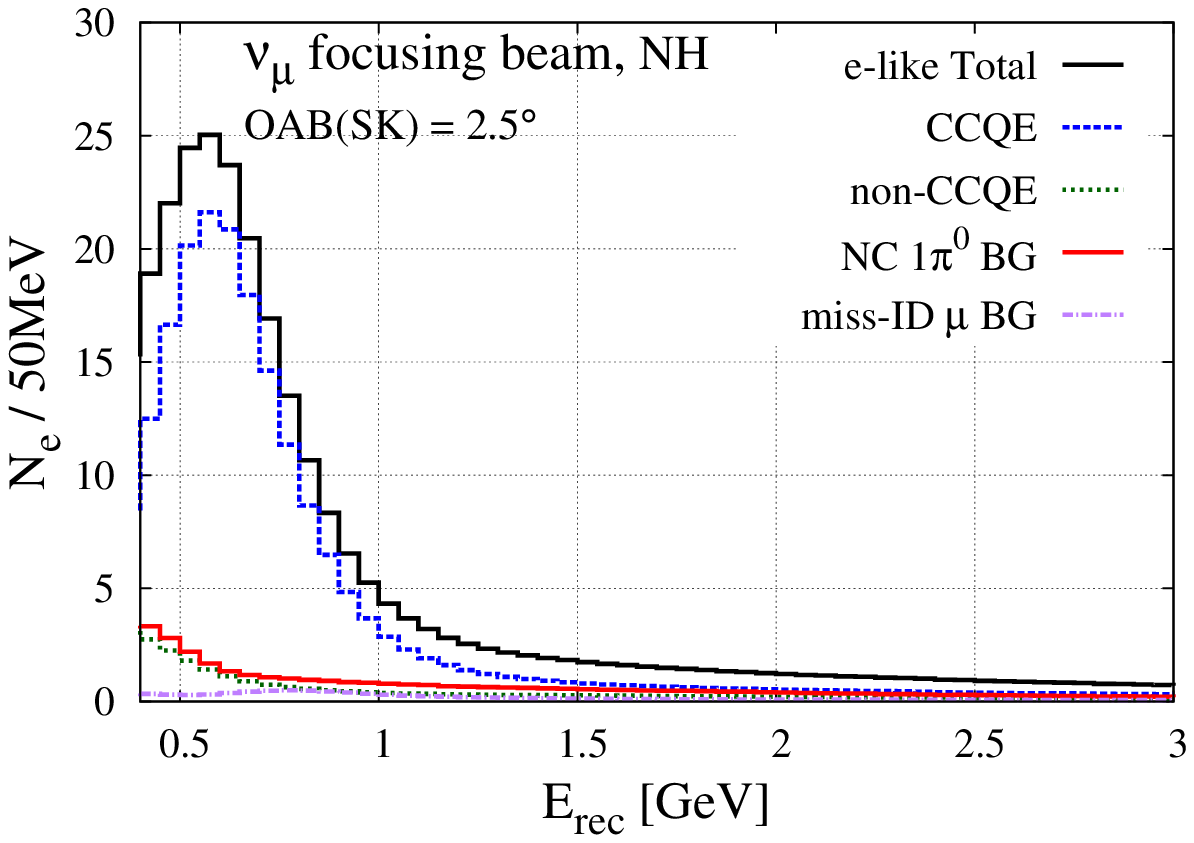}  
  \includegraphics[width=0.4\textwidth]{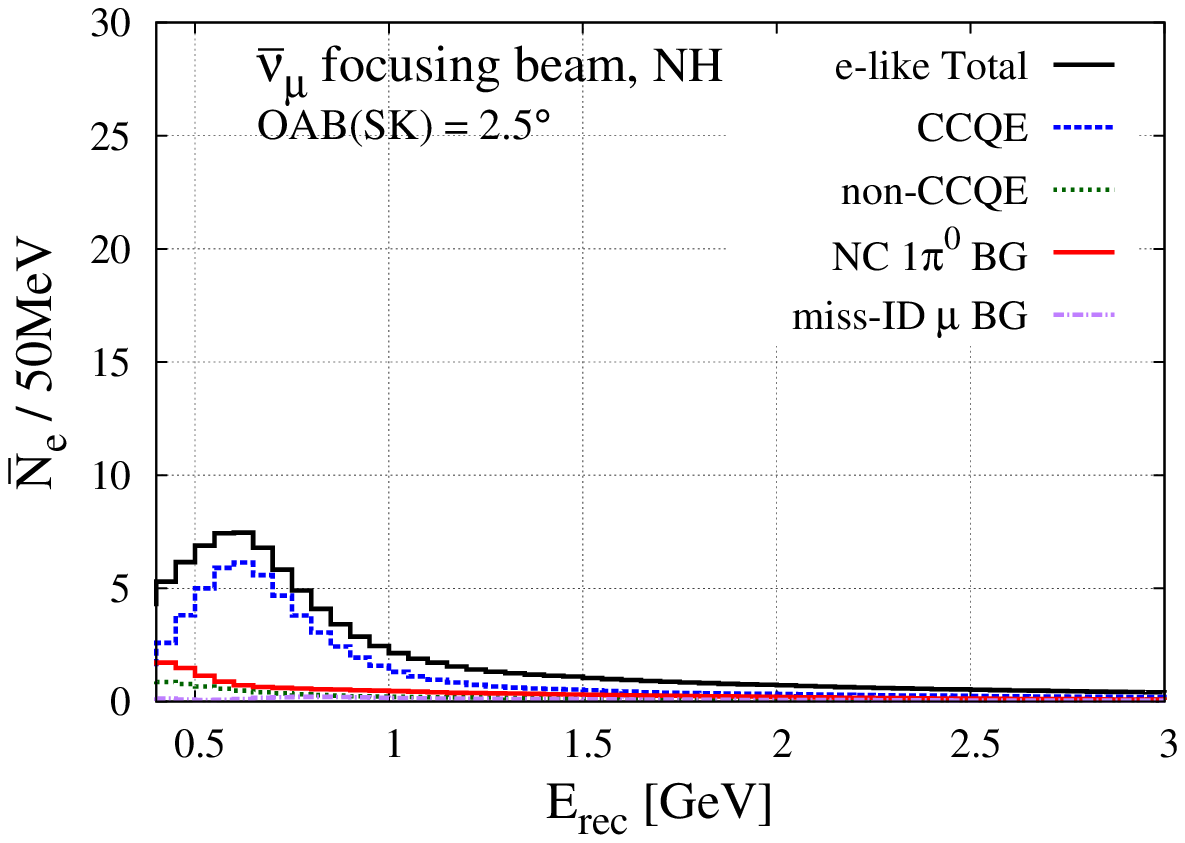}  
}
\resizebox{0.8\textwidth}{!}{
  \includegraphics[width=0.4\textwidth]{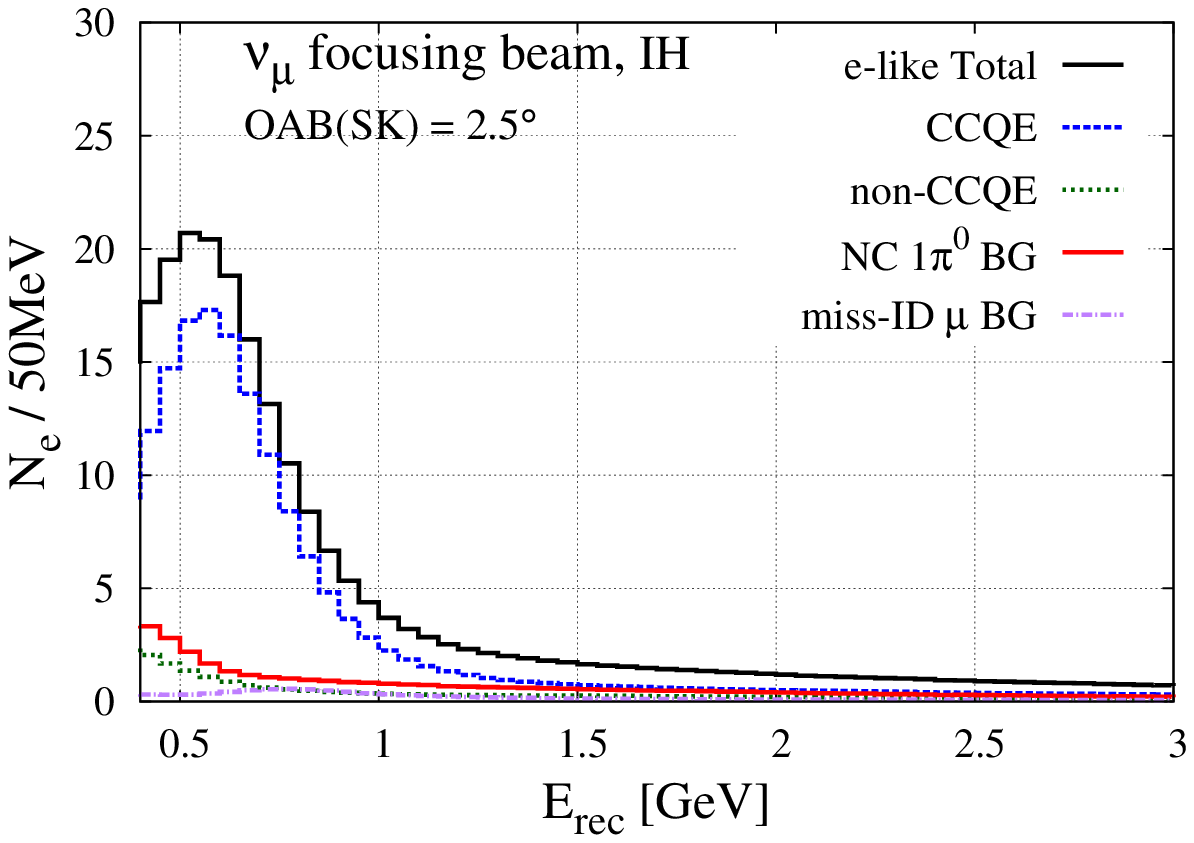}  
  \includegraphics[width=0.4\textwidth]{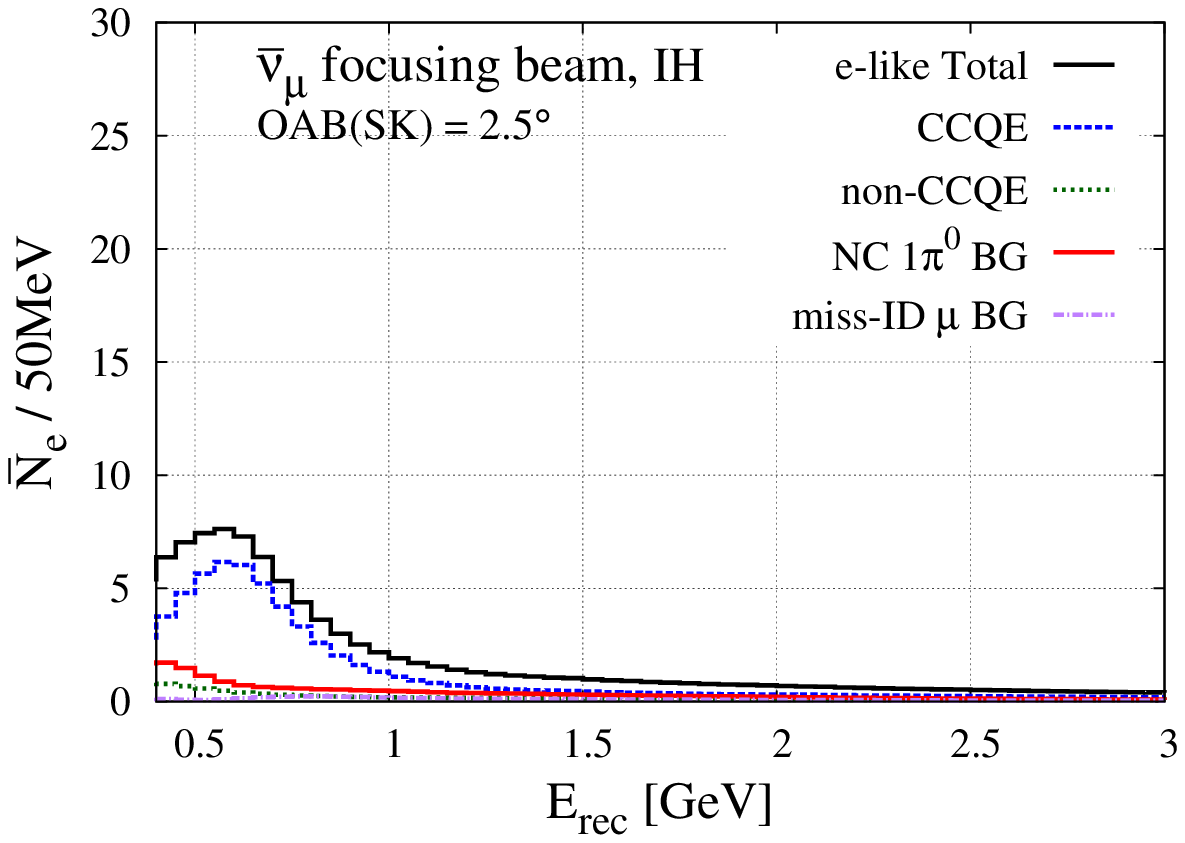}  
}
\resizebox{0.8\textwidth}{!}{
  \includegraphics[width=0.4\textwidth]{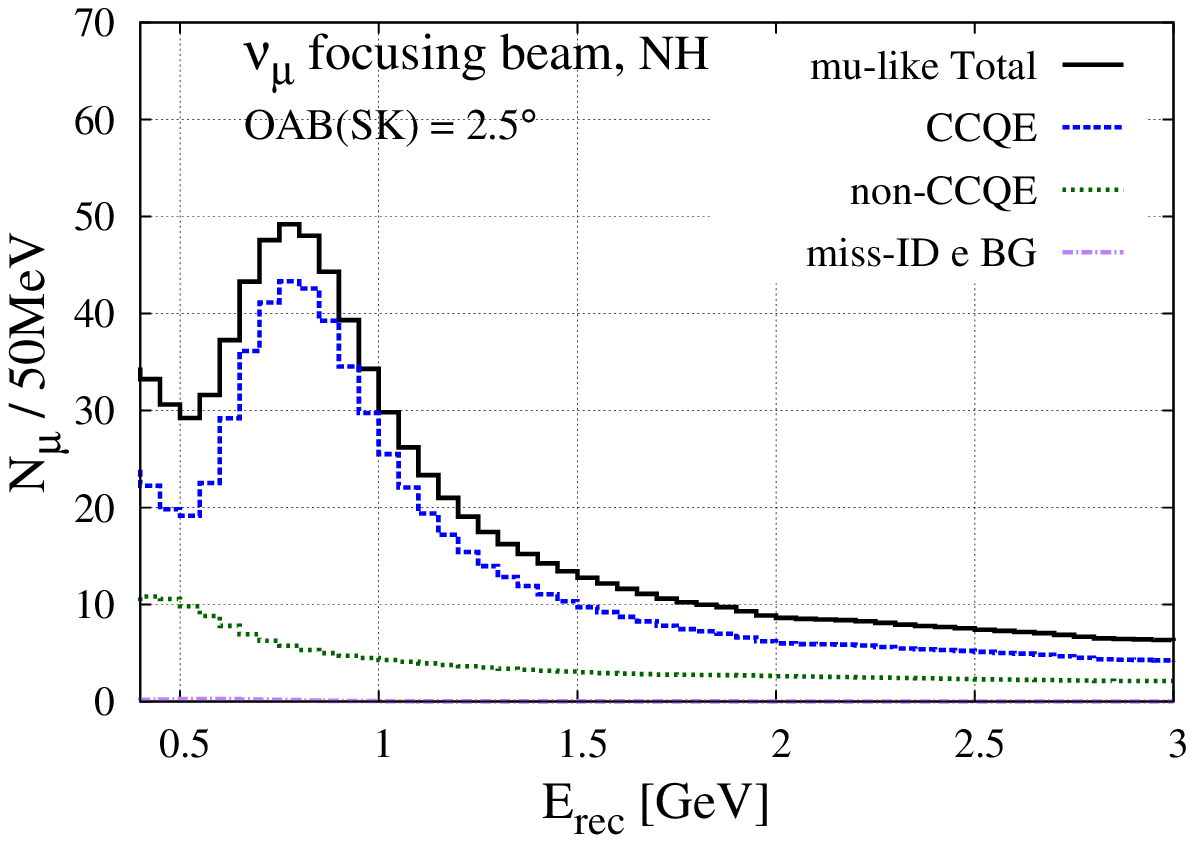}  
  \includegraphics[width=0.4\textwidth]{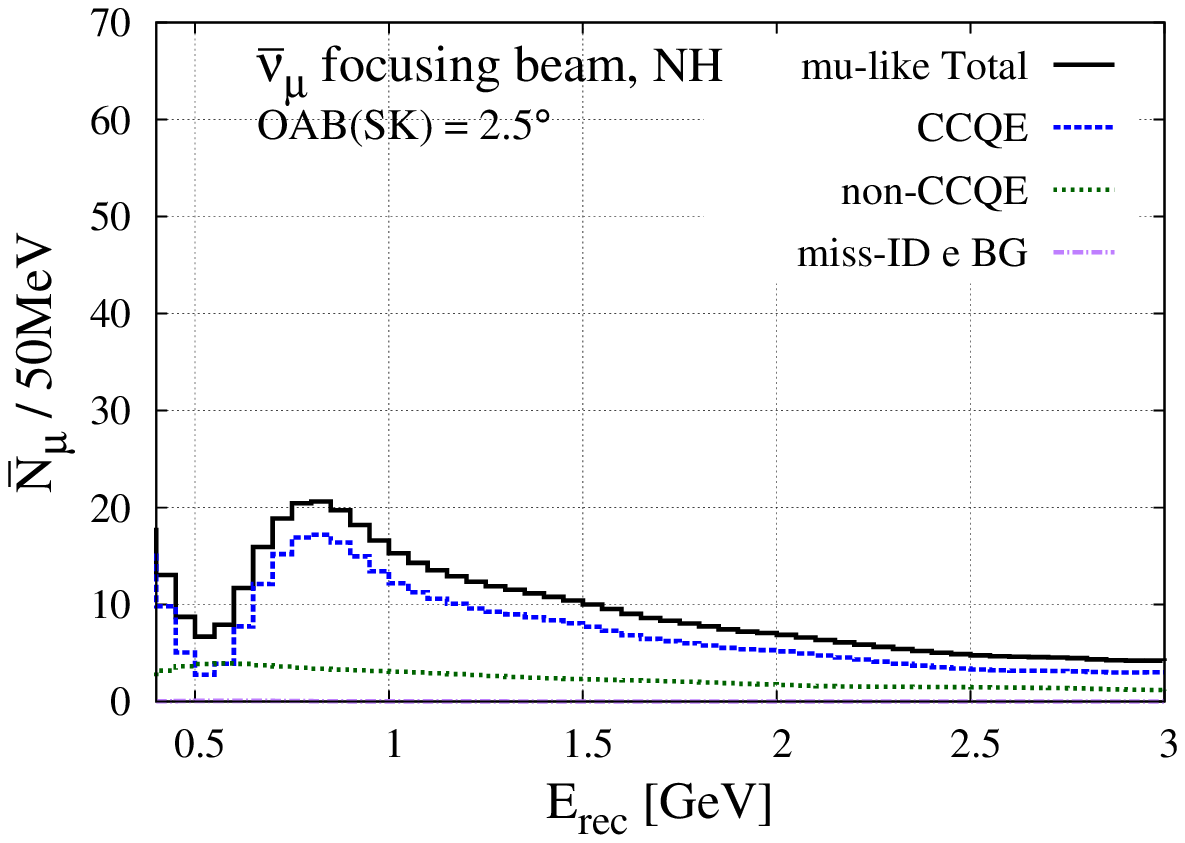}  
}
\resizebox{0.8\textwidth}{!}{
  \includegraphics[width=0.4\textwidth]{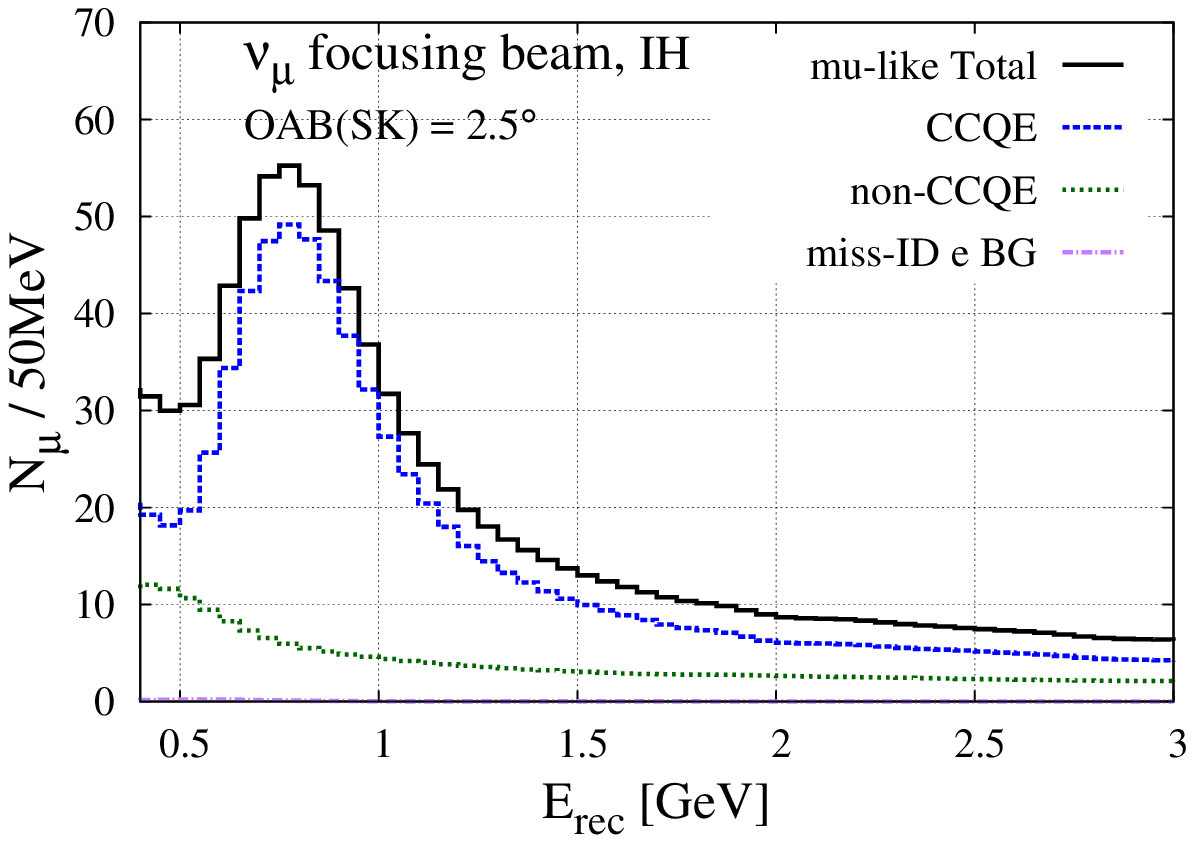}  
  \includegraphics[width=0.4\textwidth]{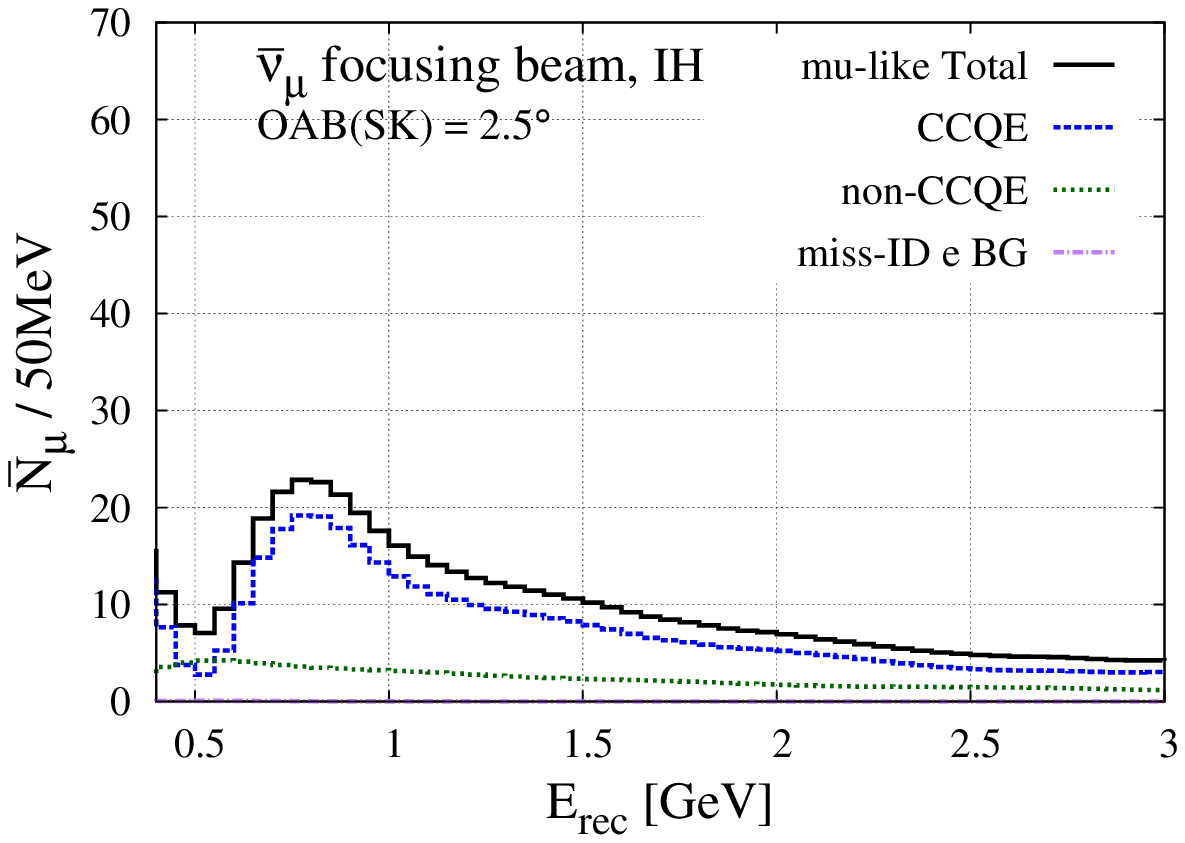}  
}
  \caption{The reconstructed energy distributions at the SK detector with
the $\nu_\mu$ (left panels) and $\bar{\nu}_\mu$ (right panels) focusing
 beams. The former four panels show distributions for $e$-like events,
 and the latter four panels are for $\mu$-like events. The 1st and 3rd
 rows are for the normal hierarchy case, while the 2nd and 4th rows are
 for the inverted hierarchy case. The dashed-blue, dotted-green, red and dash-dotted-purple
 histograms are for CCQE, non-CCQE, NC single-$\pi^0$ background and
 misidentified muon/electron background events, respectively. The black
 histogram shows the total of those contributions. The event numbers are
 calculated for T2K experiment with the $2.5^\circ$ OAB and the beam flux corresponding to $5\times 10^{21}$POT with the proton
 energy of 40 GeV.}
\label{fig:erec_sk}
\end{figure}
\clearpage
\begin{figure}[t]
 \centering
\resizebox{0.8\textwidth}{!}{
  \includegraphics[width=0.4\textwidth]{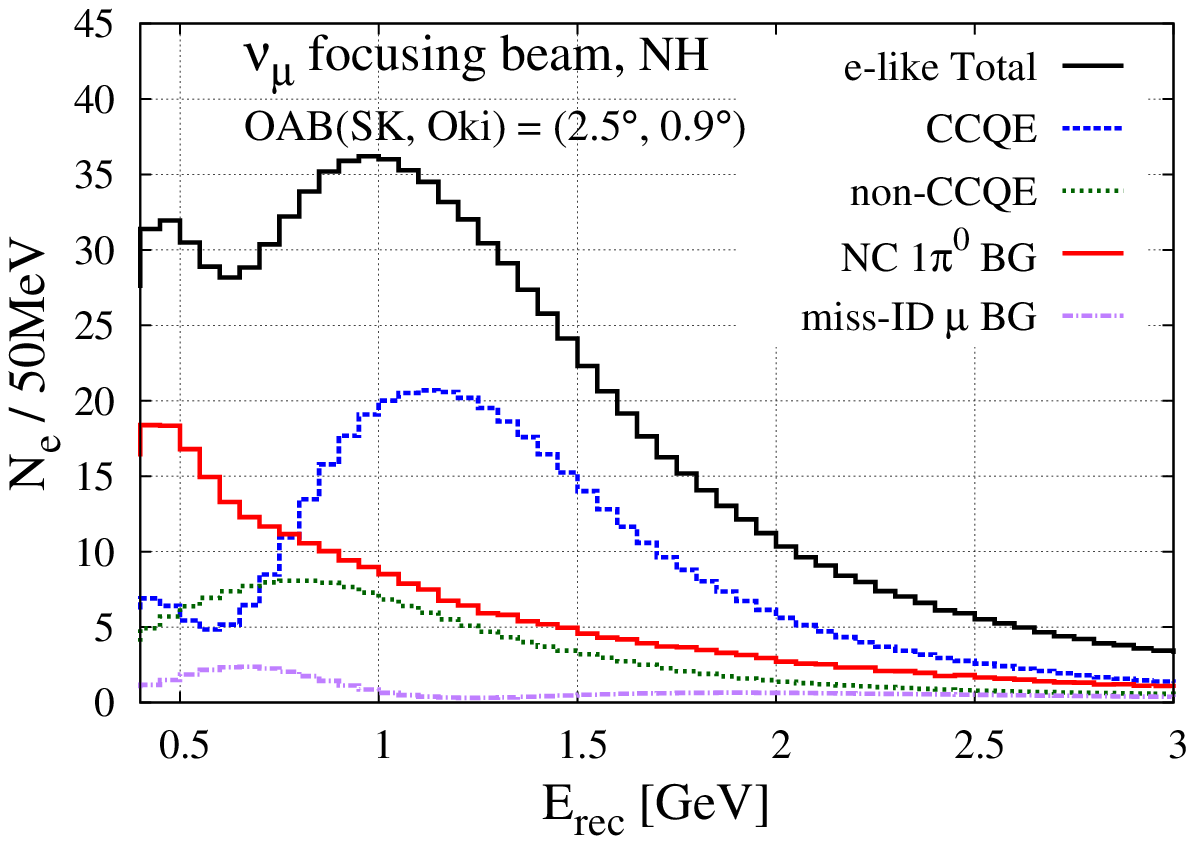}
  \includegraphics[width=0.4\textwidth]{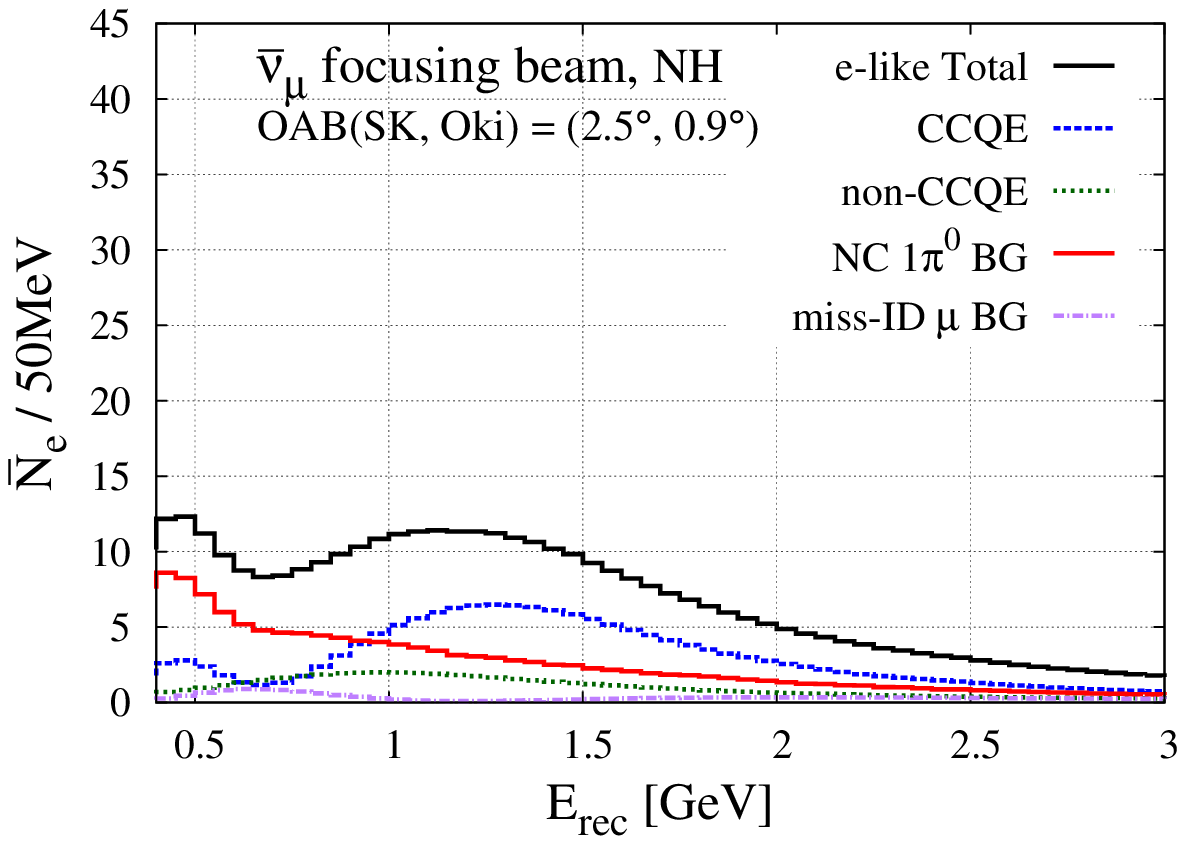}
}
\resizebox{0.8\textwidth}{!}{
  \includegraphics[width=0.4\textwidth]{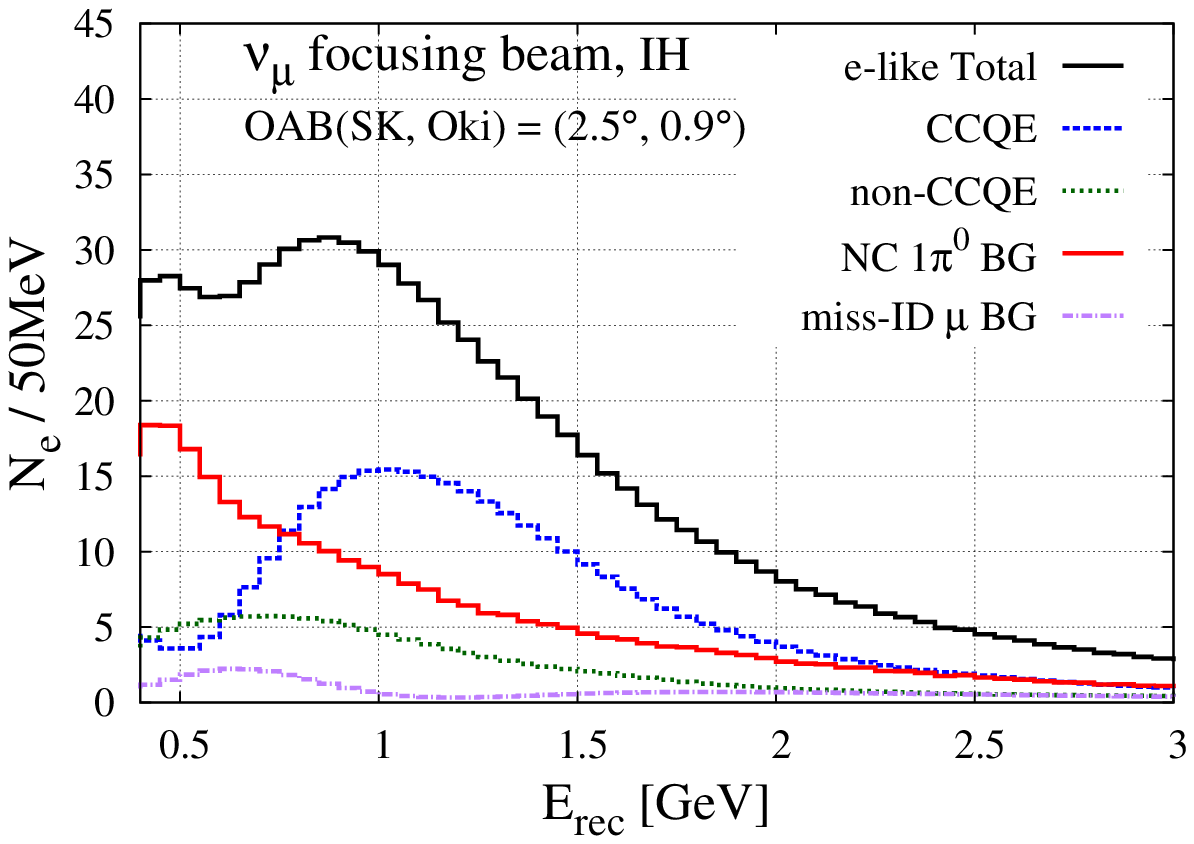}
  \includegraphics[width=0.4\textwidth]{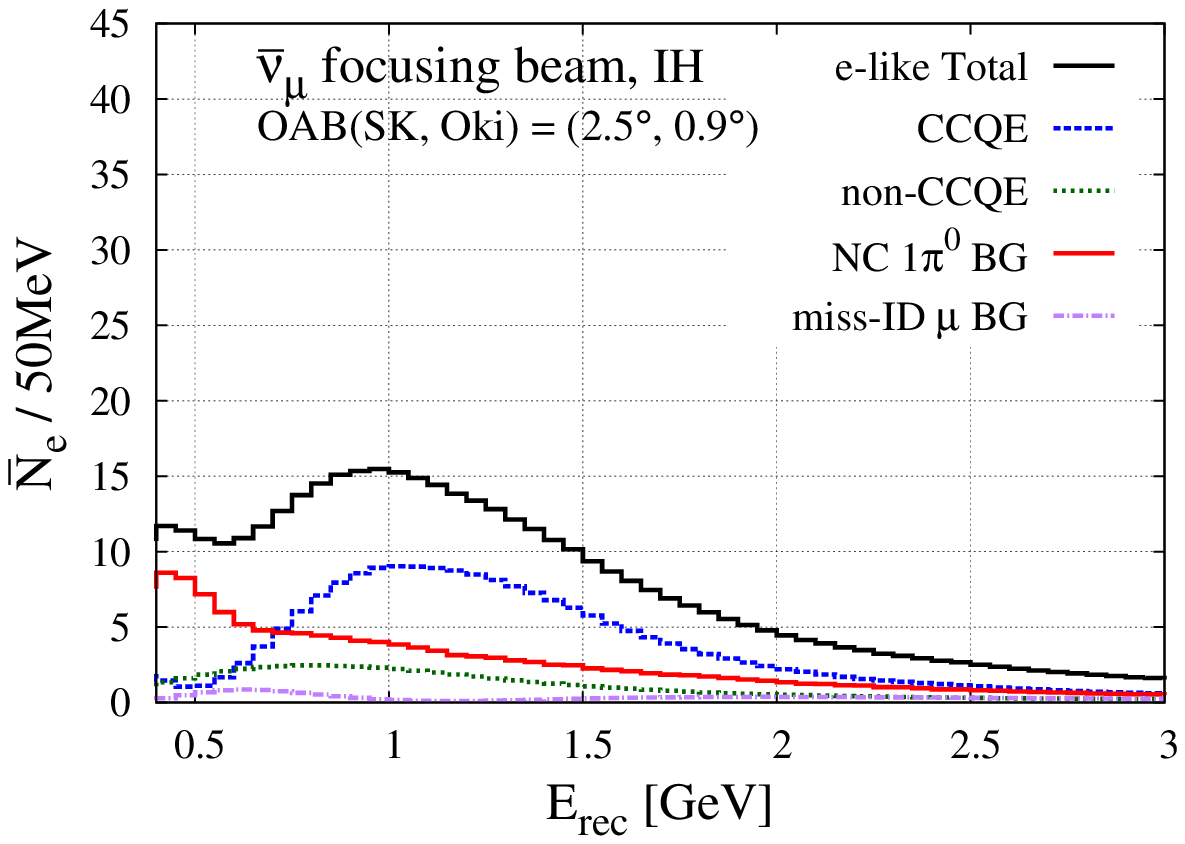}
}
\resizebox{0.8\textwidth}{!}{
  \includegraphics[width=0.4\textwidth]{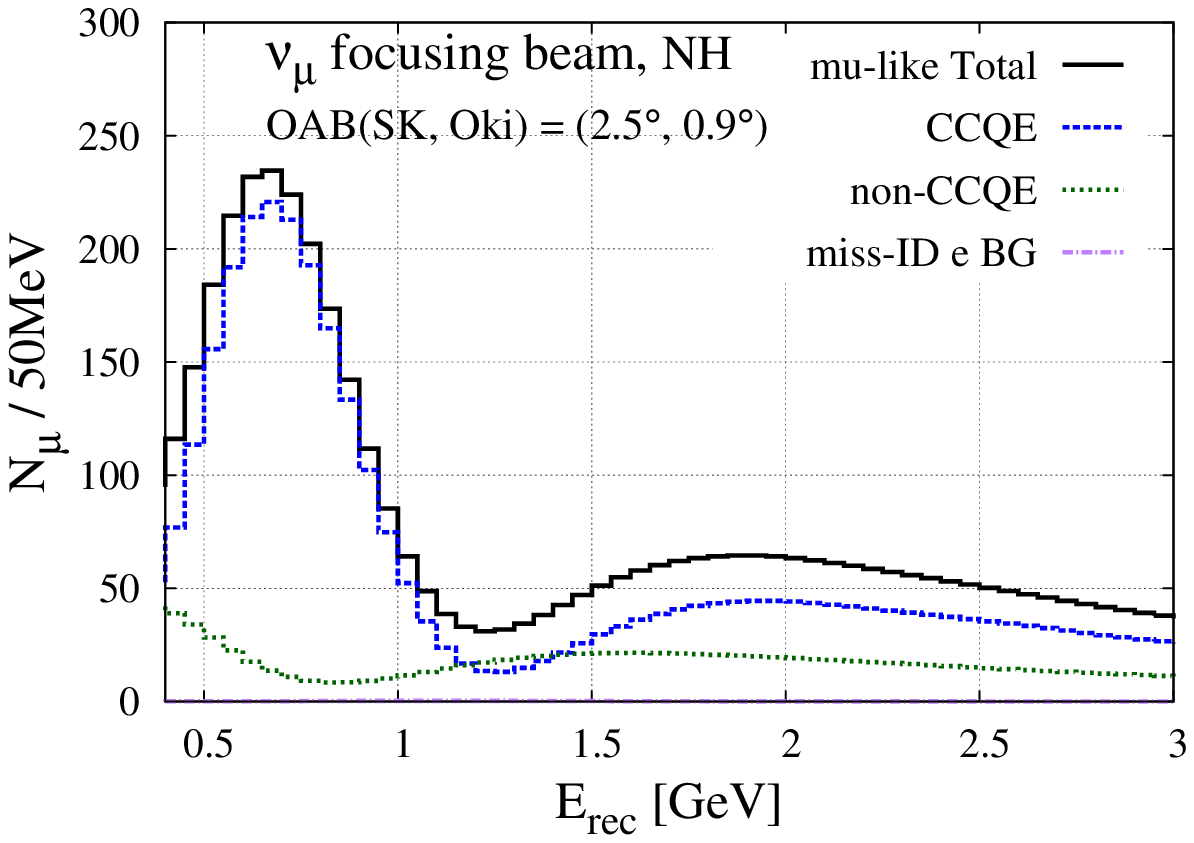}
  \includegraphics[width=0.4\textwidth]{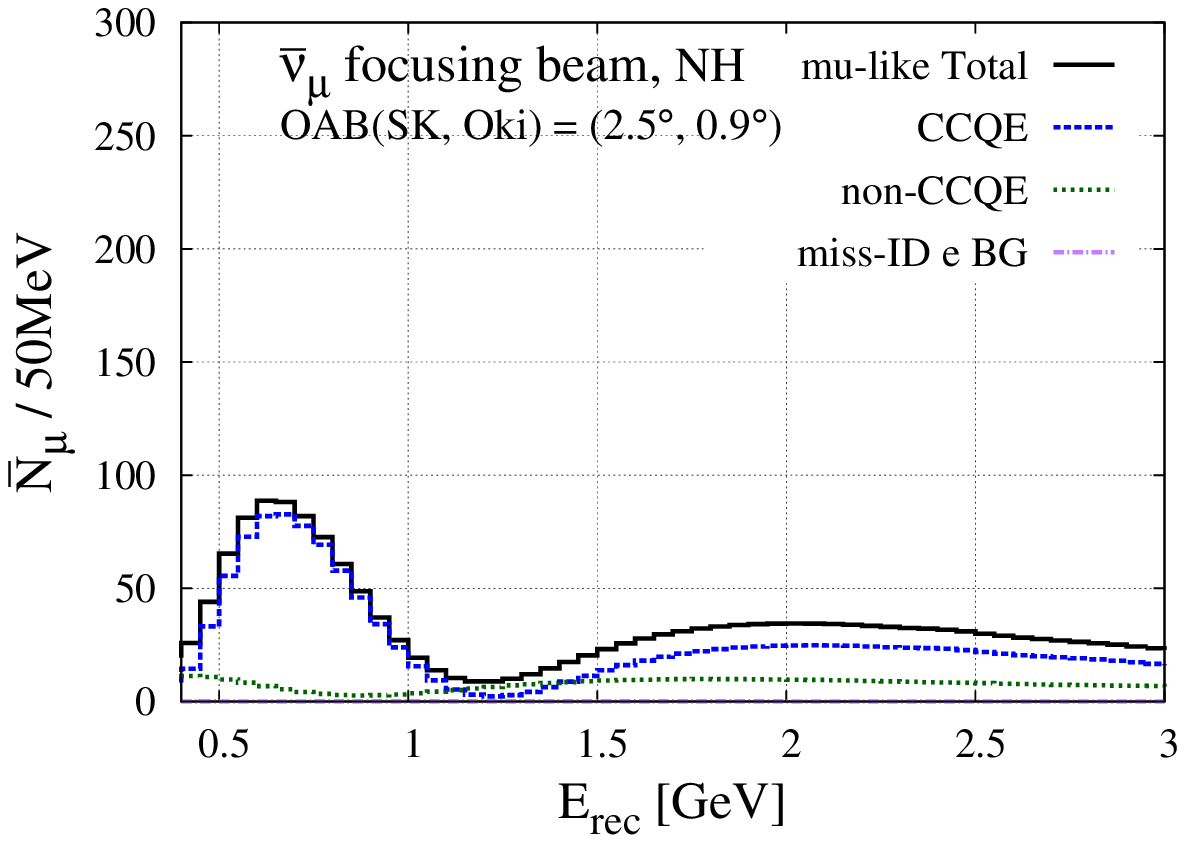}
}
\resizebox{0.8\textwidth}{!}{
  \includegraphics[width=0.4\textwidth]{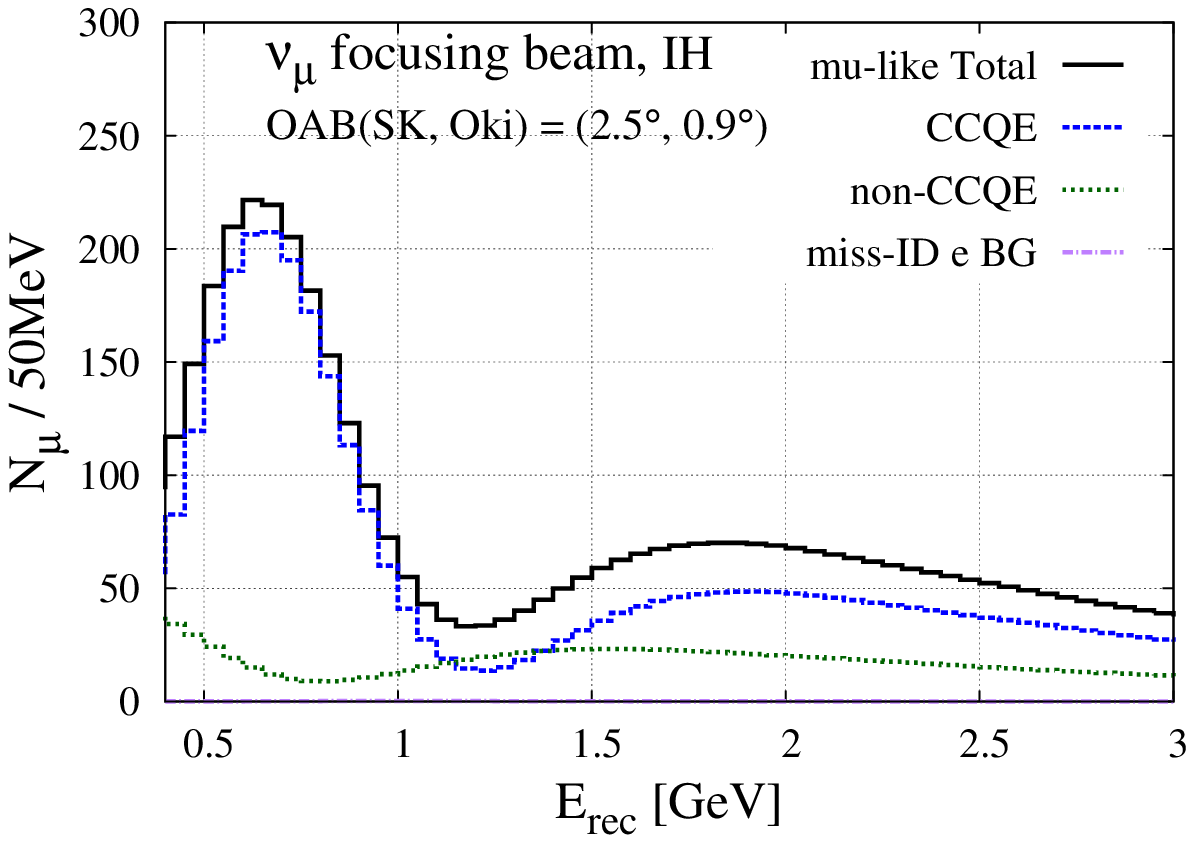}
  \includegraphics[width=0.4\textwidth]{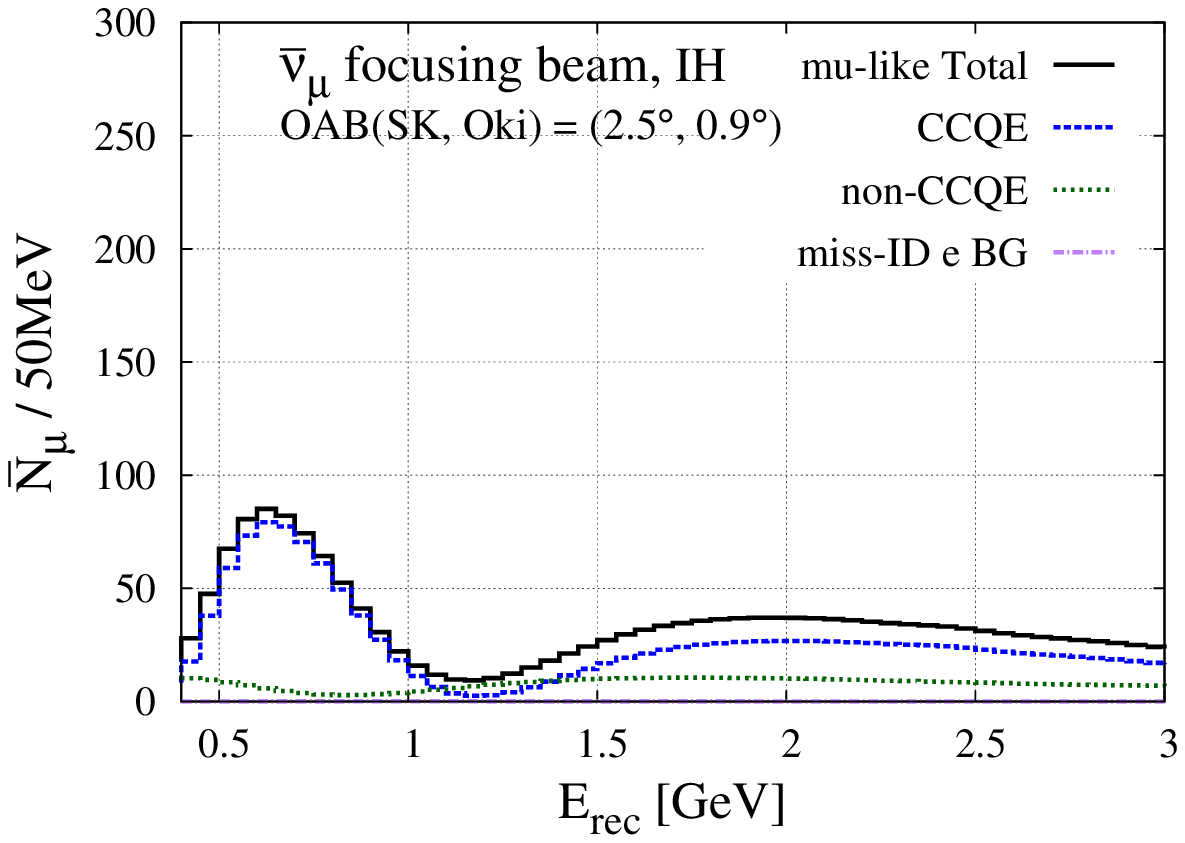}
}
  \caption{Same as Fig.~\ref{fig:erec_sk} but for the T2KO
 experiment with the $0.9^\circ$ OAB at an Oki detector with the 100 kton
 fiducial volume.
}
\label{fig:erec_t2ko}
\end{figure}
\begin{figure}[t]
 \centering
\resizebox{0.8\textwidth}{!}{
  \includegraphics[width=0.45\textwidth]{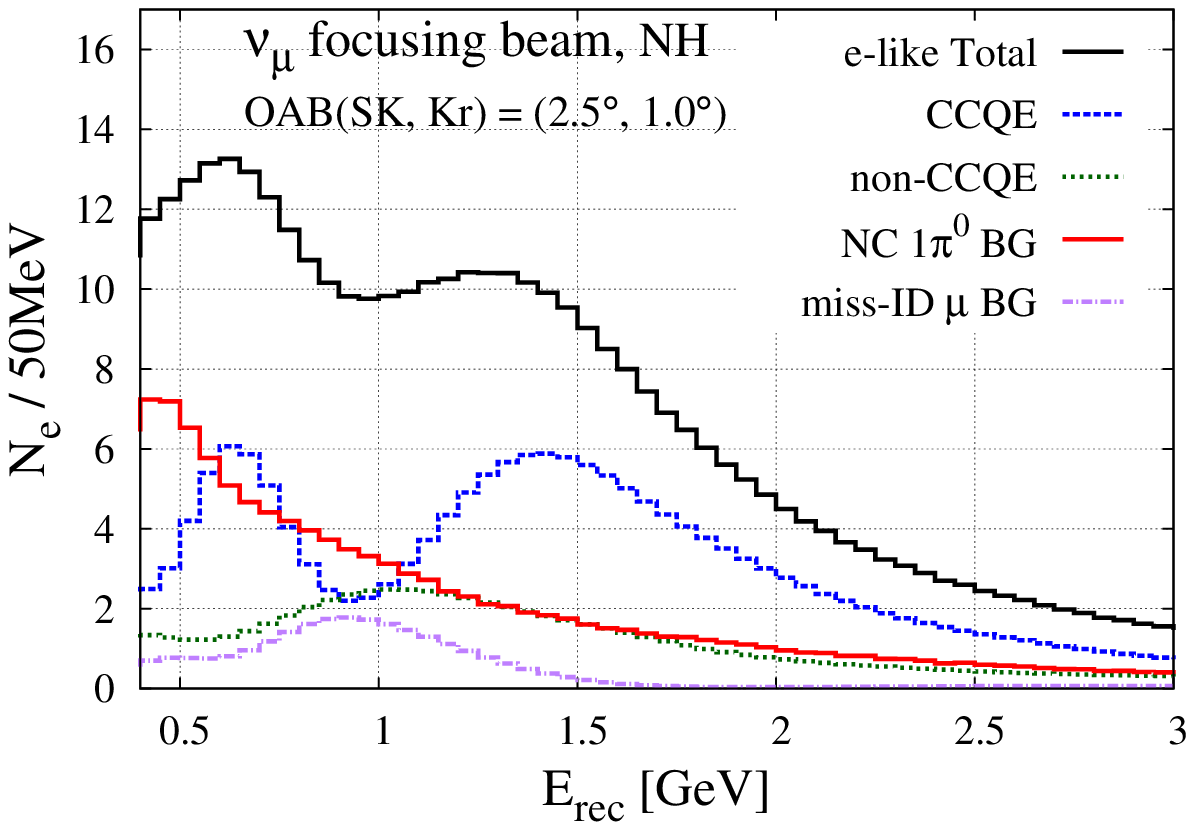}
  \includegraphics[width=0.45\textwidth]{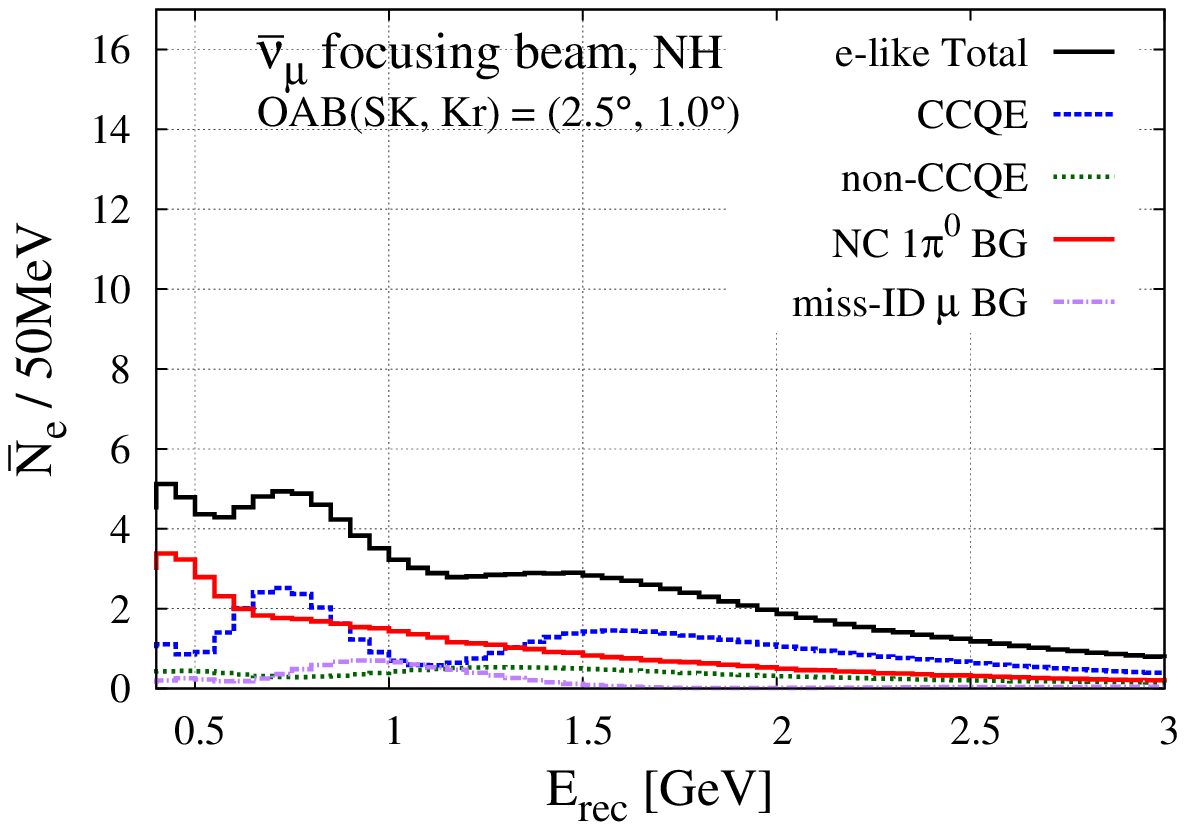}
}
\resizebox{0.8\textwidth}{!}{
  \includegraphics[width=0.45\textwidth]{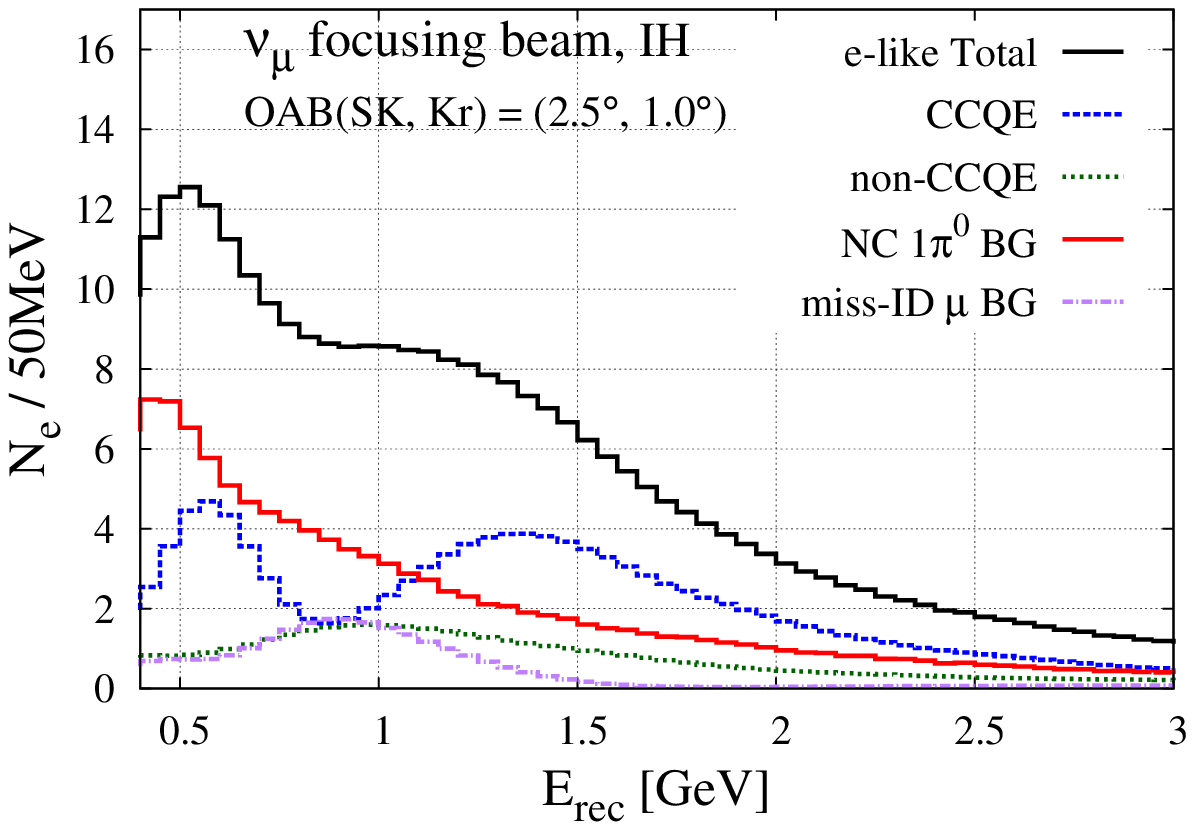}
  \includegraphics[width=0.45\textwidth]{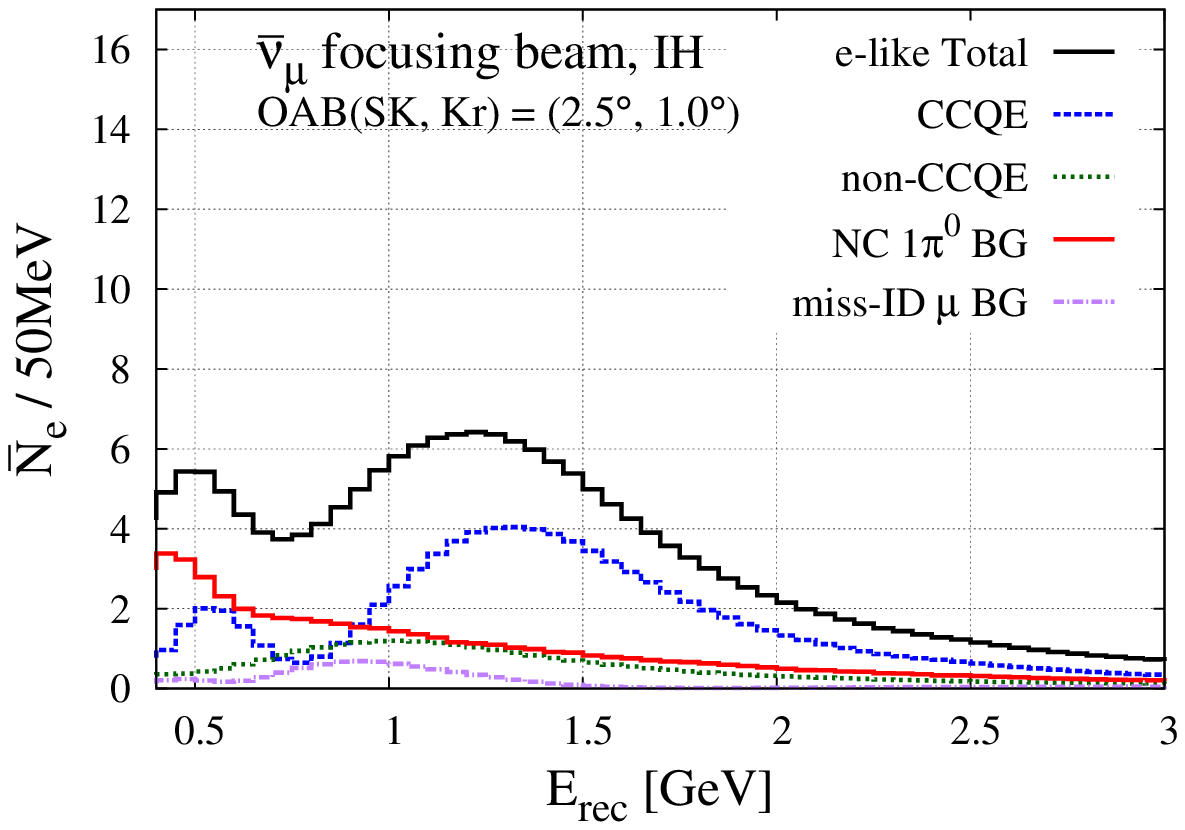}
}
\resizebox{0.8\textwidth}{!}{
  \includegraphics[width=0.45\textwidth]{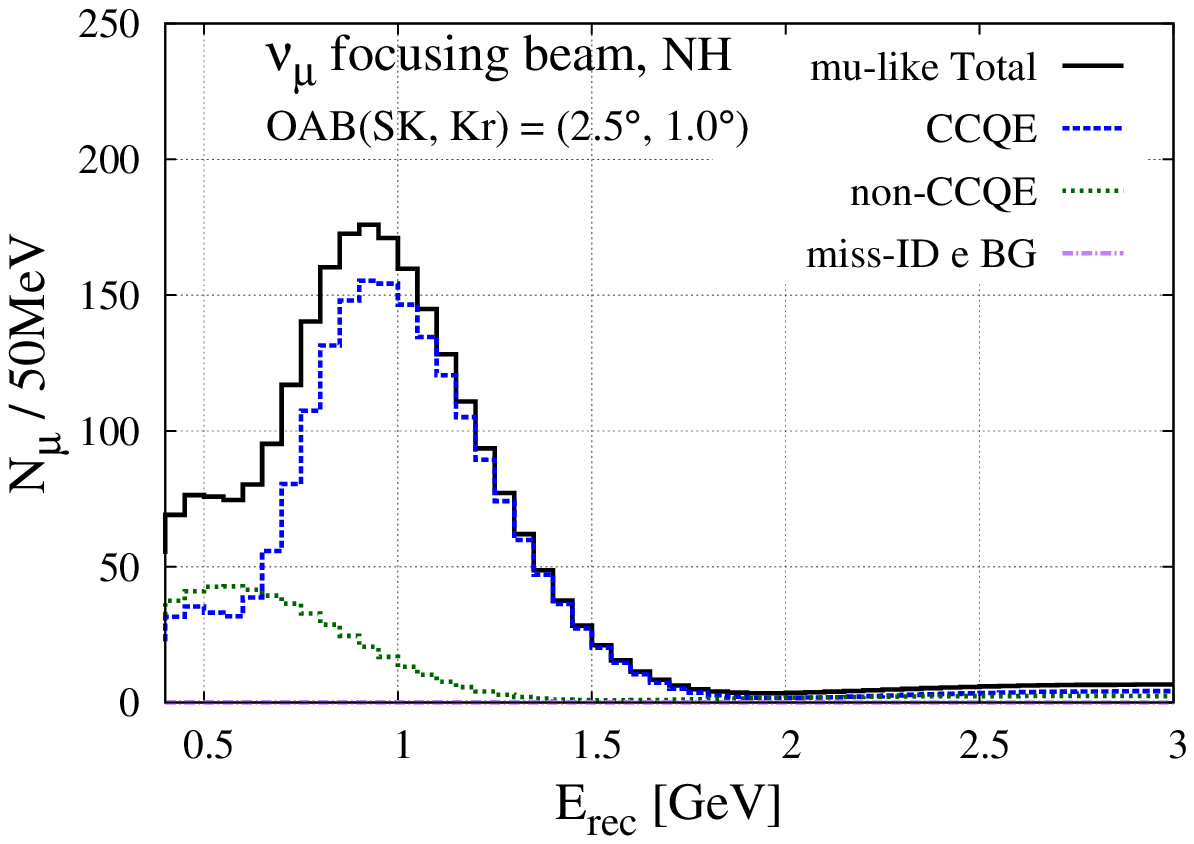}
  \includegraphics[width=0.45\textwidth]{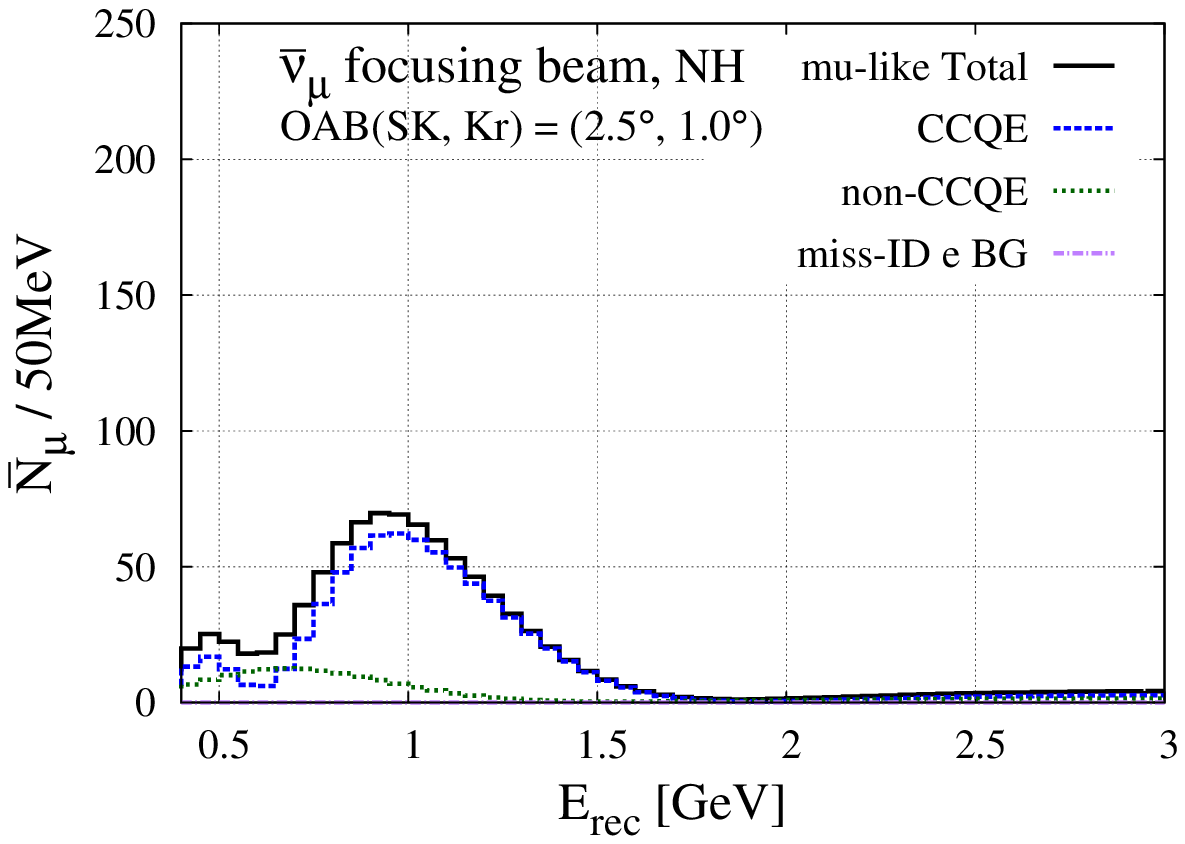}
}
\resizebox{0.8\textwidth}{!}{
  \includegraphics[width=0.45\textwidth]{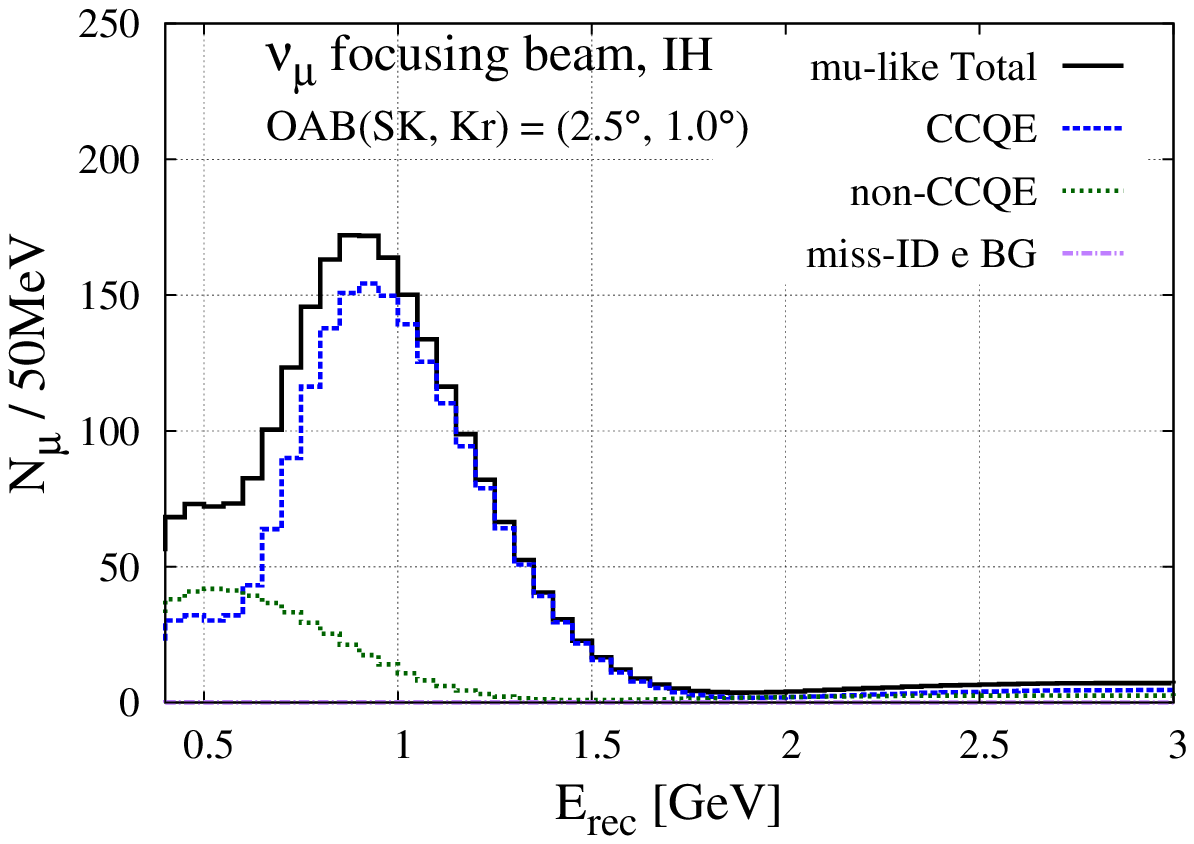}
  \includegraphics[width=0.45\textwidth]{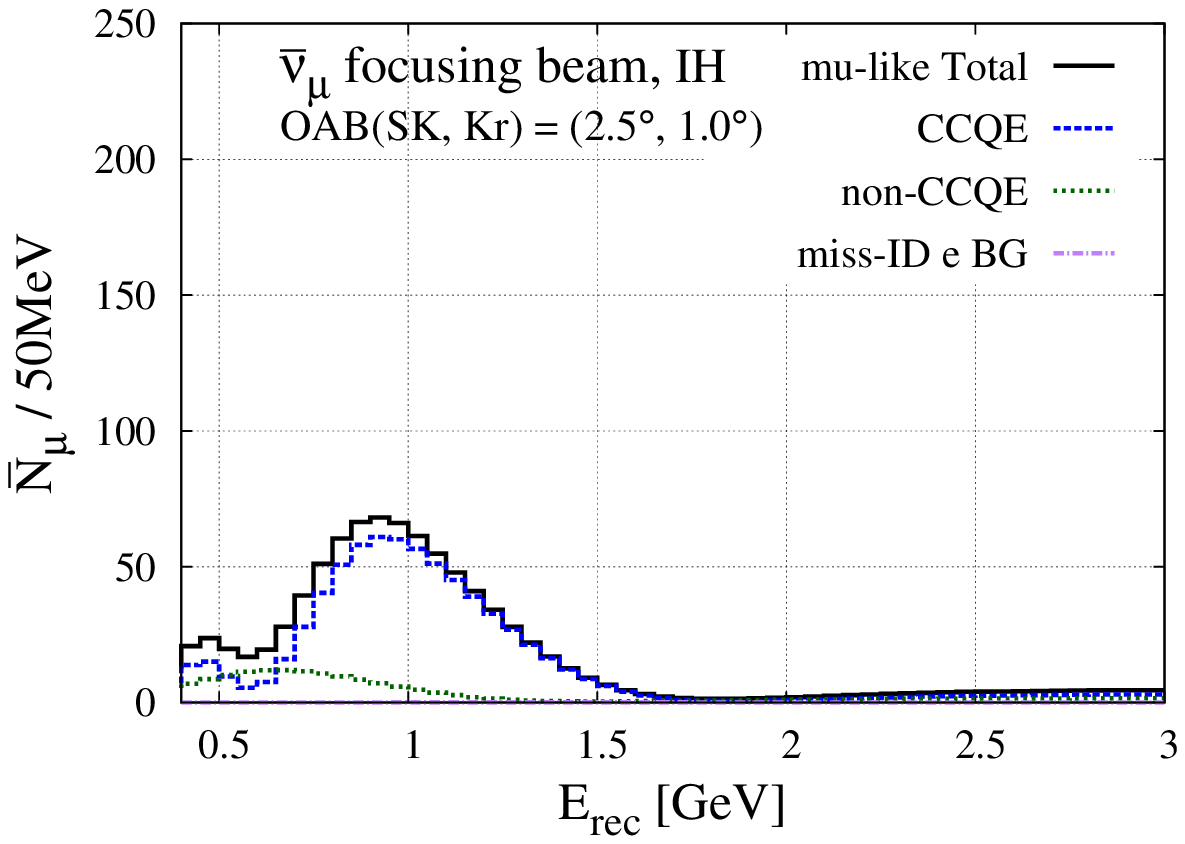}
}
  \caption{Same as Fig.~\ref{fig:erec_sk} but for the T2KK
 experiment with the $1.0^\circ$ OAB at a Kr detector with the 100 kton
 fiducial volume.
}
\label{fig:erec_t2kk}
\end{figure}
As shown in those figures, T2KK and T2KO experiments can observe
up to the second peak of the $\nu_\mu \rightarrow \nu_e$ and $\nubar_\mu
\ra \nubar_e$ oscillations due to their long baseline length, while the T2K experiment only observe the
first peak. 
Observing the several peaks of those oscillation modes has advantages especially for the accurate
CP phase measurement 
because tails of the oscillation peaks have the information of both
$\sin\dmns$ and $\cos\dmns$.

On the other hand, the theoretical predictions of the $\mu$- and
$e$-like event numbers, $(N_{\mu,D}^{i})^{\rm fit}$ and $(N_{e,D}^{i})^{\rm fit}$, are calculated as

\begin{subequations} 
 \begin{align}
\left(N_{\mu,D}^{i}\right)^{\rm fit}
=&
\,f_{\rm V}^{D}
\left[
\left(
1-P_{e/\mu}^D
\right) \effm^D   
  \sum_{X,\nu_\alpha}
f_{\nu_\alpha}^{D}
  \left\{ f_{\nu_\mu}^{X}N_D^{i,X}(\nu_\alpha\!\rightarrow\!\nu_\mu) +f_{\nubar_\mu}^{X} N_D^{i,X}(\nu_\alpha\!\rightarrow\!\nubar_\mu)\right\}  \right.\nn\\
&\left.\hspace{4.3em}+
P_{\mu/e}^D \,\effe^D
  \sum_{X,\nu_\alpha}
f_{\nu_\alpha}^{D}
  \left\{ f_{\nu_e}^{X}
  N_D^{i,X}(\nu_\alpha\!\rightarrow\!\nu_e)
+f_{\nubar_e}^{X}
  N_D^{i,X}(\nu_\alpha\!\rightarrow\!\nubar_e) \right\}
\right]\,,\label{eq:Nfitm}
\end{align}
\begin{align}
\left(N_{e,D}^{i}\right)^{\rm fit}
=&
\,f_{\rm V}^{D}
\left[\,P_{e/\mu}^D \,\effm^D
  \sum_{X,\nu_\alpha}
f_{\nu_\alpha}^{D}
    \left\{ f_{\nu_\mu}^{X}
   N_D^{i,X}(\nu_\alpha\!\rightarrow\!\nu_\mu) 
  +f_{\nubar_\mu}^{X} N_D^{i,X}(\nu_\alpha\!\rightarrow\!\nubar_\mu)\right\} \right.\nn\\
&\left.+
(1-P_{\mu/e}^D) \effe^D
  \sum_{X,\nu_\alpha}
f_{\nu_\alpha}^{D}
   \left\{ f_{\nu_e}^{X}
  N_D^{i,X}(\nu_\alpha\!\rightarrow\!\nu_e)
+f_{\nubar_e}^{X} N_D^{i,X}(\nu_\alpha\!\rightarrow\!\nubar_e) \right\} \right] \nn\\
&+
\,f_{\rm V}^{D}\sum_{\nu_\alpha}
  f_{\nu_\alpha}^{D}
  f_{\pi^0}^{\rm NC}
  \left\{ f_{\pi^0}^{\rm NCRes}N^{i,{\rm NCRes}}_{\pi^0,D}(\nu_\alpha) +f_{\pi^0}^{\rm
  NCCoh}N^{i,{\rm NCCoh}}_{\pi^0,D}(\nu_\alpha) \right.\nn\\
&\left.+N^{i,{\rm NCDI}}_{\pi^0,D}(\nu_\alpha) +N^{i,{\rm NCQE}}_{\pi^0,D}(\nu_\alpha) \right\},
\label{eq:Nfite}
 \end{align}
\label{eq:Nfit}
\end{subequations}
where $f_V^D$ and $f_{\nu_\alpha}^D$ are the normalization factors for the fiducial volume of
 and the $\nu_\alpha$ flux at a detector $D$ (= SK, Oki, Kr),
 respectively. 
$f_{\nu_\beta}^X$ is the normalization factor for the CC cross section of a
 neutrino flavor $\nu_\beta$ via a $X (=$
 CCQE or \nonqe) interaction. $f_{\pi^0}^{\rm NCRes}$ and
 $f_{\pi^0}^{\rm NCCoh}$ are the normalization factors for the NC cross
 sections of resonant and coherent single-$\pi^0$ production processes, respectively, while $f_{\pi^0}^{\rm NC}$ is the
 overall normalization factor for the NC single-$\pi^0$ backgrounds,
 mainly reflecting the uncertainty of the $\pi^0$ misidentification
 probability, Eq.~(\ref{eq:polfitfunc}). These factors are varied in the minimization
 of the $\chi^2$ function, and their deviation
 from unity measures systematic uncertainties. 

Using the above normalization factors, 
detection efficiencies ($\effe$, $\effm$)
and 
misidentification probabilities ($P_{e/\mu}$, $P_{\mu/e}$),
we take into account effects of the systematic uncertainty in the $\chi^2$
function as

\begin{eqnarray}
  \chi^2_{\rm sys} &=& 
   \sum_{S} \left(\dfrac{S_{\rm fit} -S_{\rm input}}{\delta S} \right)^2,
%
\label{eq:sys_chi2}
\end{eqnarray}
where $S_{\rm fit}$ is the systematic parameter value
used to calculate the theoretical predictions,
$S_{\rm input}$ is the one used to generate the data, and $\delta S$ is the uncertainty of the parameter.   
We summarize the systematic parameters used in our analysis in
 Table~\ref{tb:parameters}, where the uncertainties related to the NC
 $\pi^0$ backgrounds are assigned based on the discussions in
 Sec.~\ref{sec:bg}, while the other uncertainties are taken as in the
 previous study~\cite{Hagiwara:2009bb}.
%

Finally, $\chi^2_{\rm para}$ accounts for external constraints 
on the physical parameters as
\begin{eqnarray}
\label{eq:para_chi2}
 \chi^2_{\rm para}
 &=&
 \sum_{P} \left(\dfrac{P_{\rm fit} -P_{\rm input}}{\delta P} \right)^2,
%
\label{eq:chi-para}
\end{eqnarray}
where $P_{\rm fit}$ is the parameter value
used to calculate the theoretical predictions, $P_{\rm input}$ is the
one used to generate the data, and $\delta P$ is the uncertainty of the parameter.   
We summarize the physical parameters used in our analysis in
 Table~\ref{tb:parameters} as well, 
where the parameter values are based on Ref.~\cite{Beringer:1900zz}, except the
uncertainty of $\ssq213$, for which we use the uncertainty achieved by
DayaBay collaboration~\cite{Zhang:2015fya}, and the matter densities,
which are taken from the Ref.~\cite{Hagiwara:2011kw,Hagiwara:2012mg}.
%

The sensitivities to the mass hierarchy is then estimated using the test statistic defined as
\begin{align}
\dchisqmh = \chi^2_{\rm min}\left.\right|_{\rm IH} -\left.\chi^2_{\rm min}\right|_{\rm NH},
\label{eq:dchi2}
\end{align} 
where $\left.\chi^2_{\rm min}\right|_{\rm NH (IH) }$ is the minimum of the
$\chi^2$ function under the assumption of the normal (inverted)
hierarchy. 
The distribution of
the $\dchisqmh$ due to the fluctuation of data can be
approximated as~\cite{Ge:2012wj,Qian:2012zn,Blennow:2013oma}
\begin{equation}
 \dchisqmh \sim {\cal N}\left(\odchisqmh, \,2\sqrt{|\odchisqmh|}\right),
\label{eq:dchisqMH}
\end{equation}
where $\cal N(\mu,\sigma)$ denotes the normal distribution with mean
$\mu$ and standard deviation $\sigma$;
%
%
$\odchisqmh$ is the
$\dchisqmh$ obtained with the average experiment (the
Asimov data set~\cite{Cowan:2010js}).
%
%
%
It has been shown with explicit Monte Carlo studies that this
approximation holds with good accuracy for long baseline
experiments which give $|\odchisqmh| \gg 1$~\cite{Blennow:2013oma}.
With this approximation, we may calculate the probability that an
experiment rejects the wrong mass hierarchy hypothesis with a given
confidence level. For example, $\sim 50\%$ is the probability
for an experiment to reject the wrong mass hierarchy
 with the $\sqrt{|\odchisqmh|}$-$\sigma$ confidence level.
%
%

The sensitivity to the CP phase measurement is estimated in terms of $\dchisqmin$ defined as
\begin{equation}
 \dchisqmin ({\cal \theta}) = \min_{\theta^{'}}\chi^2(\theta,\theta^{'})\left.\right|_{\rm true MH} -\min_{\theta,\,\theta^{'}}\chi^2(\theta,\theta^{'})\left.\right|_{\rm true MH},
\label{eq:dchisqmin}
\end{equation}
where the minimum of the $\chi^2$ function in the first term is found by
fixing some model parameters $\theta$ and marginalizing the other parameters $\theta^{'}$, assuming that the true
mass hierarchy is known, while the minimum of the $\chi^2$ function in the
second term is found by marginalizing the whole parameters, $\theta$ and $\theta^{'}$.
Under certain conditions (especially linear dependence of the
theoretical prediction on the
parameters $\theta$), the $\dchisqmin(\theta_{\rm true})$ is known to be approximately distributed as the
 $\chi^2$ distribution of $N_\theta$ degrees of freedom (d.o.f)
 when the data size is large, where $\theta_{\rm true}$ is the true
values of the parameters $\theta$, and $N_\theta$ is the number of the fixed
 parameters~\cite{wilks1938}.

Since the CP phase is a cyclic parameter in the oscillation
probabilities, the linearity condition is not satisfied in general,
and deviation 
of the distribution of $\dchisqmin(\delta_{\rm CP_{\rm true}})$  
form the 
$\chi^2$ distribution of 1 {\rm d.o.f} would be expected~\cite{Blennow:2014sja,Elevant:2015ska}.
However, this deviation is not so significant for experiments with
sufficiently high sensitivity such that the 1-$\sigma$ uncertainty of $\delta_{\rm CP}$ measurement
is less than $\sim 20^\circ$ for $\delta_{\rm CP} = 0^\circ$~\cite{Blennow:2014sja}.
%
This is the case for the T2KK and T2KO experiments as we will see later,
and we estimate sensitivity to the
CP phase of those experiments based on the $\chi^2$ distribution approximation of the
$\dchisqmin(\delta_{\rm CP_{\rm true}})$ distribution\footnote{In this
approximation, the resultant sensitivities would be slightly
overestimated, as discussed in Ref.~\cite{Blennow:2014sja}}. 
The $n$-$\sigma$ confidence interval of the CP phase
measurement, $[\dCP^a,\dCP^b]_{n\sigma} \,\,(\dCP^{a} < \dCP^{b})$, is
then estimated such that
%
\begin{equation}
 \dchisqmin(\dCP^{a}) = \dchisqmin(\dCP^{b}) = n^2  \Longleftrightarrow
  [\dCP^{a}, \dCP^{b}]_{n\sigma}.
\end{equation}

\section{Sensitivity to the mass hierarchy determination}
\label{sec:MH}
In this section, we present the results for the sensitivity
studies on the mass hierarchy determination by the T2KK and T2KO
experiments, discussing the sensitivity dependence on the $\nu_\mu$ - $\bar{\nu}_\mu$ focusing beam
ratio and $\ssq{}23$. 

In Fig.~\ref{fig:mh_t2kk}, the sensitivity of the T2KK experiments to the mass
hierarchy determination is shown.
\begin{figure}[t]
\centering
\resizebox{1.0\textwidth}{!}{
\includegraphics[width=0.5\textwidth]{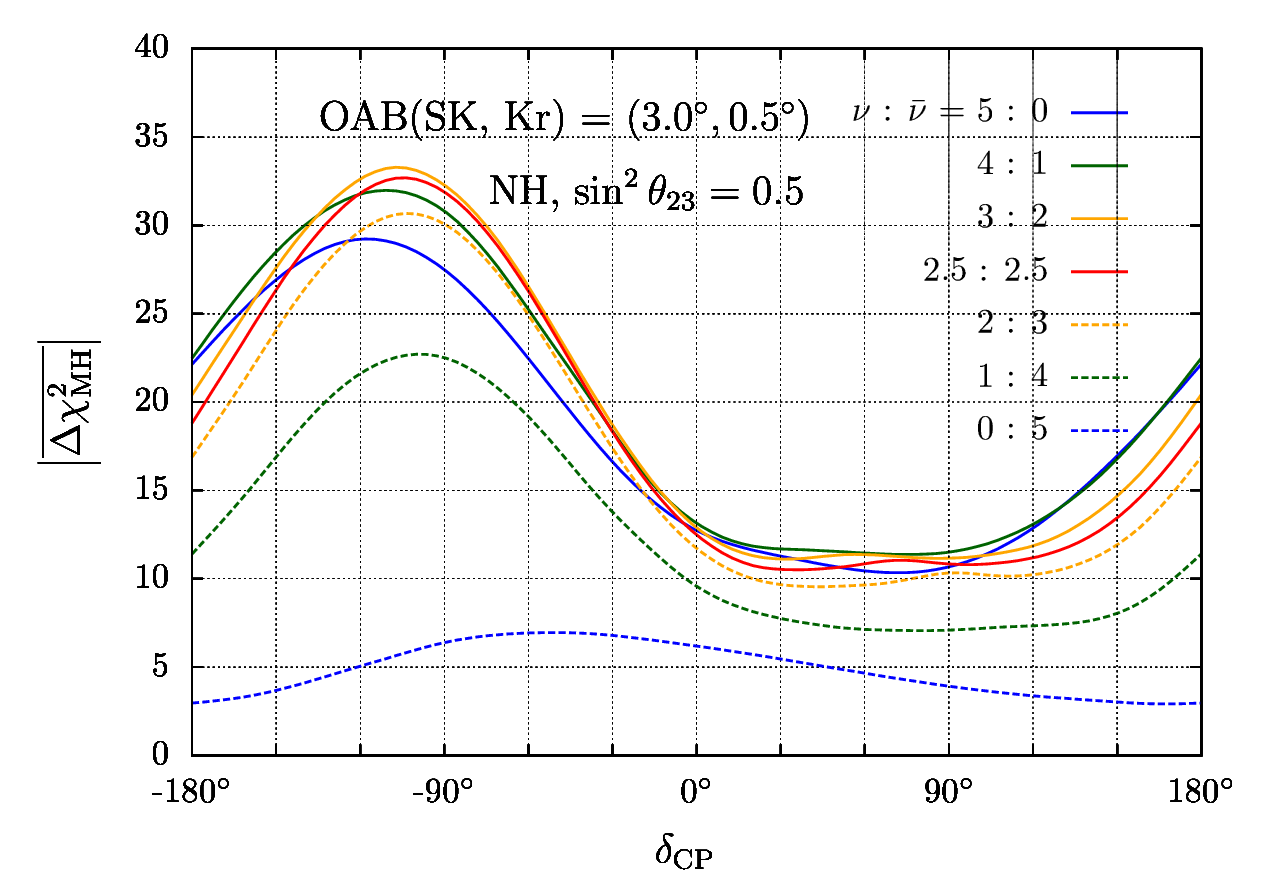} 

\includegraphics[width=0.5\textwidth]{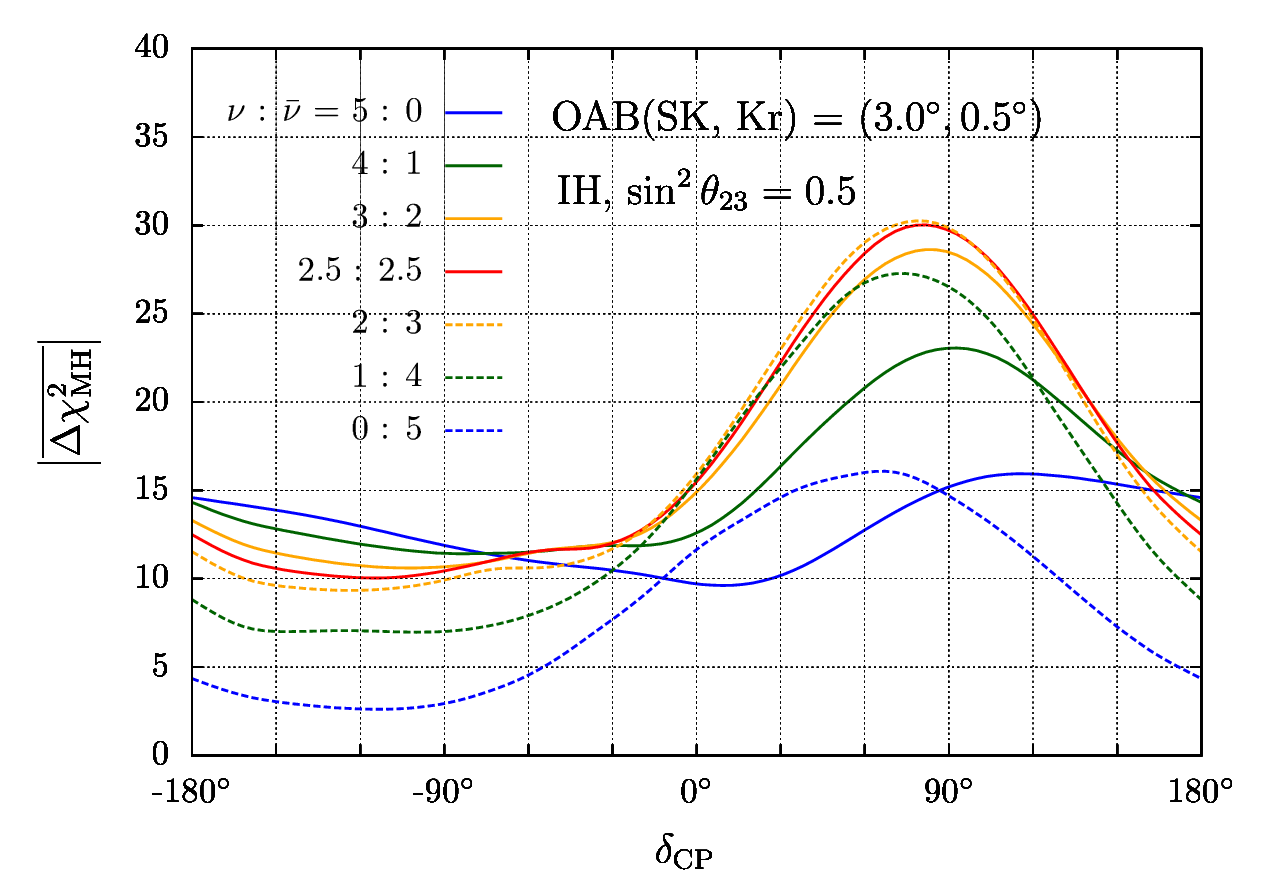} 
}
\resizebox{1.0\textwidth}{!}{
\includegraphics[width=0.5\textwidth]{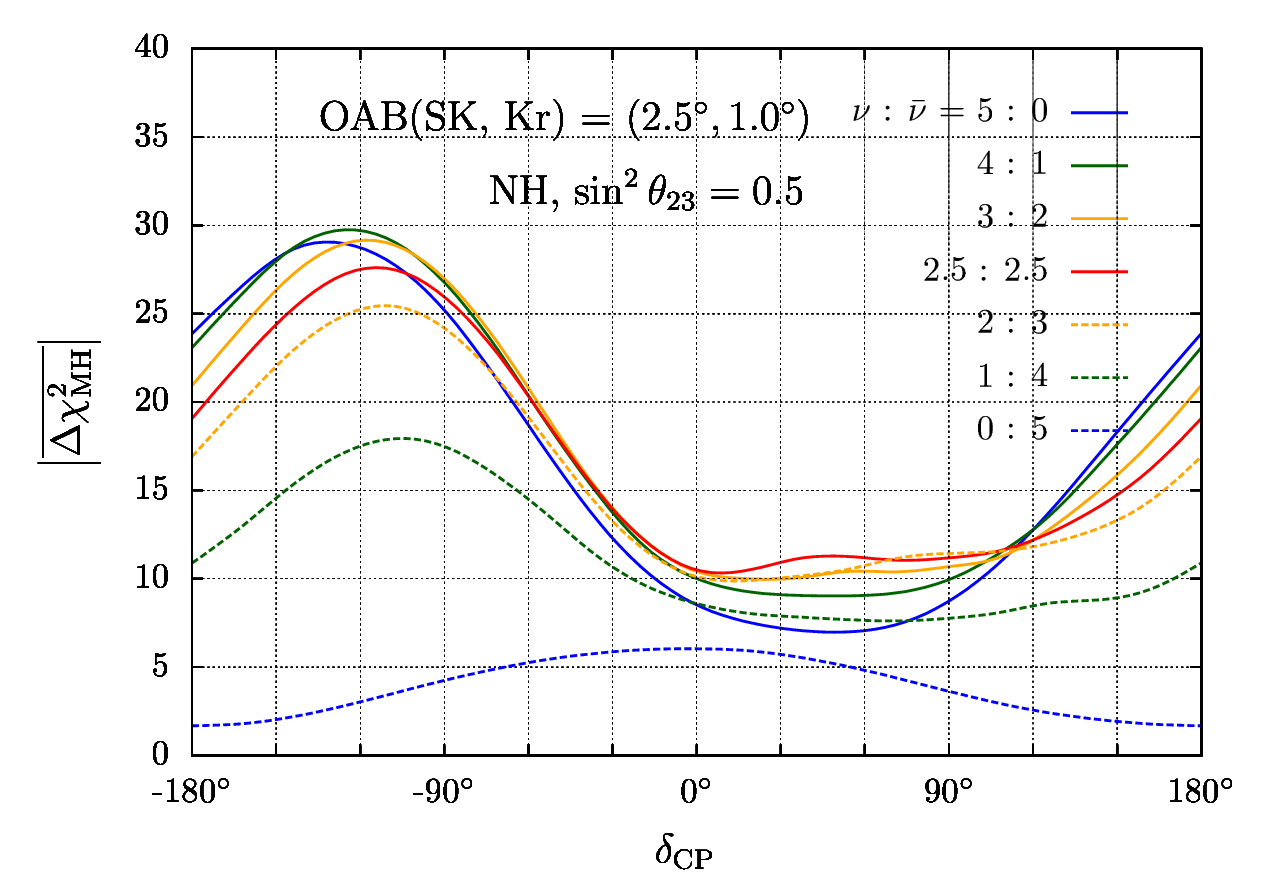} 
\includegraphics[width=0.5\textwidth]{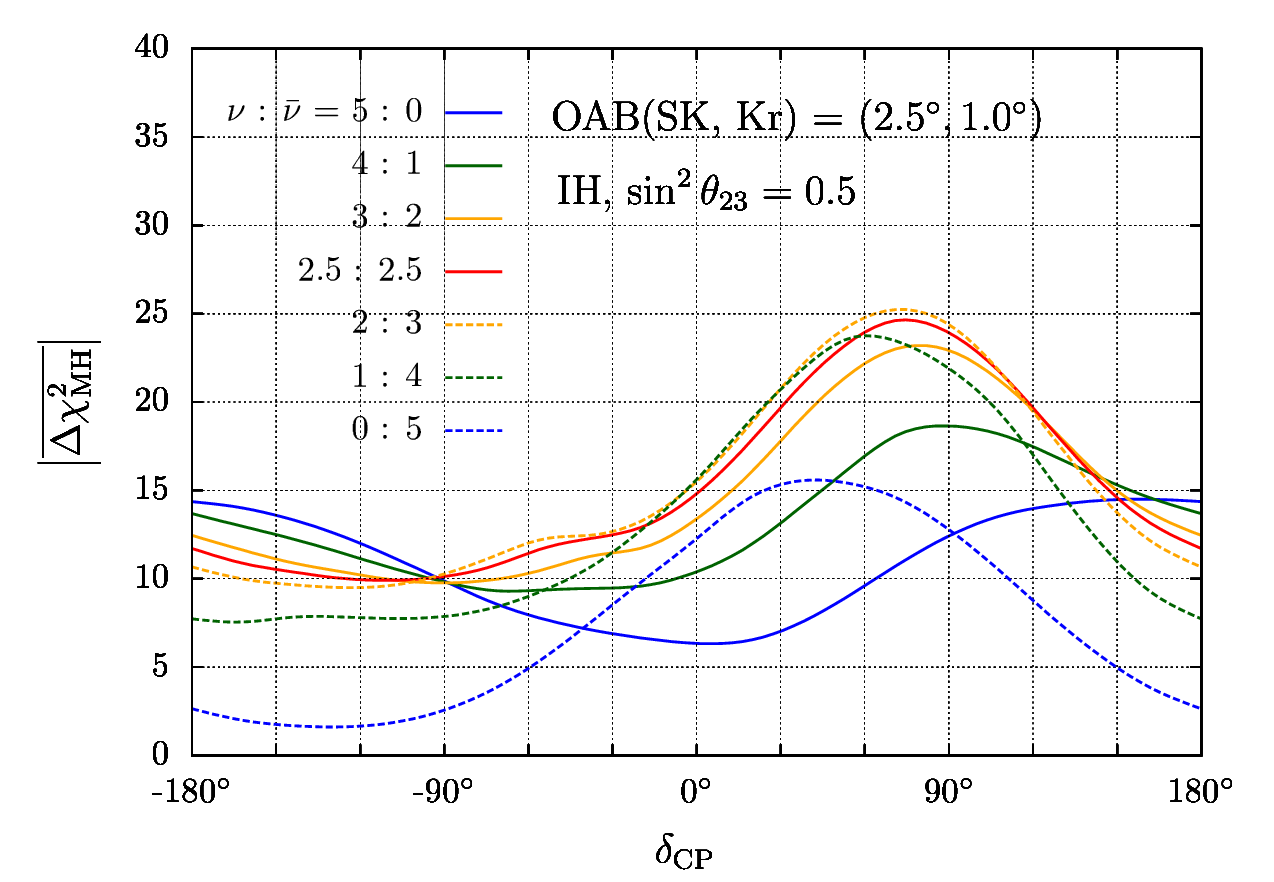} 
}
\caption{The $|\odchisqmh|$ for the T2KK experiment to
 reject the wrong mass hierarchy as a
 function of the CP phase, $\dmns$, with $\sin^2\theta_{23} = 0.5$. The left and right panels
 are for the normal and inverted hierarchy cases, while the upper and lower
 panels are for the $3.0^{\circ}$ ($0.5^\circ$) and $2.5^{\circ}$
 ($1.0^\circ$) off-axis beams 
at the SK (Kr) detector,
 respectively. The blue (dashed-blue), green (dashed-green), orange
 (dashed-orange) and red curves show the sensitivities
 when 
the $\nu_\mu$ - $\bar{\nu}_\mu$ focusing beam ratio is
5\,:\,0 (0\,:\,5), 4\,:\,1 (1\,:\,4), 3\,:\,2 (2\,:\,3) and 2.5\,:\,2.5 $\times 10^{21}$ POT, respectively, with the
 proton energy of 40 GeV.}
\label{fig:mh_t2kk}
\end{figure} 
The left and right panels
 are for the normal and inverted hierarchy cases, while the upper and lower
 panels are for the $3.0^{\circ}$ ($0.5^\circ$) and $2.5^{\circ}$
 ($1.0^\circ$) off-axis beams, OAB, at the SK (Kr) detector, respectively. The true value of $\sin^2\theta_{23}$ is assumed to
 be 0.5. The blue (dashed-blue), green (dashed-green), orange
 (dashed-orange) and red curves show the absolute value of the $\odchisqmh$, Eq.~\eqref{eq:dchisqMH}, for rejecting the wrong mass hierarchy
 when the ratio of the $\nu_\mu$ and $\bar{\nu}_\mu$ focusing beams is
 5\,:\,0 (0\,:\,5), 4\,:\,1 (1\,:\,4), 3\,:\,2 (2\,:\,3) and 2.5\,:\,2.5
 ($\times 10^{21}$ POT with the
 proton energy of 40 GeV), respectively.
It is shown that including $\bar{\nu}_\mu$ focusing beam can improve
the sensitivity, especially in
 high sensitivity regions. Although inclusion of the $\nubar_\mu$
 focusing beam causes reduction of sensitivities
for some
 $\dmns$, this can be alleviated by adjusting the beam ratio
 appropriately. 

 In order to minimize the reduction of the
 sensitivities, $\nu_\mu : \nubar_\mu =$ 4\,:\,1 is the best ratio for
 both OAB cases.   
Comparing the lowest $|\odchisqmh|$ in the whole range of the CP
phase, 
$\nu_\mu : \nubar_\mu =$ 4\,:\,1 is the
best ratio for the $3.0^\circ$ OAB
at the SK, and 3\,:\,2 - 2\,:\,3 are the best for the $2.5^\circ$ OAB
at the SK.
In terms of the highest sensitivity,
4\,:\,1, 3\,:\,2 and 2.5\,:\,2.5 beam ratios give comparable sensitivity for the normal hierarchy,
but 3\,:\,2 - 2.5\,:\,2.5 are significantly better than 4\,:\,1
 for the inverted hierarchy case. Thus, around 3\,:\,2 - 2.5\,:\,2.5 would be a preferred choice for
3.0$^\circ$ OAB at the SK. For the $2.5^\circ$ OAB at the SK, the beam
ratio of 3\,:\,2 - 2.5\,:\,2.5 would be a preferred
choice. 
Although there is not such a $\nu_\mu$ - $\nubar_\mu$ focusing beam
ratio that gives the best sensitivity for any $\dmns$ values and mass
hierarchies, the beam ratio between 4\,:\,1 and 2.5\,:\,2.5 would be
a reasonable choice for both $2.5^\circ$ and $3.0^\circ$ OAB at the SK.

In Fig.~\ref{fig:mh_t2ko}, we show the sensitivities of the T2KO experiment.
\begin{figure}[t]
\centering
\resizebox{1.0\textwidth}{!}{
\includegraphics[width=0.5\textwidth]{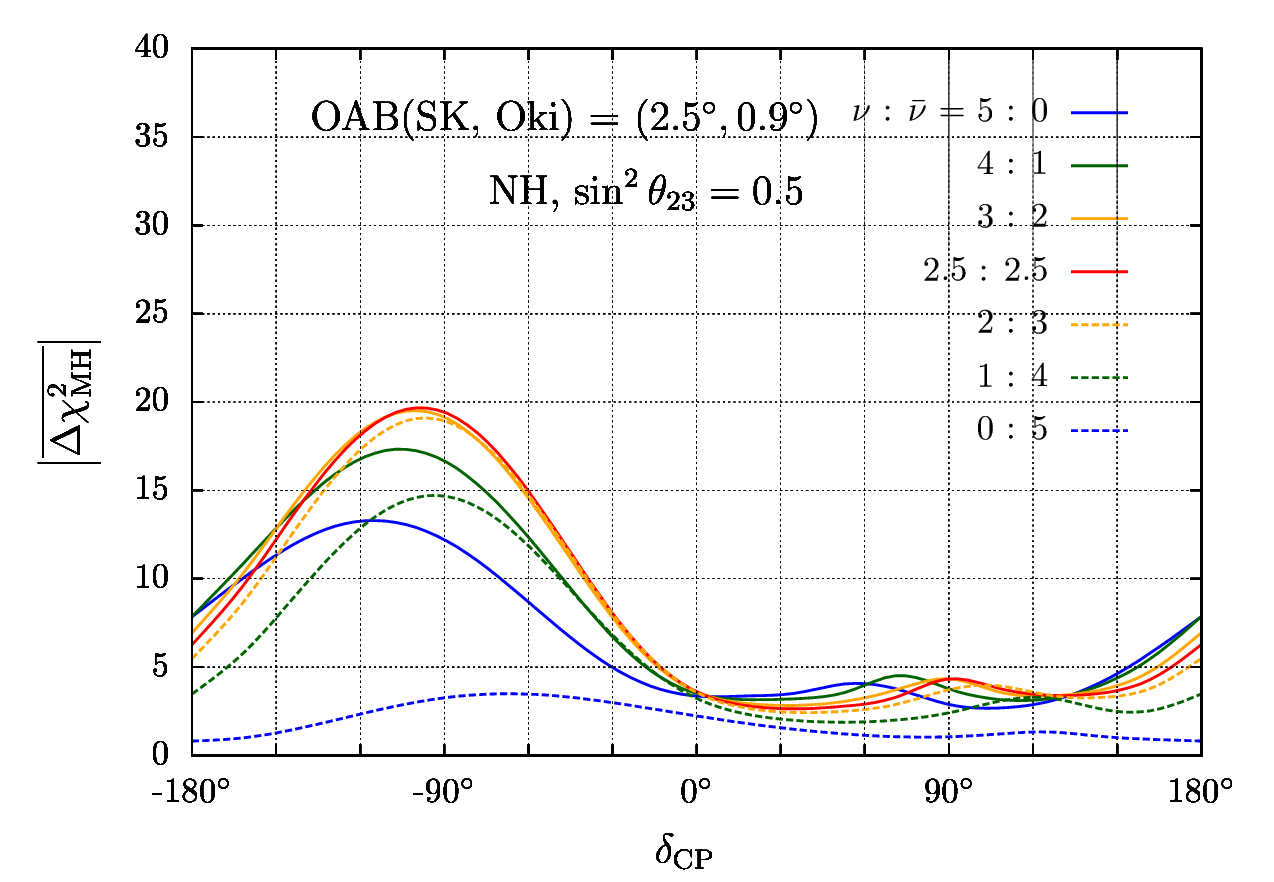} 
%
\includegraphics[width=0.5\textwidth]{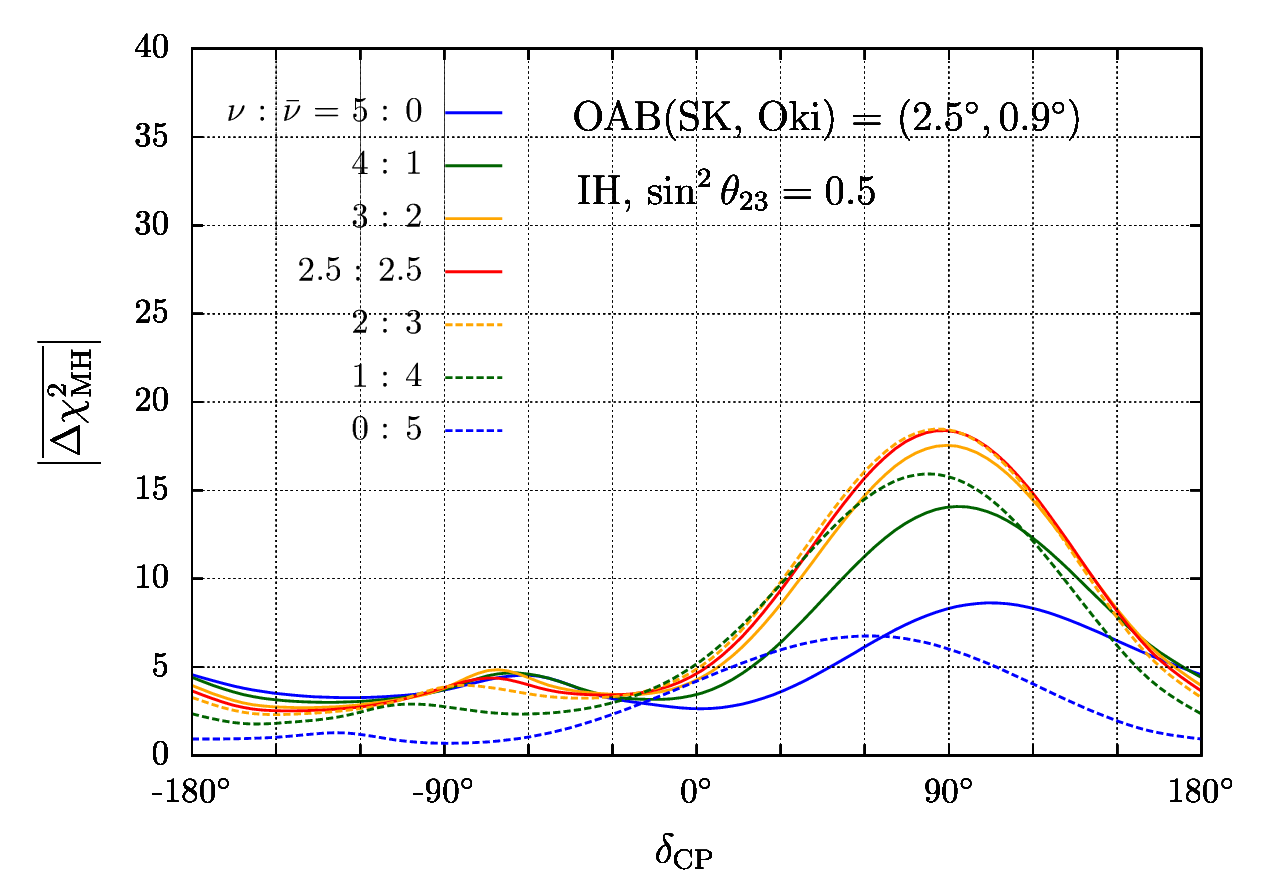} 
}
\caption{Same as Fig.~\ref{fig:mh_t2kk}, but for the T2KO experiment
 with the $2.5^\circ$ ($0.9^\circ$) off-axis beam at the SK (Oki).}
\label{fig:mh_t2ko}
\end{figure} 
Improvement of the sensitivities by including the $\nubar_\mu$ focusing beam is significant
in the high sensitivity region, preferring the running ratio of 3\,:\,2
- 2\,:\,3, while improvement in the low
sensitivity region is not so evident.
Comparing to the T2KK experiments, the sensitivity is lower by $20\%$
- $80\%$ in $|\odchisqmh|$.
The lower sensitivity in the T2KO experiment is
basically due to the smaller matter effects. The difference between the
mass hierarchies mainly shows up in $A^e$ and $B^e$ in
Eq.~\eqref{eq:Pmue} through the matter effects, and enhanced by the baseline length. Because of the
shorter baseline length to the Oki detector than the detector in Korea, this
enhancement is reduced more easily by adjusting the CP phase,
resulting in the lower
sensitivity to the mass hierarchy in the T2KO experiment.

We show the $\sin^2\theta_{23}$ dependence
 of the sensitivity
to the mass hierarchy determination in Fig.~\ref{fig:mh_cp_th23}.  
\begin{figure}[t]
\centering
\resizebox{1.0\textwidth}{!}{
 \includegraphics[width=0.5\textwidth]{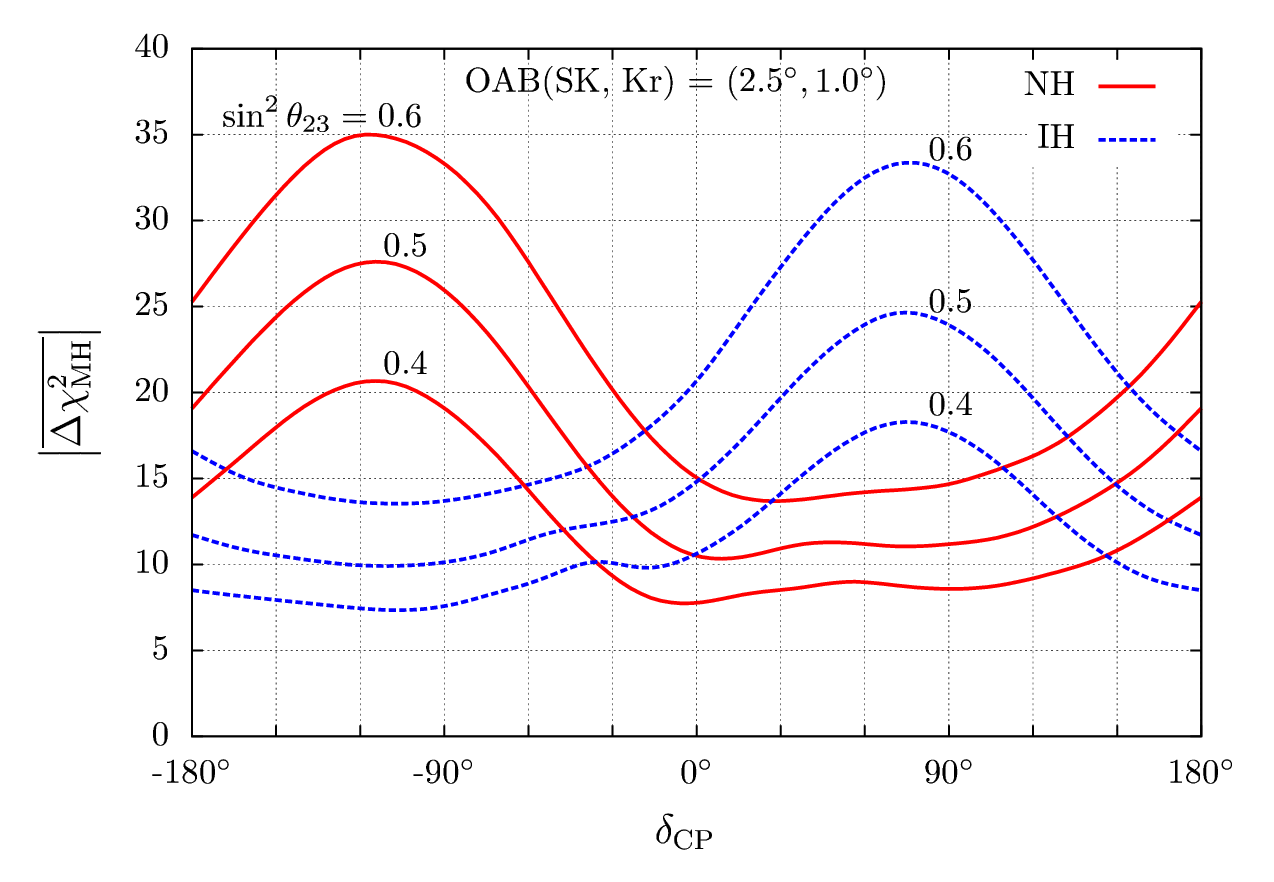} 
 \includegraphics[width=0.5\textwidth]{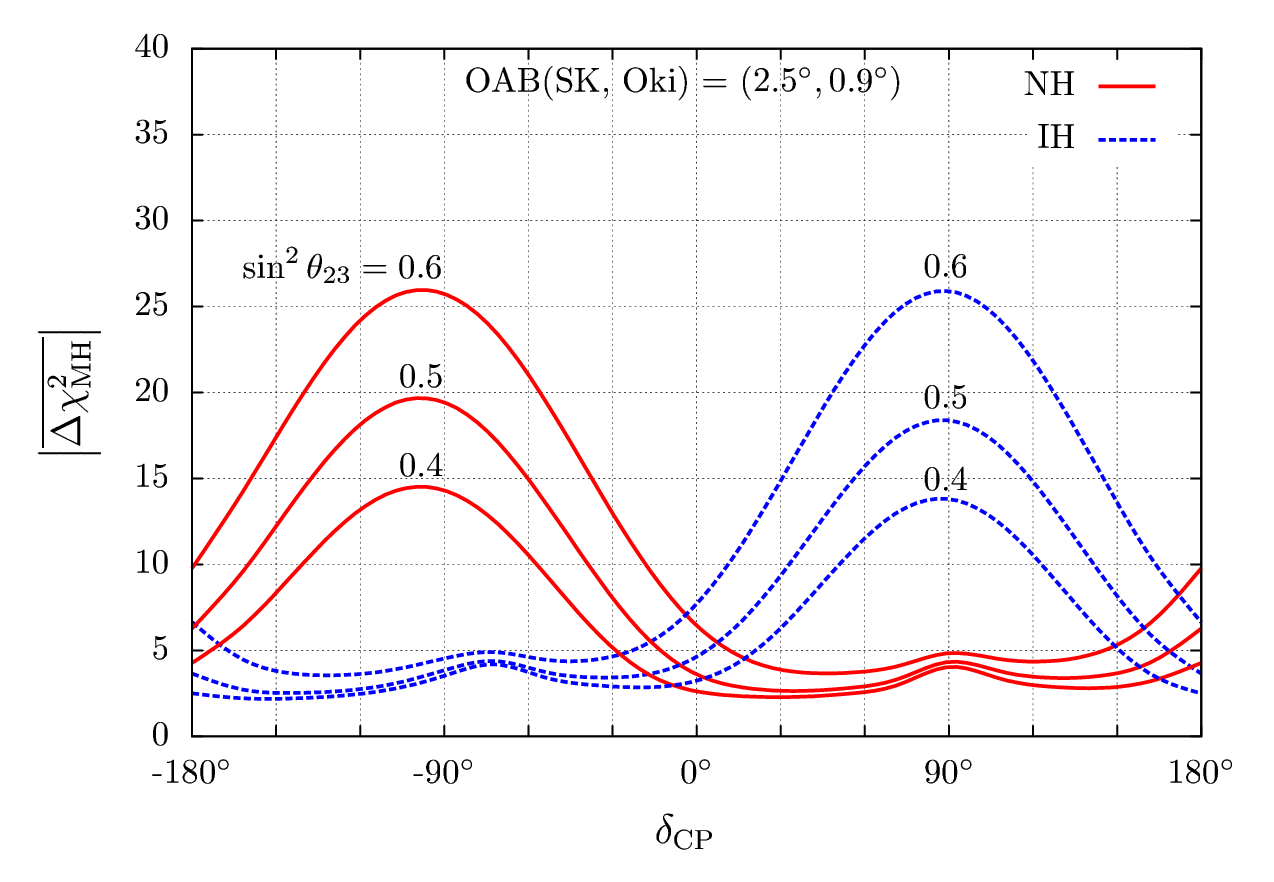} 
}
\caption{The dependence of the $|\odchisqmh|$ on $\sin^2\theta_{23}$ for the
 T2KK and T2KO experiments to reject the wrong mass
 hierarchy as functions of the CP phase. The left and
 right plots are for the T2KK and T2KO experiments with the $2.5^\circ$ OAB at
 the SK, respectively. The
 red and dashed-blue curves are for the normal and inverted hierarchy
 cases. $\sin^2\theta_{23}$ is assumed to be 0.6, 0.5 and
 0.4 from the top to the bottom curves. The $\nu_\mu$ - $\bar{\nu}_\mu$ focusing beam ratio is
 fixed at $\nu_\mu : \bar{\nu}_\mu =$ 2.5\,:\,2.5 $\times 10^{21}$ POT with the
 proton energy of 40 GeV.}
\label{fig:mh_cp_th23}
\end{figure}     
The left and right plots are for the T2KK 
and the T2KO experiments with the $2.5^\circ$ OAB at the SK, respectively. The red
and dashed-blue curves are for the normal and inverted hierarchy cases. $\ssq{}23$ is assumed to be
0.6, 0.5 and 0.4 from the top to the bottom curves for both mass hierarchy cases. We fix the
$\nu_\mu$ - $\bar{\nu}_\mu$ beam ratio at $2.5:2.5$, but dependence on
$\ssq{}23$ are similar in the other beam ratios. The $|\odchisqmh|$ is
reduced by up to $30\%$ - $40\%$ when
$\ssq{}23$ decreases by 0.1 since the number of the $\nu_e$ appearance
signal decreases.
We will reject the wrong mass
hierarchy with $|\odchisqmh| > 8$ (T2KK) and $> 3$ (T2KO) for any CP phases
and $\ssq{}23 > 0.4$. 
In the most sensitive
region around $\dmns = -90^\circ$ for the normal hierarchy case, we may reject
the wrong mass hierarchy with $|\odchisqmh| > 20$ in the T2KK
experiment and $> 14$ in the T2KO experiment
with $\sin^2\theta_{23} > 0.4$.
For the inverted hierarchy case, the most sensitive region is around
$\dmns = 90^\circ$, and we may reject
the wrong mass hierarchy with $|\odchisqmh| > 18$ in the T2KK
experiment and $> 14$ in the T2KO experiment
with $\sin^2\theta_{23} > 0.4$.

\section{Sensitivity to the CP phase measurement}
\label{sec:CP}
In this section, we discuss the sensitivities of the T2KK and T2KO experiments
to the CP phase measurement, comparing to an experiment where a 100
kton detector is placed at the Kamioka site in addition to the 22.5 kton
SK detector, which is called as the $\T2K122$ experiment in this study.
Comparison with this experiment will clearly show the
dependence of the CP phase sensitivity on the baseline length. 
We put emphasis on the effects of
including the $\bar{\nu}_\mu$ focusing beam.

In Fig.~\ref{fig:CPmeasure_nh}, we show the uncertainties of the CP phase
measurements as functions of the true CP phase, $\delta^{\rm true}_{\rm CP}$, for the four experiments: (a) T2KK with
$3.0^\circ$ OAB at the SK and $0.5^\circ$ OAB at the Kr, (b) T2KK with
$2.5^\circ$ OAB at the SK and $1.0^\circ$ OAB at the Kr, (c) T2KO and (d) ${\rm
T2K}_{122}$.
\begin{figure}[t]
\centering
\resizebox{1.0\textwidth}{!}{
\includegraphics[width=0.5\textwidth]{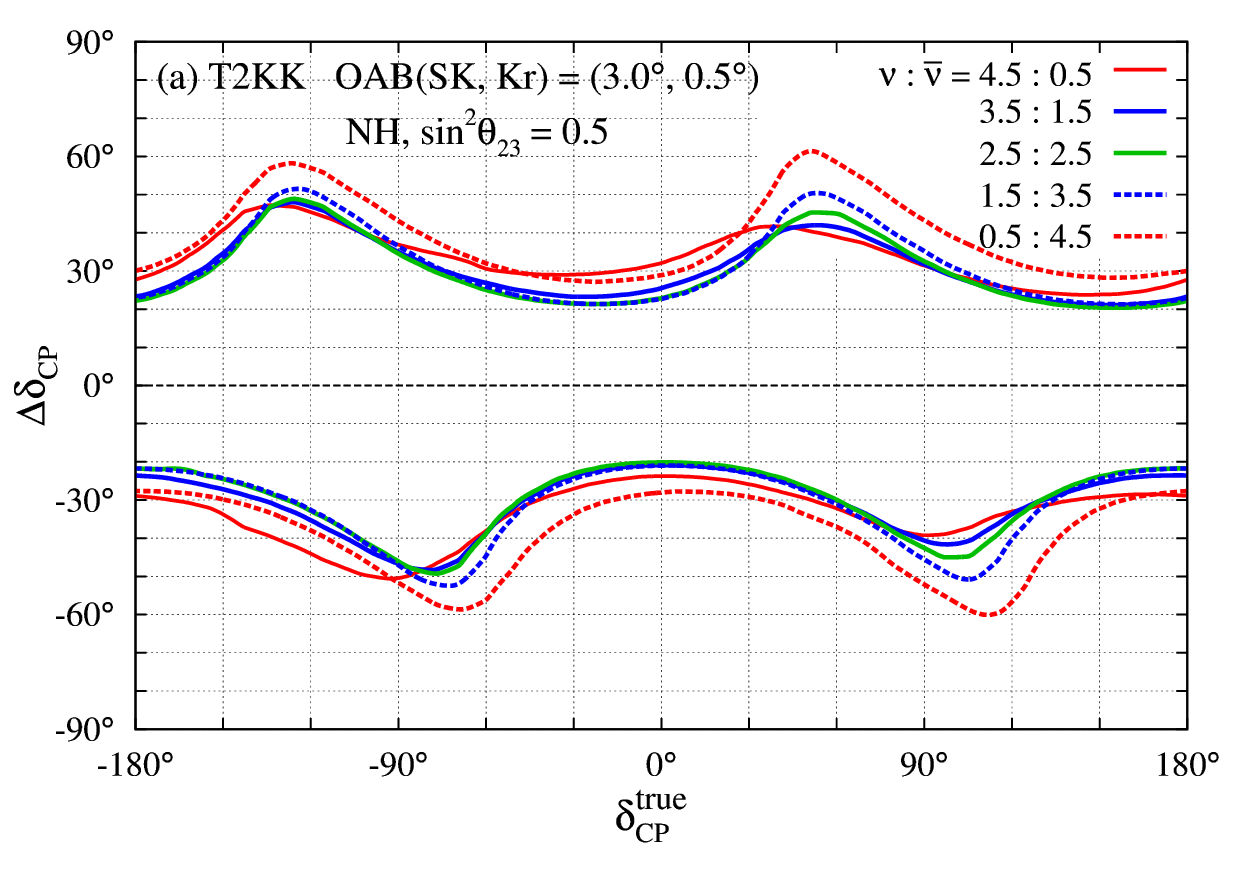} 
\includegraphics[width=0.5\textwidth]{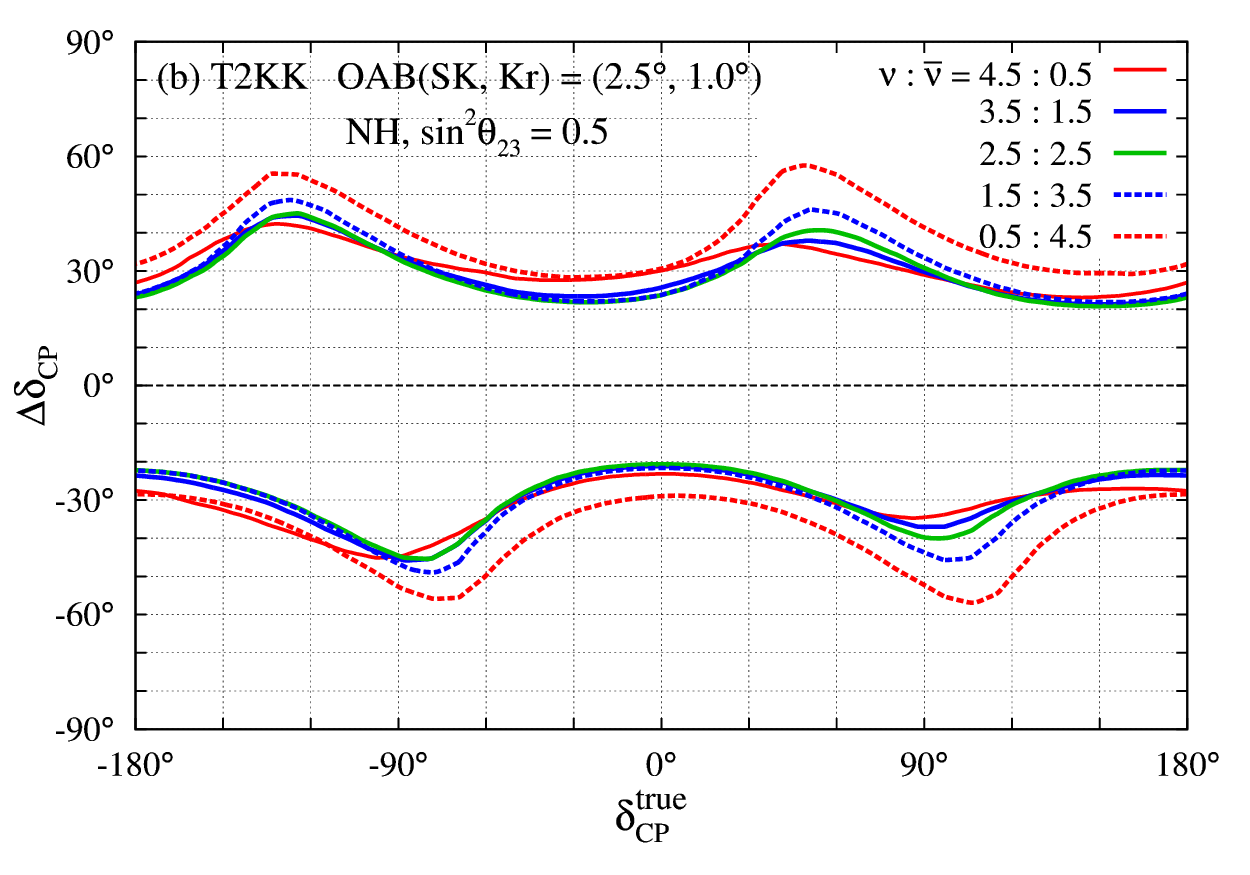} 
}
\resizebox{1.0\textwidth}{!}{
\includegraphics[width=0.5\textwidth]{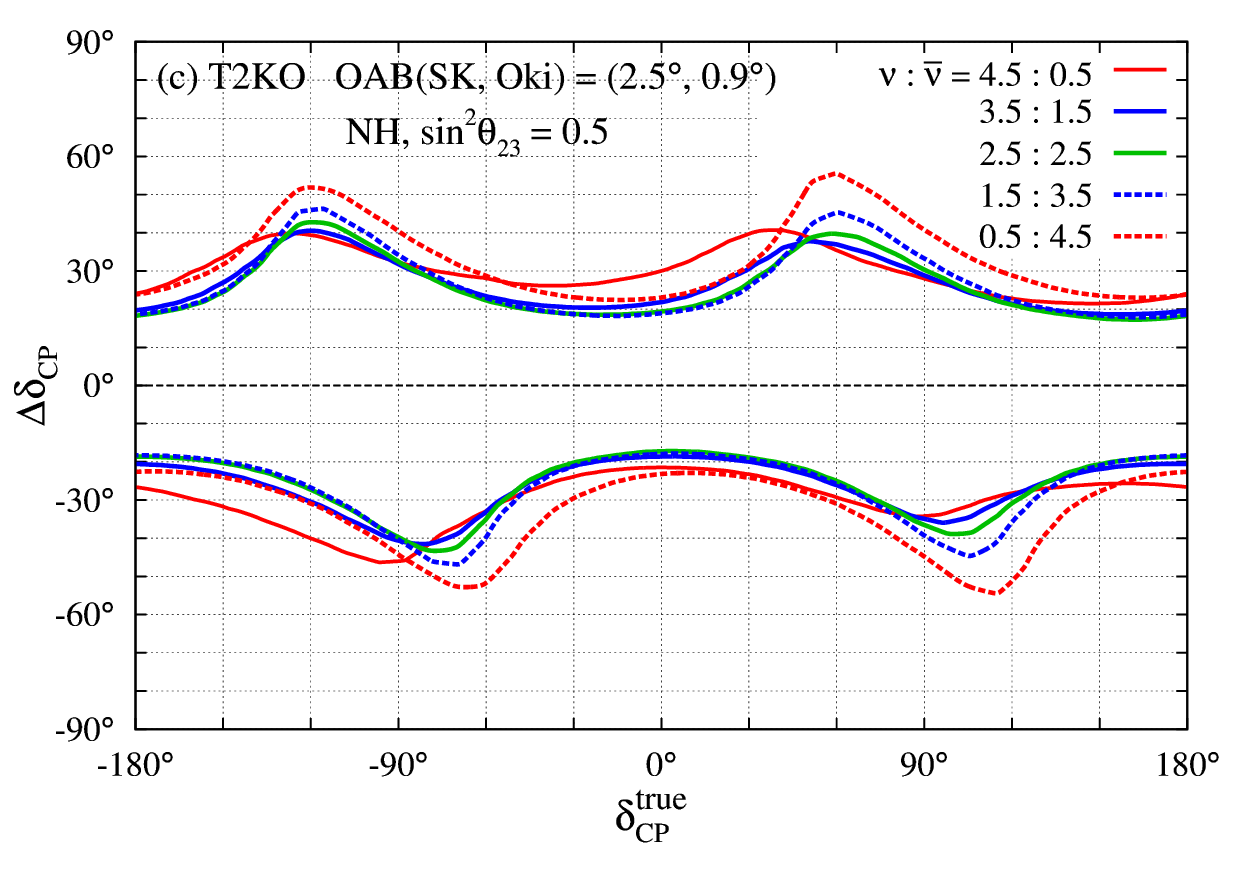} 
\includegraphics[width=0.5\textwidth]{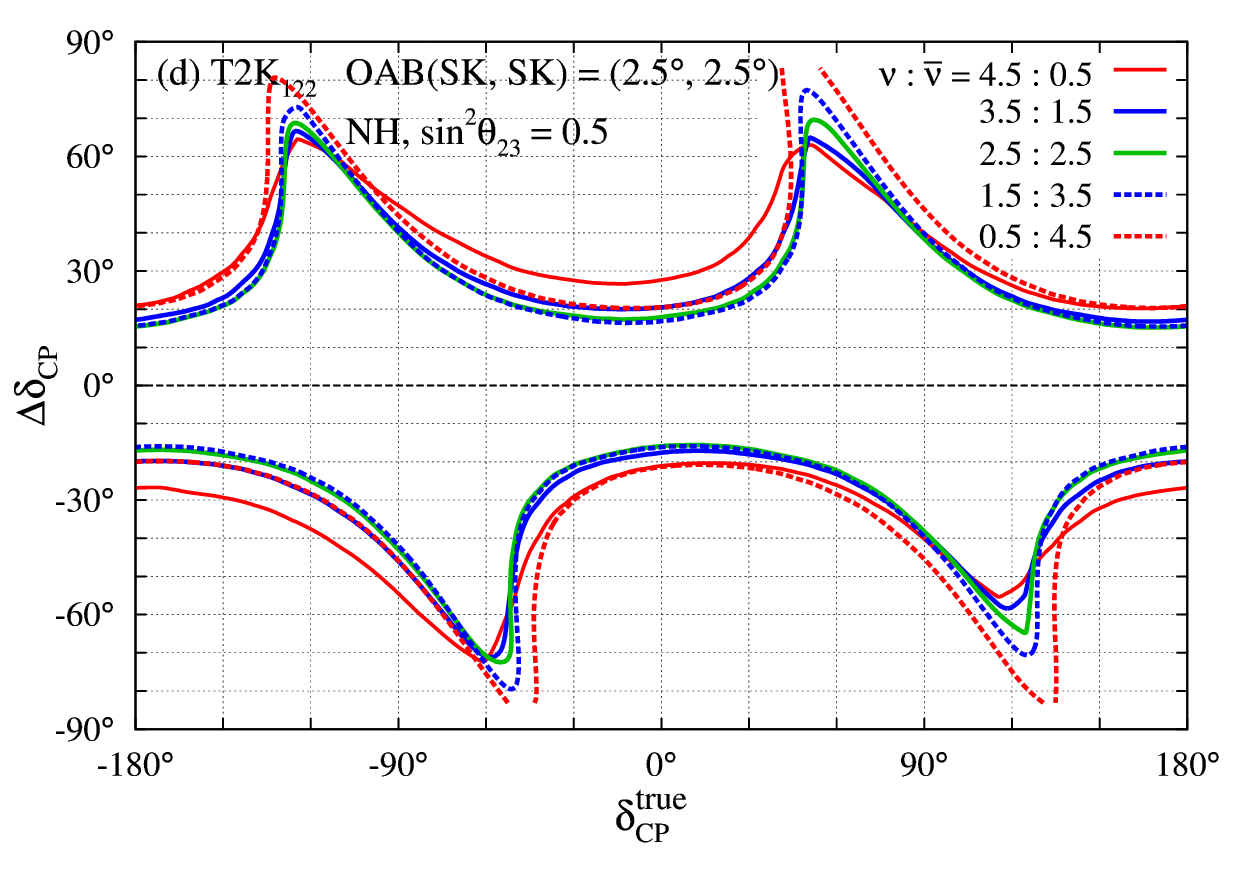} 
}
\caption{The uncertainty of CP phase measurements as functions of the
 CP phase when $\sin^2\theta_{23} = 0.5$, and mass hierarchy is
 known to be the normal hierarchy. 
The
 solid-red, solid-blue, solid-green, dashed-blue and dashed-red curves
 are for the $\nu_\mu$ - $\bar{\nu}_\mu$ focusing beam ratio
 of 4.5\,:\,0.5, 3.5\,:\,1.5, 2.5\,:\,2.5, 1.5\,:\,3.5 and 0.5\,:\,4.5 $\times 10^{21}$ POT with the proton energy of 40 GeV, respectively.}
\label{fig:CPmeasure_nh}
\end{figure}
The uncertainty is defined by the deviation of the test $\dmns$ from
the true $\delta_{\rm CP}$ which gives $\dchisqmin = 1$. The curves
correspond to the different $\nu_{\mu}$ - $\bar{\nu}_{\mu}$ focusing beam
ratios: $4.5:0.5$ (solid-red), $3.5:1.5$ (solid-blue), $2.5:2.5$
(solid-green), $1.5:3.5$ (dashed-blue) and $0.5:4.5$ (dashed-red) in the
unit of $10^{21}$ POT. 
The uncertainty of the
 CP phase measurements is
 smallest around $\dmns = 0^\circ$ and $180^\circ$. 
 This is
  because the uncertainty mainly reflects the
  $\sin\dmns$ dependence of the signal event number since
the magnitude of
 the $\sin\dmns$ term is larger than that of the $\cos\dmns$ term in
 Eq.~\eqref{eq:Pmue} on average.

On the other hand, the uncertainty is largest around $\dmns = \pm
60^\circ$ and $\pm 120^\circ$ as clearly shown in
the $\T2K122$ experiment, Fig.~\ref{fig:CPmeasure_nh}(d); for the T2KK and T2KO experiments, the low
 sensitivity regions slightly shift from $\pm 60^\circ$ and $\pm
 120^\circ$ due to the matter effects~\cite{Coloma:2012wq}. 
This low sensitivity reflects the degeneracy
 between $\dmns$ and $\pi -\dmns$ in $\sin\dmns$. To resolve the degeneracy, we need
 information of the $\cos\dmns$ term, which becomes large around tails
 of oscillation peaks.
 The
T2KK and T2KO experiments observe up to the second peak of the $\nu_\mu
\rightarrow \nu_e$ and $\nubar_\mu \ra \nubar_e$ oscillations, while the $\T2K122$
experiment only observes the first peak (see Figs.~\ref{fig:erec_sk} -
 \ref{fig:erec_t2kk}). Therefore, the former experiments are more sensitive to the
$\cos\dmns$ term
and can measure the CP phase more accurately around those low
sensitive regions.

As for the $\nu_\mu$ - $\bar{\nu}_\mu$ focusing beam ratio, $\nu_\mu:\nubar_\mu =$ 3.5\,:\,1.5 - 1.5\,:\,3.5 give the smallest
uncertainty for most of the CP phases, except for the low sensitivity
region, where the ratio of $4.5:0.5$ gives the best
accuracy. Using the 2.5\,:\,2.5 beam ratio, for example, the T2KK and T2KO experiments measure the CP phase with
the uncertainty of $\sim 20^\circ$ - $50^\circ$ (T2KK with $3.0^\circ$
OAB at the SK), $\sim 20^\circ$ - $45^\circ$ (T2KK with $2.5^\circ$ OAB at
the SK and T2KO)
and $\sim 15^\circ$ - $70^\circ$ ($\T2K122$), depending on the CP
phase.

The effects of including the $\bar{\nu}_\mu$ focusing beam can be
understood in terms of correlations of the oscillation parameters between the $\nu_\mu$ and $\bar{\nu}_\mu$ focusing beams. 
We illustrate this point taking the $\T2K122$ experiment as an example. In Fig.~\ref{fig:chi2-pulls}, we show the $\dchisq$ minimums and pull
 factors of the oscillation parameters for
the different $\nu_\mu$ - $\bar{\nu}_\mu$ focusing beam ratios as
 functions of the test $\dmns$.
\begin{figure}[t]
\centering
\resizebox{1.0\textwidth}{!}{
 \includegraphics[width=0.5\textwidth]{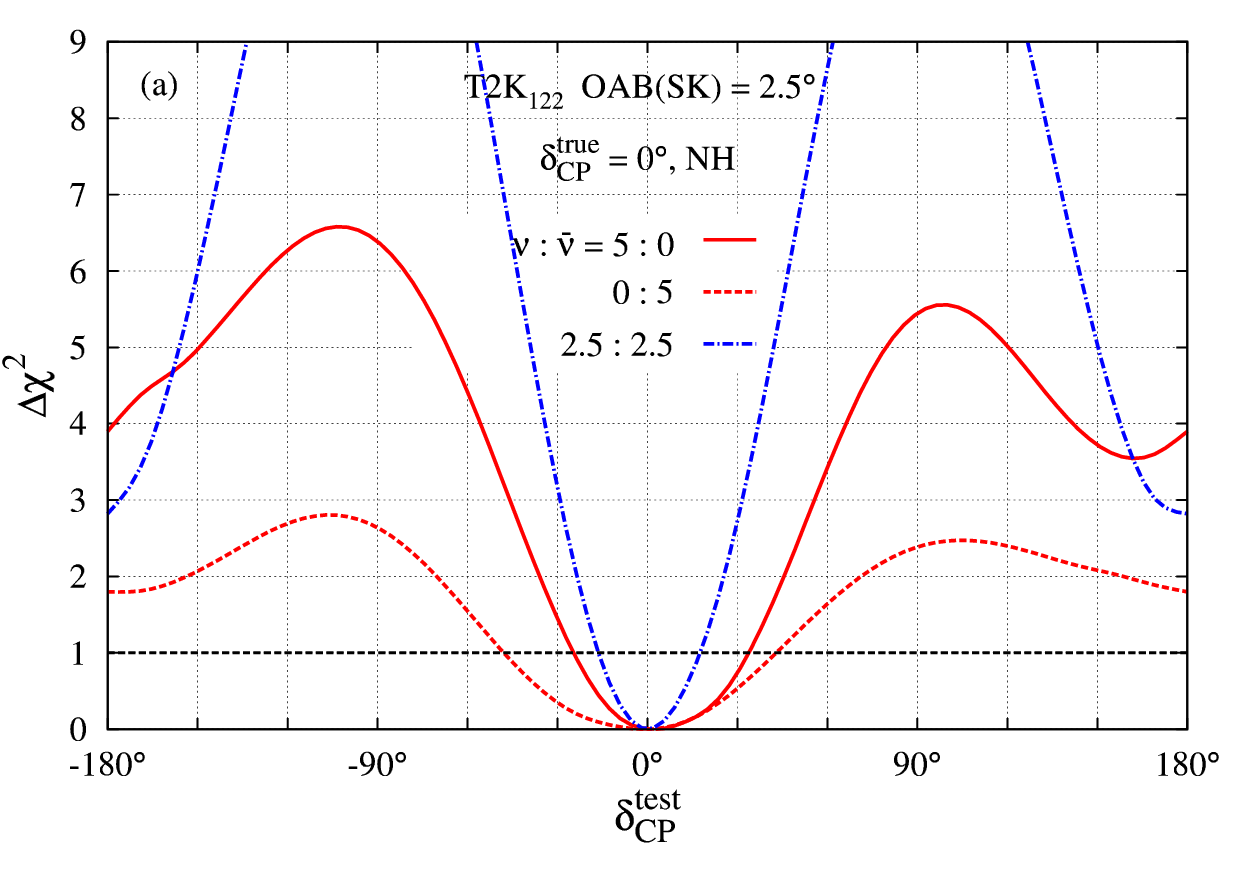} 
 \includegraphics[width=0.5\textwidth]{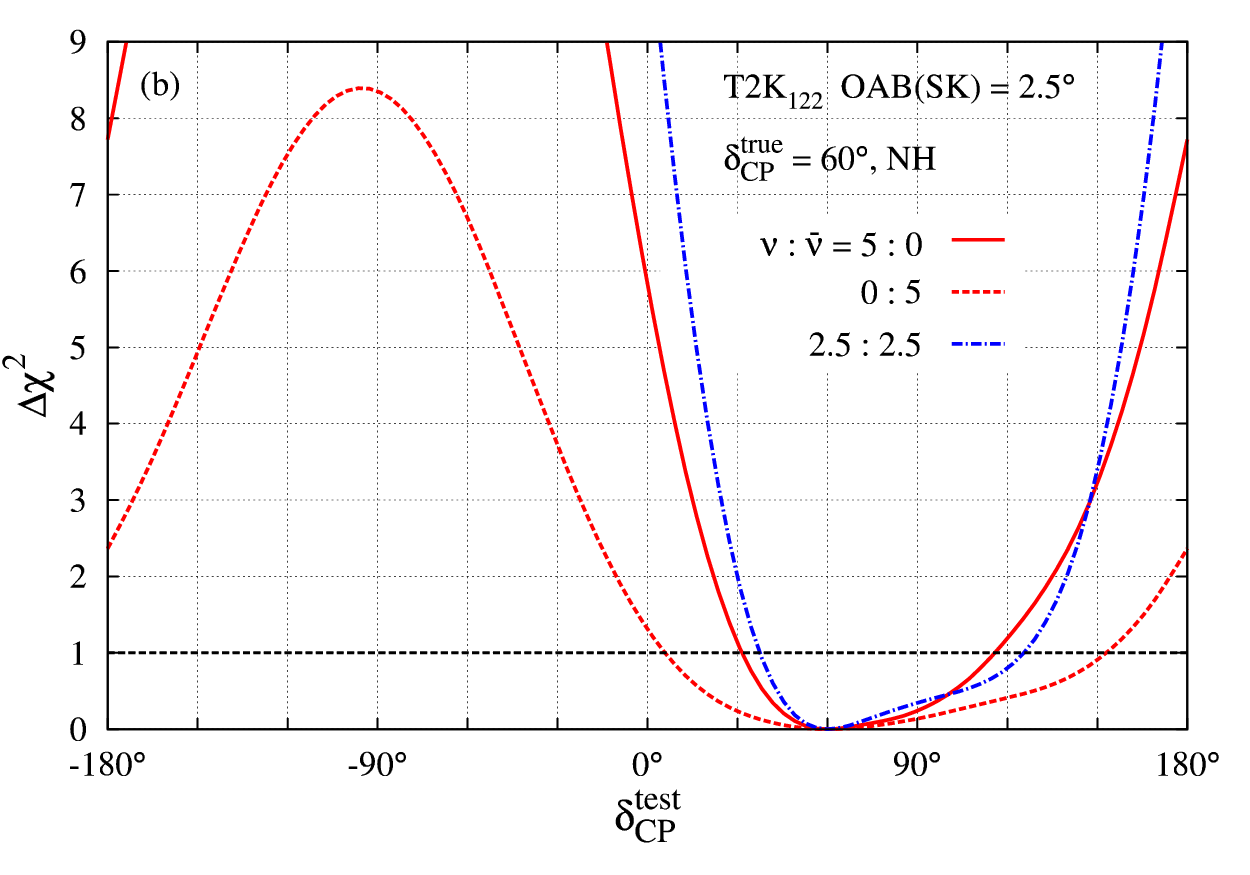} 
}
\resizebox{1.0\textwidth}{!}{
 \includegraphics[width=0.5\textwidth]{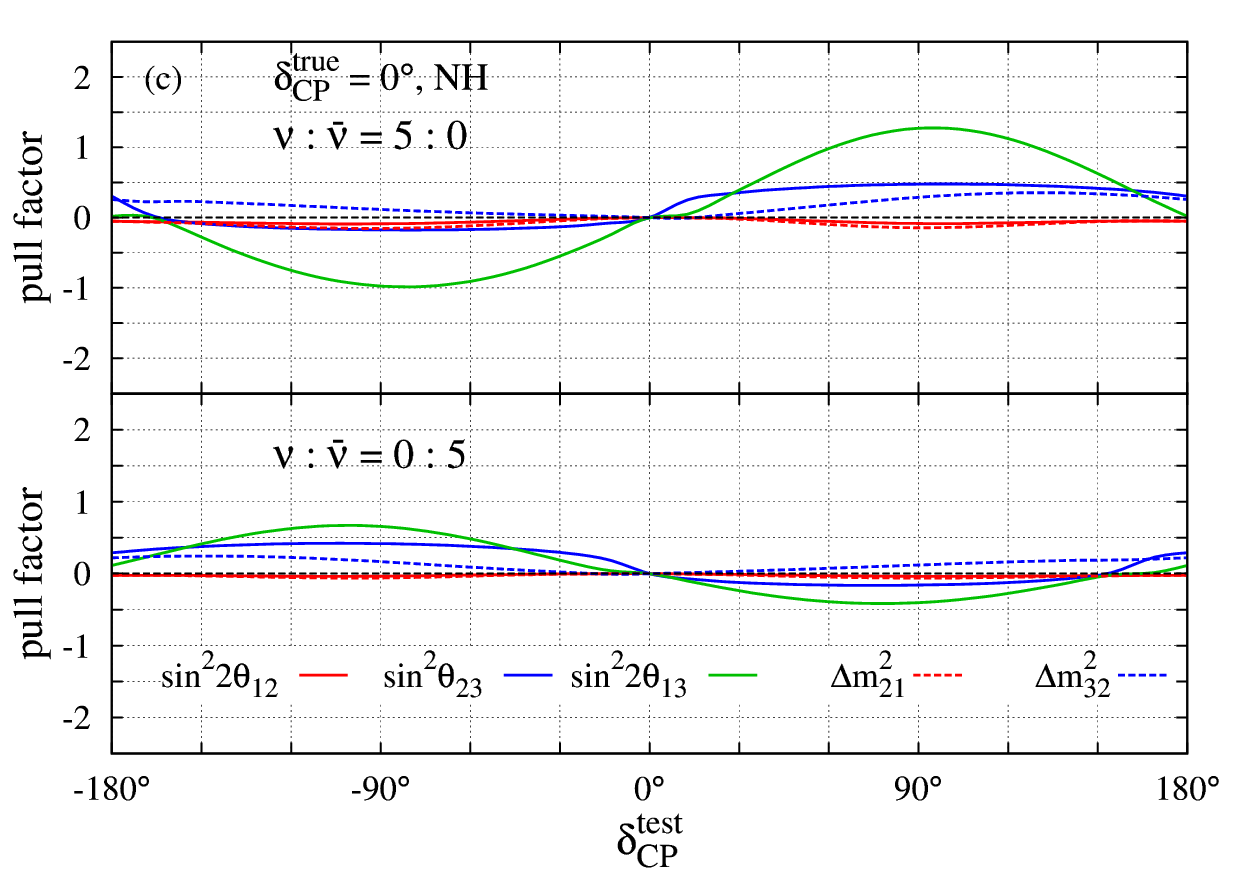} 
 \includegraphics[width=0.5\textwidth]{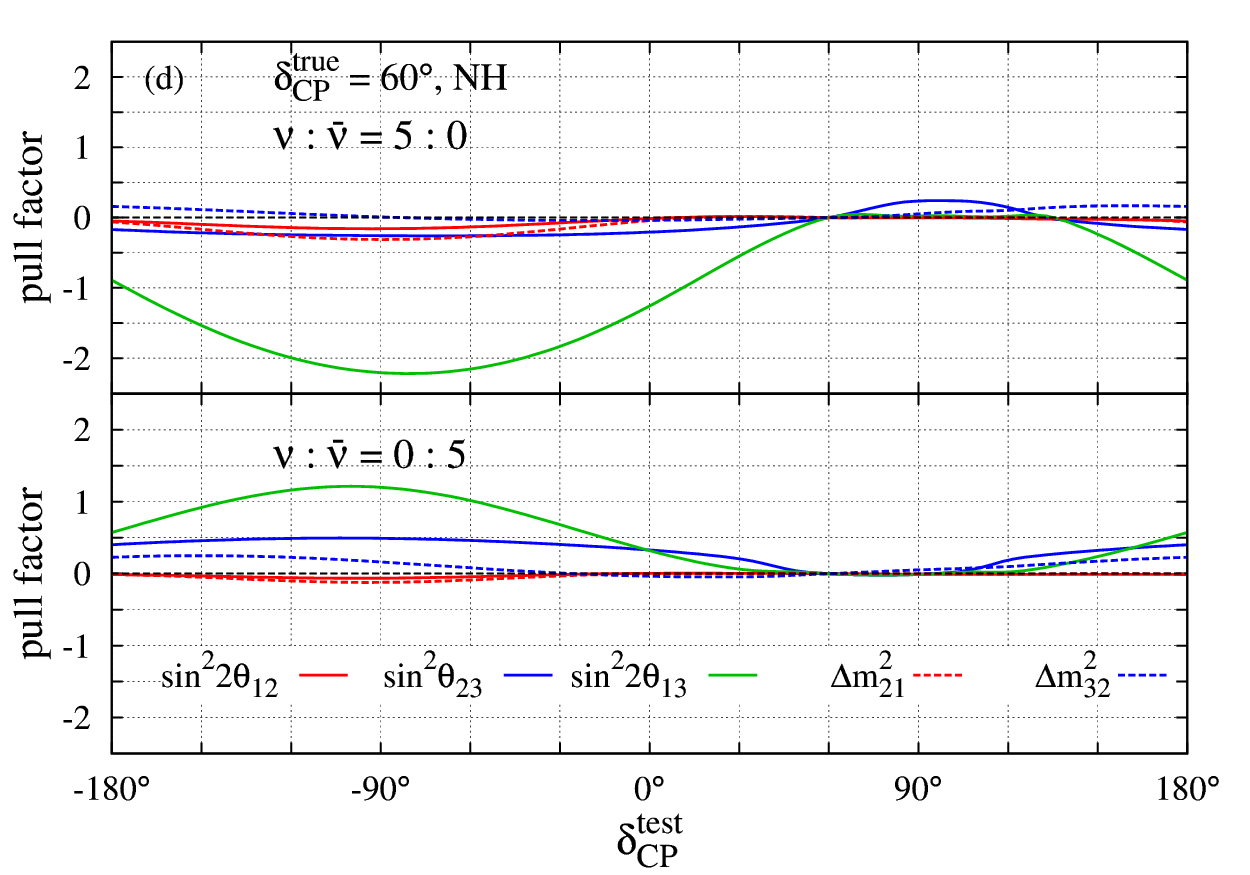} 
}
\caption{The $\dchisqmin$ ((a), (b)) and pull
 factors of the oscillation parameters ((c), (d)) as
 functions of the test $\dmns$ in the $\T2K122$ experiment. The left
 ((a), (c)) and right ((b), (d)) panels are for $\dmns^{\rm true} = 0^\circ$ and $60^\circ$, respectively. Solid-red, dashed-red and
 dash-dotted blue curves in the panels (a) and (b) show the sensitivity with $\nu_\mu$ - $\bar{\nu}_\mu$
 focusing beam ratio of 5\,:\,0, 0\,:\,5 and 2.5\,:\,2.5 $\times
 10^{21}$ POT with the proton energy of 40 GeV, respectively. The upper and lower half
 figures in the panels (c) and (d) show pull factors with $\nu_\mu :
 \bar{\nu}_\mu =$ 5\,:\,0 and 0\,:\,5, respectively. It is assumed that
 the mass hierarchy is known to be the normal hierarchy and $\ssq{}23 = 0.5$.}
\label{fig:chi2-pulls}
\end{figure}
The pull factor of a fitting parameter $X$ is defined as
 $(X^{\rm fit} -X^{\rm input})/\delta X$, where $\delta X$ is the
 uncertainty of the parameter. 
In the upper-left panel (a), solid-red, dashed-red and
 dash-dotted blue curves show the
 $\dchisq$ minimums for the $\nu_\mu$ - $\bar{\nu}_\mu$
 focusing beam ratios of 5\,:\,0, 0\,:\,5 and 2.5\,:\,2.5 $(\times 10^{21}$ POT
 with the proton energy of 40 GeV), respectively. Here, the true CP phase is
 assumed to be $0^\circ$, and the mass hierarchy is known
 to be the normal
 hierarchy.  In the lower plot (c), we show the
 corresponding pull factors of the oscillation parameters for $\nu_\mu :
 \bar{\nu}_\mu =$ 5\,:\,0 (upper-half panel) and 0\,:\,5 (lower-half panel). We see
 that each pull factor of $\ssq213$ and $\ssq{}23$ shows clear anti-correlation
 between the $\nu_\mu$ and $\bar{\nu}_\mu$ focusing beams, i.e., the
 sign of the each pull factor is opposite between those
 focusing beams. This is
 because the sign of those pull factors is mainly related to the sign of the
 $\sin\dmns$ term in the $\nu_\mu \rightarrow
 \nu_e$ oscillation probability (Eq.~\eqref{eq:Pmue}), which is inverted
 for the anti-neutrino beam case. Thus,
 inclusion of the $\bar{\nu}_\mu$ focusing beam would restrict
 the deviations of $\ssq213$ and $\ssq{}23$, resulting in the larger 
 $\dchisq$ minimum for the 2.5\,:\,2.5 beam ratio than 5\,:\,0
 in the upper panel (a). 
For the $\dmns^{\rm true} = 60^\circ$ case (right
panels in Fig.~\ref{fig:chi2-pulls}), on the other hand, the anti-correlation of
the pull-factors is not so evident for $60^\circ \lesssim \delta_{\rm CP}^{\rm test} \lesssim
120^\circ$ 
resulting in the rather reduction of the
accuracy of the CP phase measurement when including $\bar{\nu}_\mu$
focusing beams. 
Similar situation occurs for $\dmns^{\rm true} \sim -60^\circ$ and $\pm 120^\circ$
as well.

In Fig.~\ref{fig:CPmeasure_ih}, we show the uncertainties of the CP phase
measurements for the inverted hierarchy case.
\begin{figure}[t]
\centering
\resizebox{1.0\textwidth}{!}{
\includegraphics[width=0.5\textwidth]{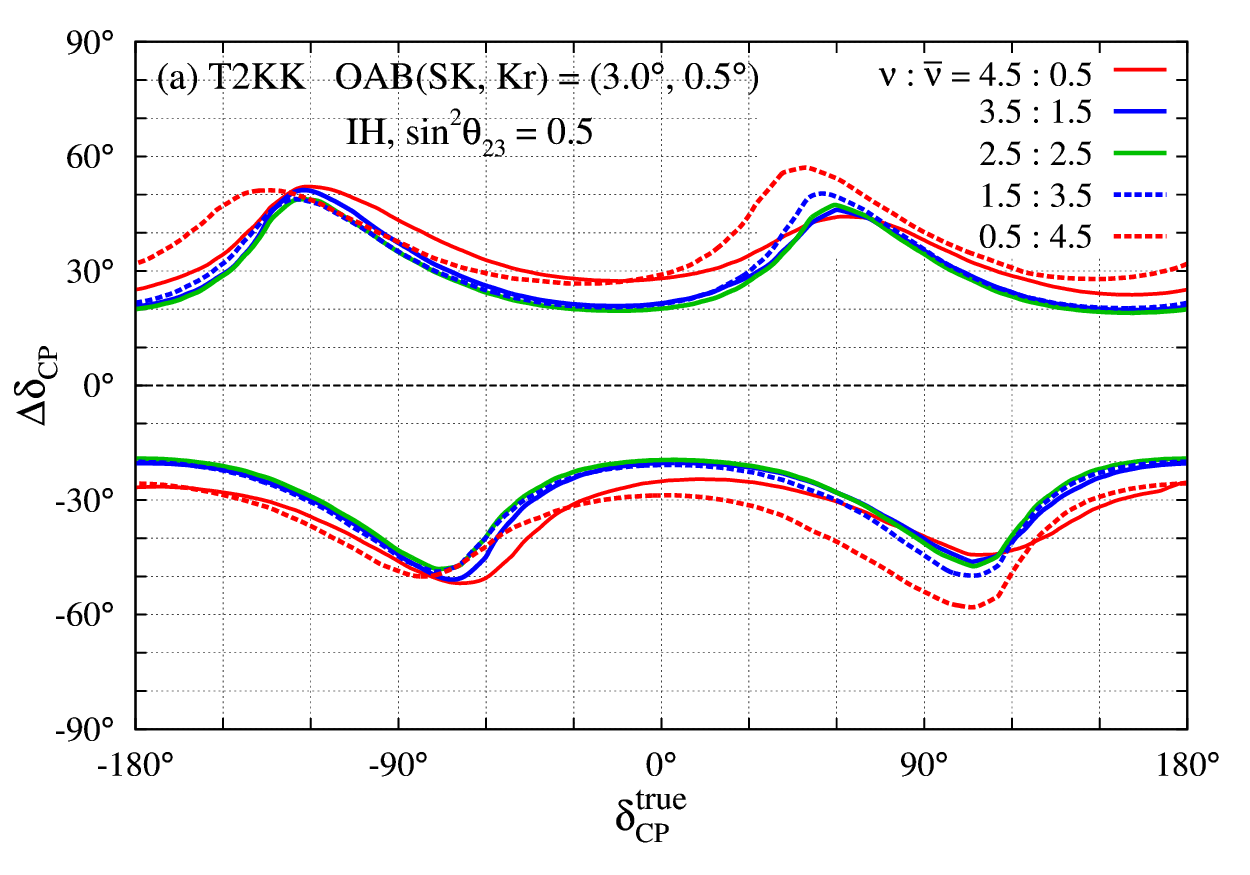} 
\includegraphics[width=0.5\textwidth]{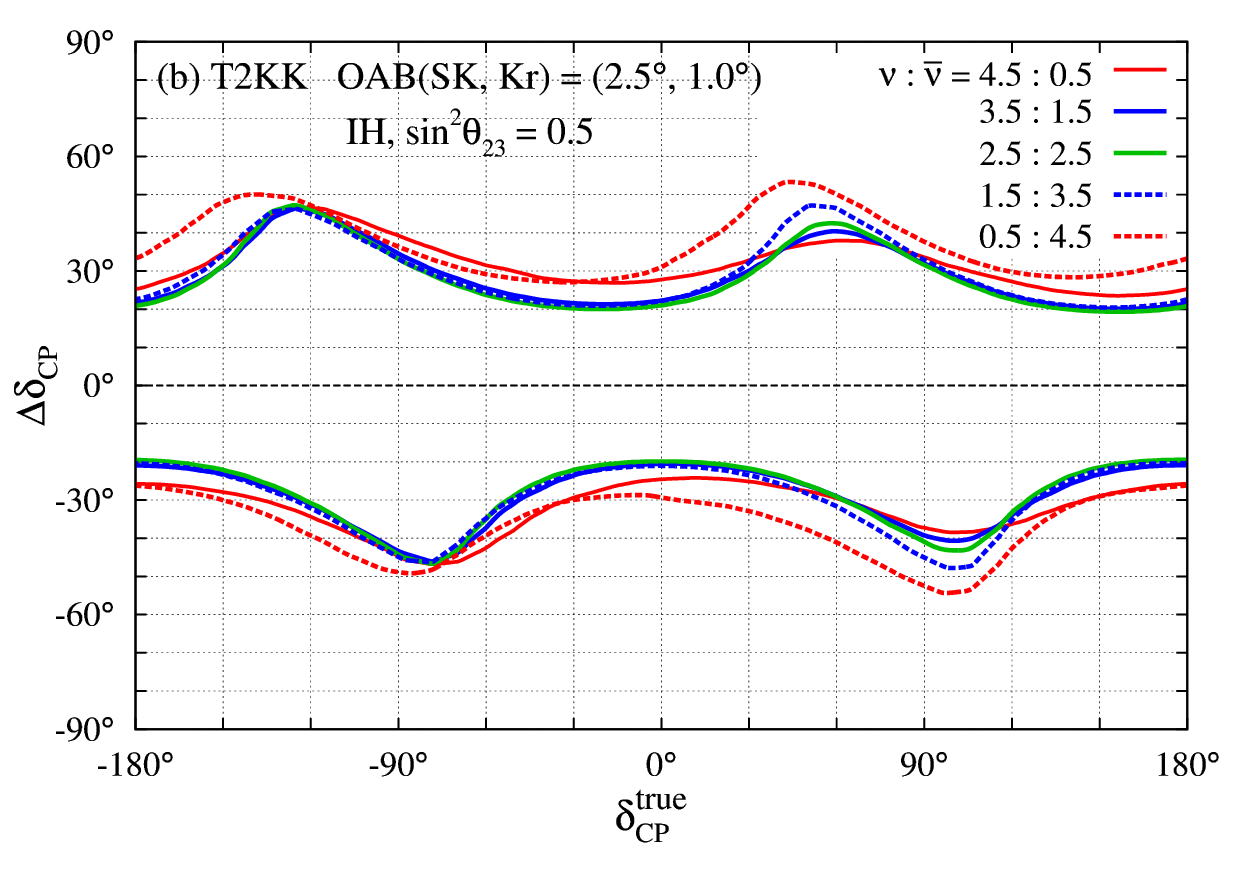} 
}
\resizebox{1.0\textwidth}{!}{
\includegraphics[width=0.5\textwidth]{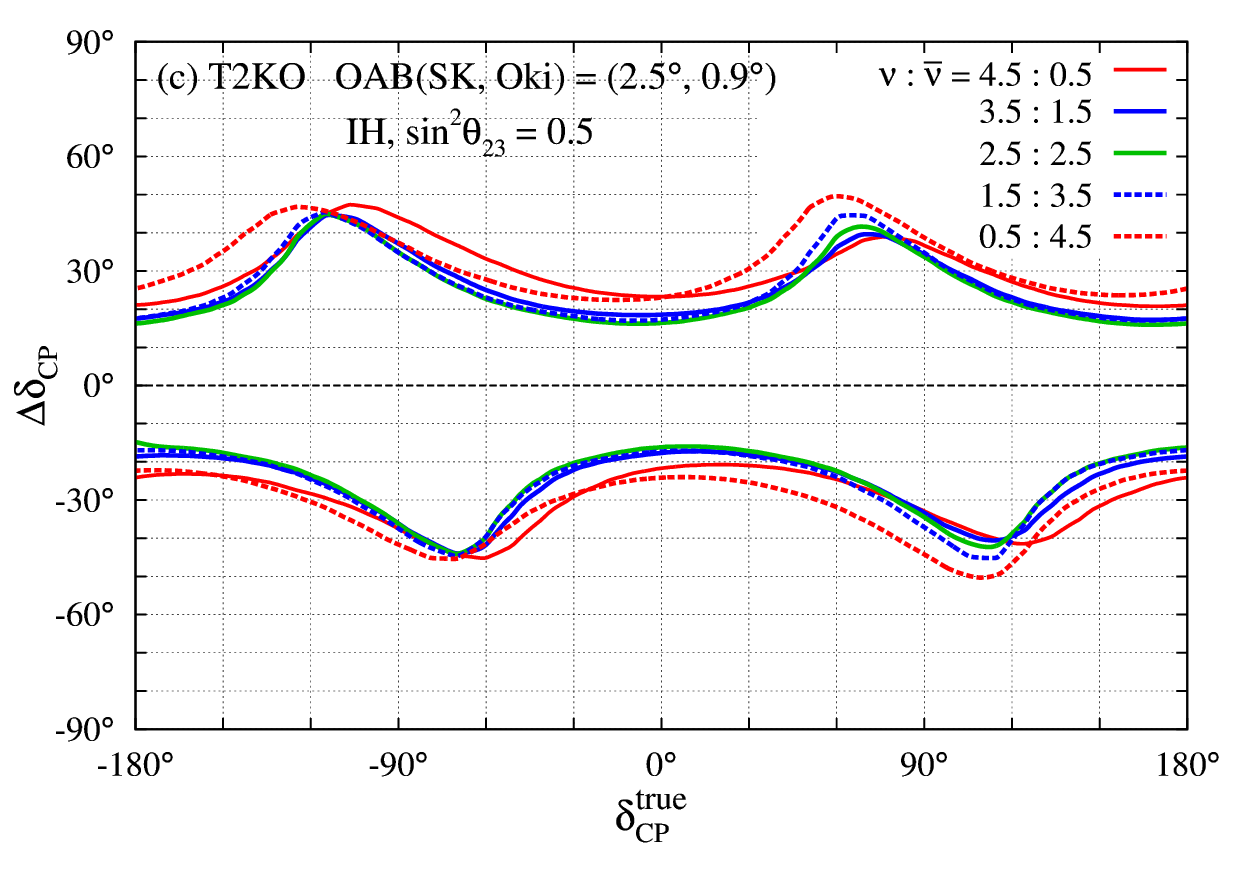} 
\includegraphics[width=0.5\textwidth]{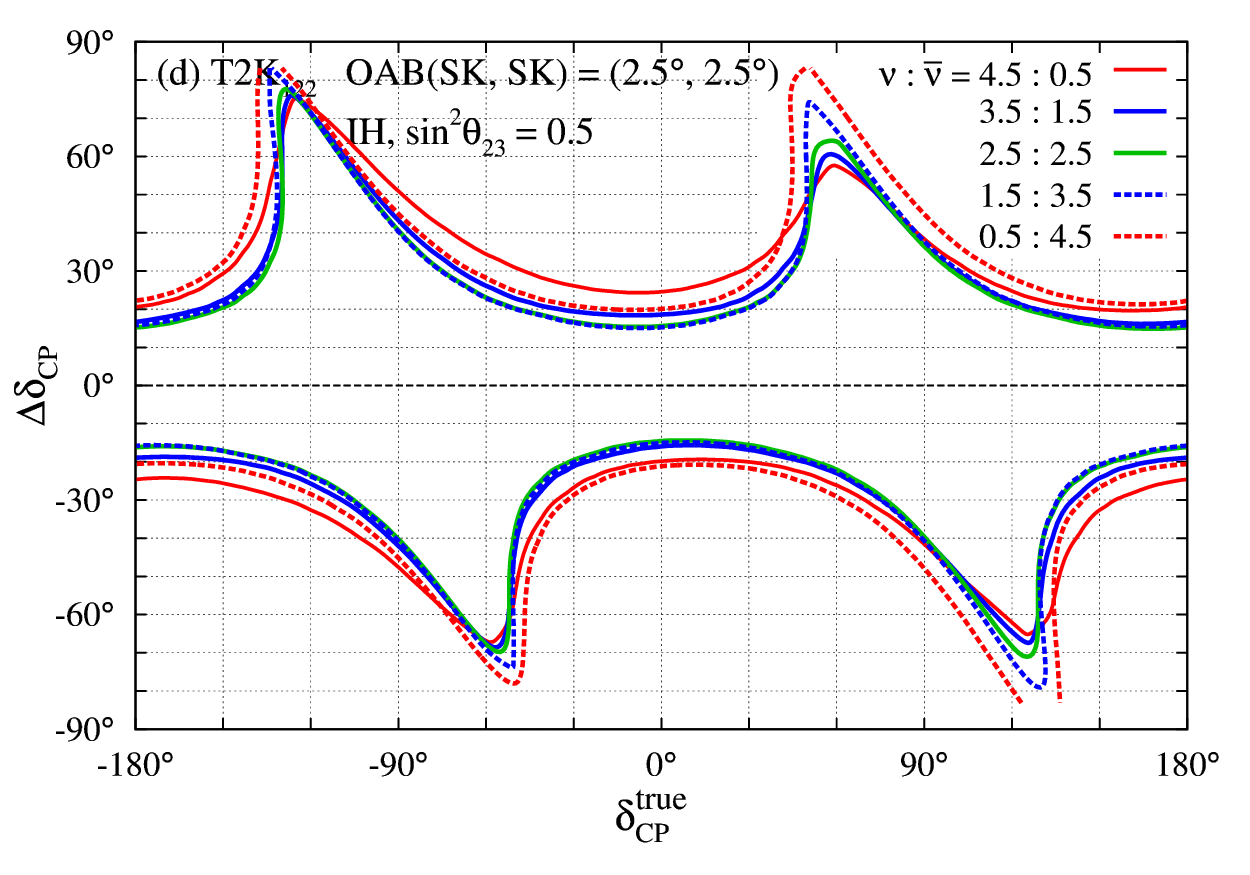} 
}
\caption{Same as Fig.~\ref{fig:CPmeasure_nh} but for the inverted hierarchy case.}
\label{fig:CPmeasure_ih}
\end{figure}
The uncertainties show similar dependences on the $\nu_\mu$ - $\bar{\nu}_\mu$ focusing beam ratio as in the normal hierarchy case,
showing that the 3.5\,:\,1.5 - 1.5\,:\,3.5 beam ratio give the smallest
uncertainty except for $ \delta_{\rm CP} \sim
\pm 60^\circ$ and $\pm 120^\circ$. Using the 2.5\,:\,2.5 beam ratio, the T2KK, T2KO
and $\T2K122$ experiments measure the CP phase with
the uncertainty of $\sim 20^\circ$ - $50^\circ$ (T2KK with $3.0^\circ$
OAB), $\sim 20^\circ$ - $45^\circ$ (T2KK with $2.5^\circ$ OAB), $\sim 15^\circ$ - $45^\circ$ (T2KO)
and $\sim 15^\circ$ - $75^\circ$ ($\T2K122$), depending on the CP
phase. 

We also show the sensitivities to the CP phase measurements in the test-$\dmns$ vs. true-$\dmns$
 plane  in Fig.~\ref{fig:3sigma_nh} when $\sin^2\theta_{23} = 0.5$ and the mass hierarchy is known to be
 the normal hierarchy.
\begin{figure}[t]
\centering
\resizebox{1.0\textwidth}{!}{
\hspace{-5em}
\includegraphics[width=0.68\textwidth]{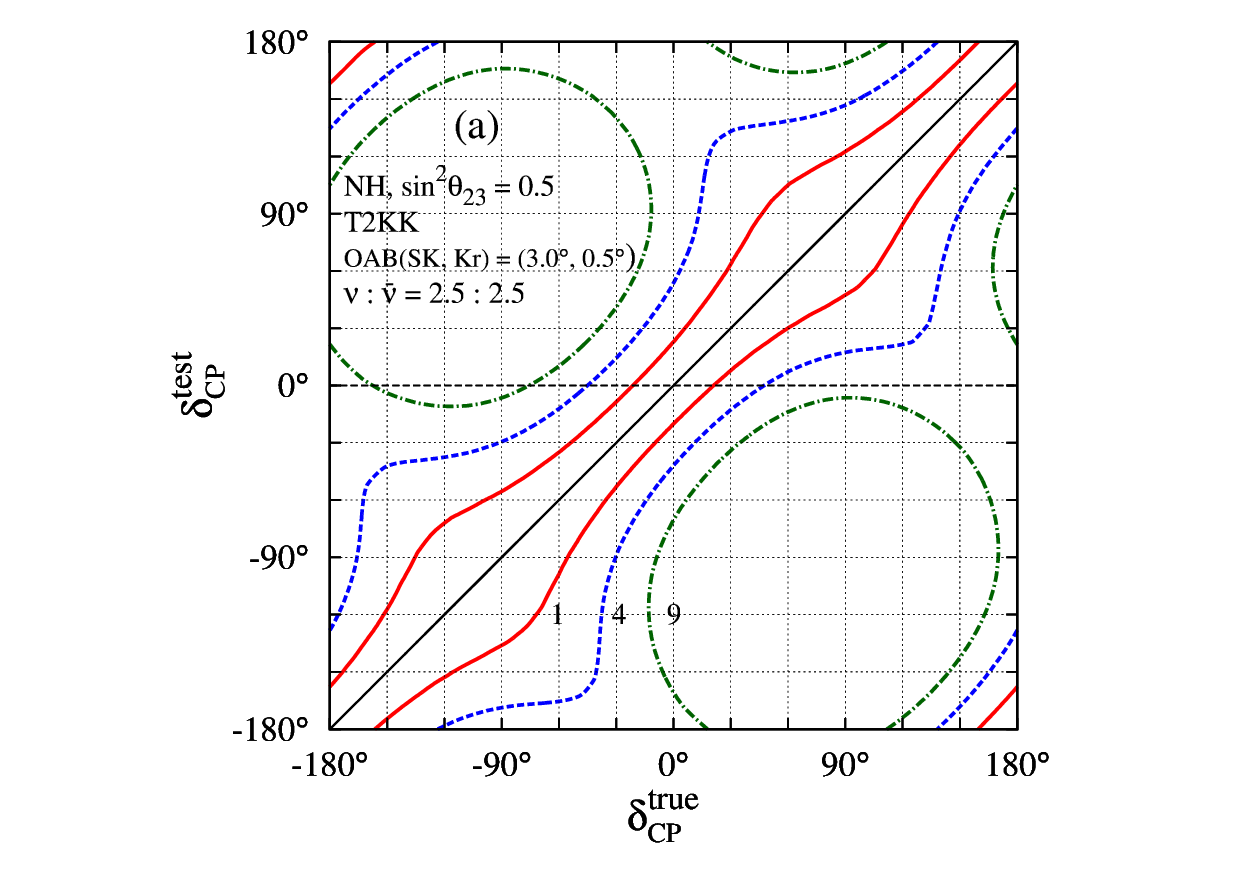}
\hspace{-8em}
\includegraphics[width=0.68\textwidth]{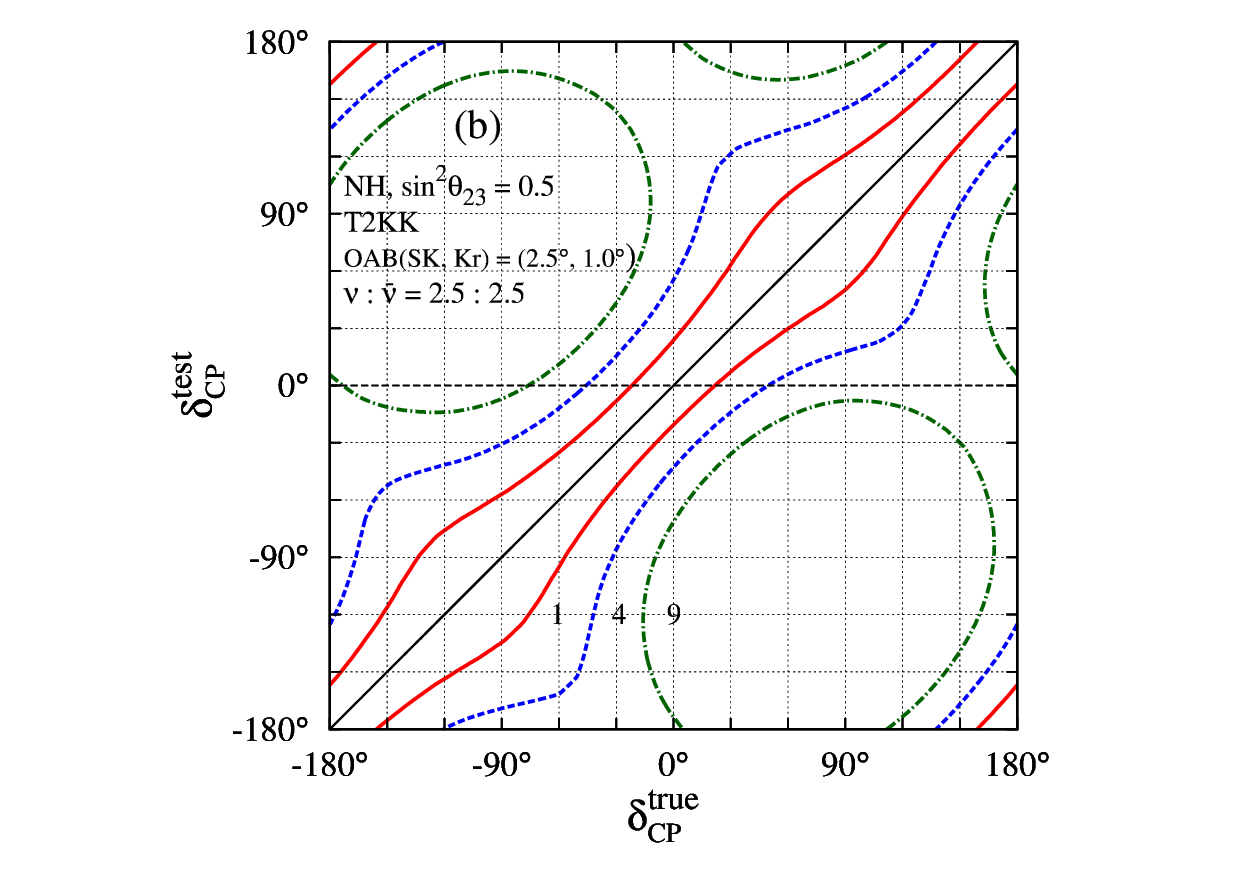} 
}
\resizebox{1.0\textwidth}{!}{
\hspace{-5em}
\includegraphics[width=0.68\textwidth]{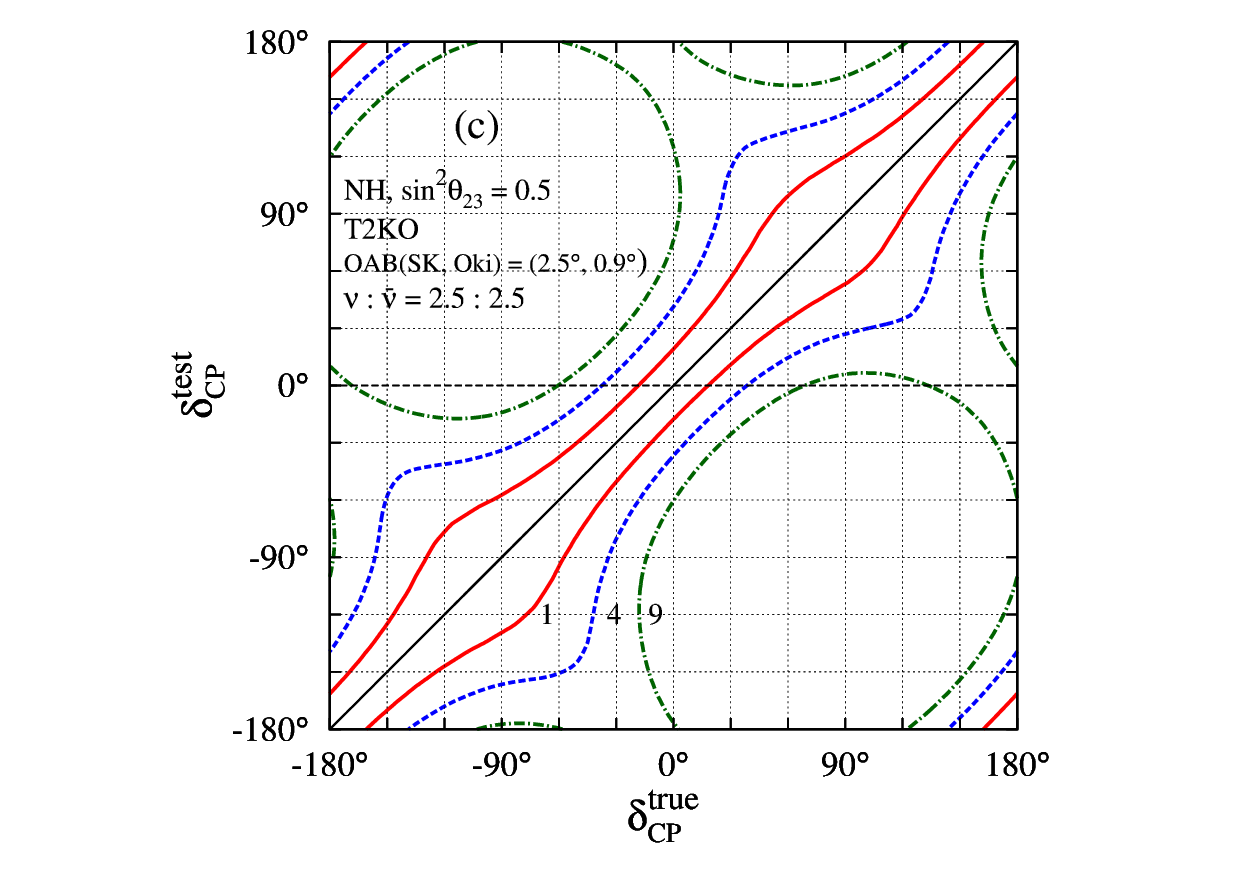}
\hspace{-8em}
\includegraphics[width=0.68\textwidth]{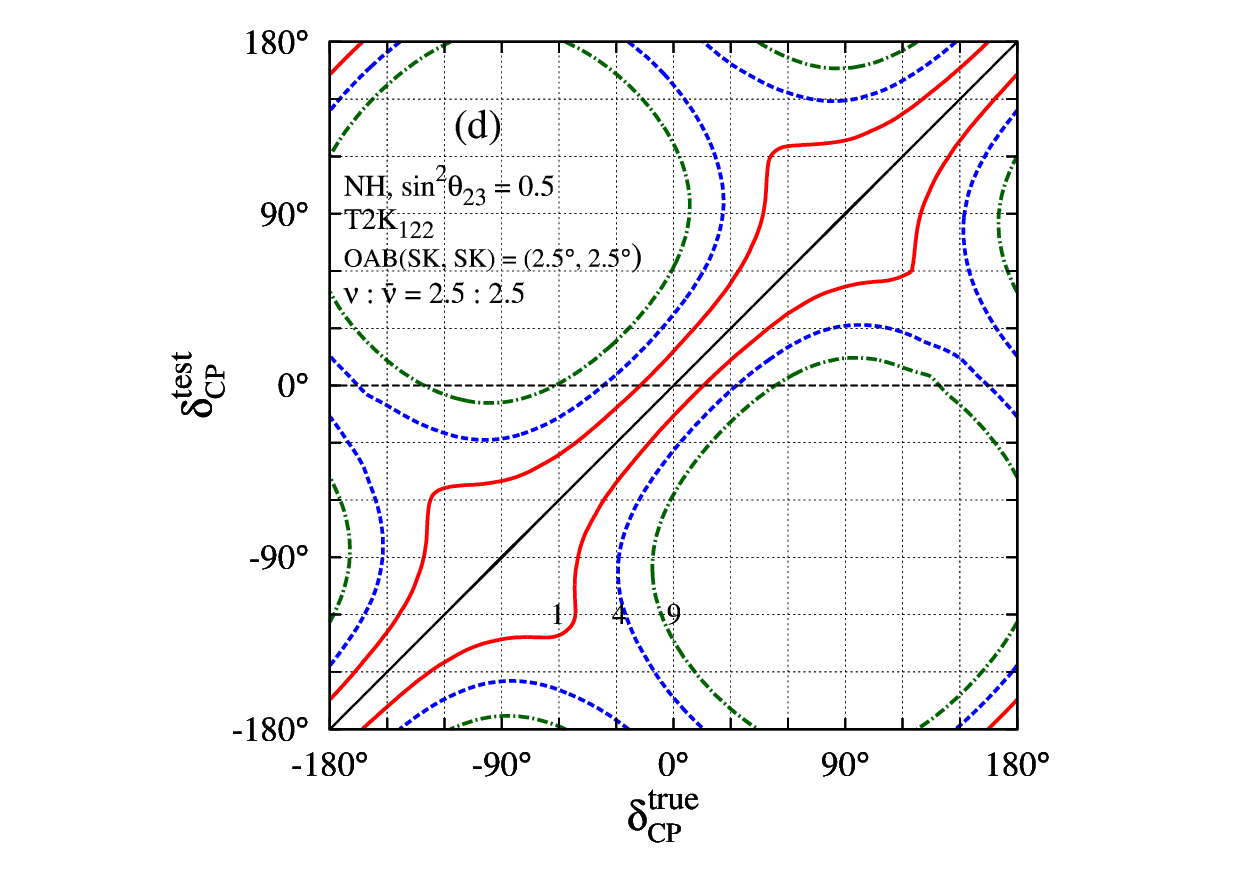} 
}
\caption{The sensitivities to the CP phase measurements in the test-$\dmns$ vs. true-$\dmns$
 plane when $\sin^2\theta_{23} = 0.5$, and mass hierarchy is known to be
 the normal hierarchy. (a) T2KK with $3.0^\circ$ OAB at the SK, (b) T2KK
 with $2.5^\circ$ OAB at the SK, (c) T2KO and (d) $\T2K122$. The $\nu_\mu$ - $\bar{\nu}_\mu$ focusing beam ratio is fixed at
 2.5\,:\,2.5 $\times 10^{21}$POT with the proton energy of 40 GeV. The solid-red, dashed-blue and dash-dotted-green contours show
 $\dchisqmin = 1, 4, 9$, respectively.}
\label{fig:3sigma_nh}
\end{figure}
The $\nu_\mu$ - $\bar{\nu}_\mu$ focusing beam ratio is fixed at
 2.5\,:\,2.5. The solid-red, dashed-blue and dash-dotted-green contours show
 $\dchisqmin = 1, 4, 9$, respectively, for (a) T2KK with 3.0$^\circ$
 OAB at the SK and 0.5$^\circ$ OAB at the Kr, (b) T2KK with 2.5$^\circ$
 OAB at the SK and 1.0$^\circ$ OAB at the Kr, (c) T2KO and (d) $\T2K122$
 experiments.
We see that both $\dmns^{\rm test} = 0^\circ$
and $180^\circ$ are rejected with the significance of $\dchisqmin > 9$ for $-100^\circ < \dmns
<-60^\circ$ and $70^\circ < \dmns < 120^\circ$ by the T2KO experiment
and for $-120^\circ < \dmns <-60^\circ$ and $60^\circ < \dmns <
120^\circ$ by the $\T2K122$ experiment. 
The T2KK experiment has less sensitivity to the CP violation than the T2KO and $\T2K122$ experiments
and rejects both $\dmns^{\rm test} = 0^\circ$
and $180^\circ$ with the significance of $9 > \dchisqmin > 4$ for $-120^\circ
(-135^\circ) < \dmns <-45^\circ$ and $45^\circ < \dmns < 140^\circ$ for
the normal (inverted) hierarchy case.
%
The low sensitivity regions due to the $\dmns$ and $\pi -\dmns$
degeneracy of $\sin\dmns$ can be seen around $\dmns^{\rm test} = \pi -\dmns$. We also show the sensitivity
plots for the inverted hierarchy case in Fig.~\ref{fig:3sigma_ih}. The
sensitivity is similar to that of the normal hierarchy case with slight
difference due to the relative sign change between the $\sin\dmns$ and
$\cos\dmns$ terms in the $\nu_\mu \rightarrow \nu_e$ and $\nubar_\mu \ra \nubar_e$
oscillation probability, Eq~\eqref{eq:Pmue}. 
\begin{figure}[t]
\centering
\resizebox{1.0\textwidth}{!}{
\hspace{-5em}
\includegraphics[width=0.68\textwidth]{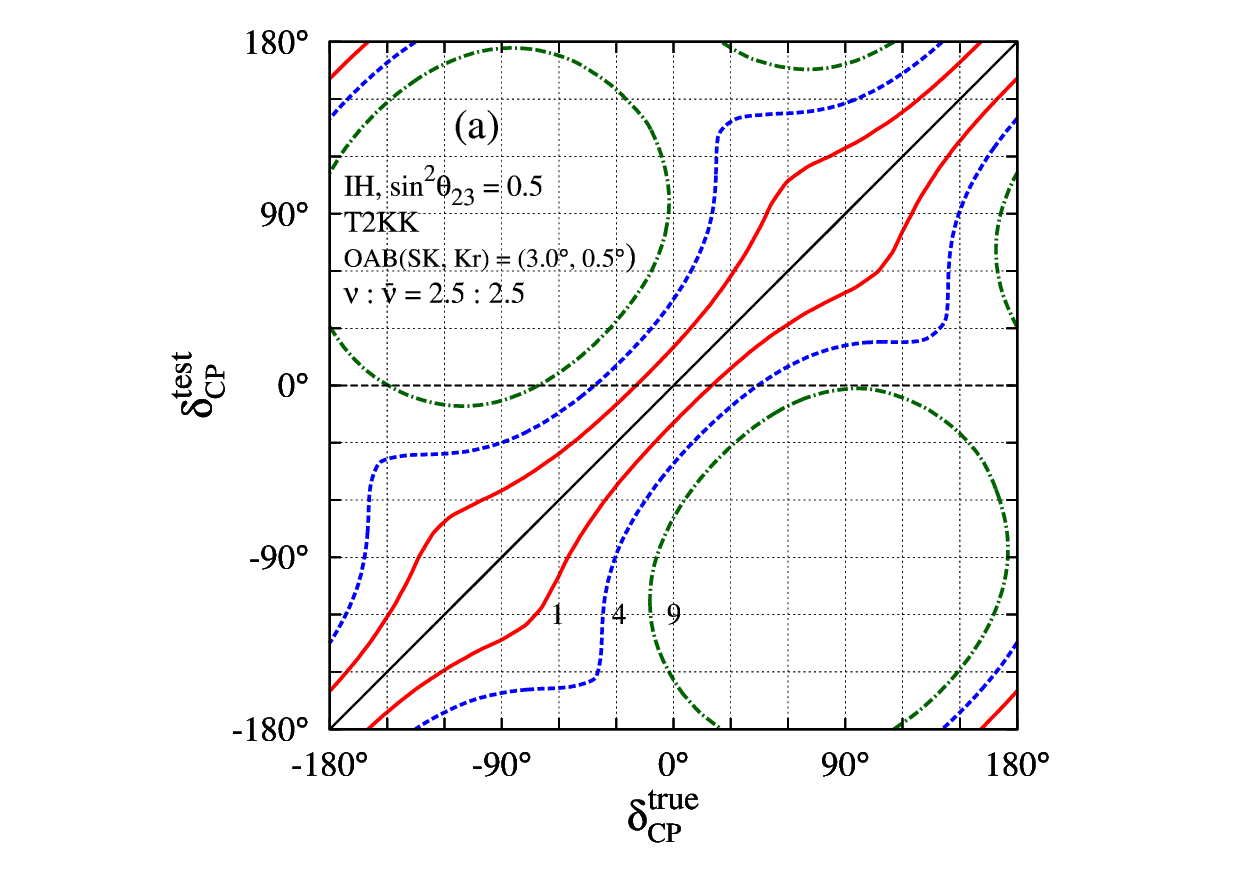} 
\hspace{-8em}
\includegraphics[width=0.68\textwidth]{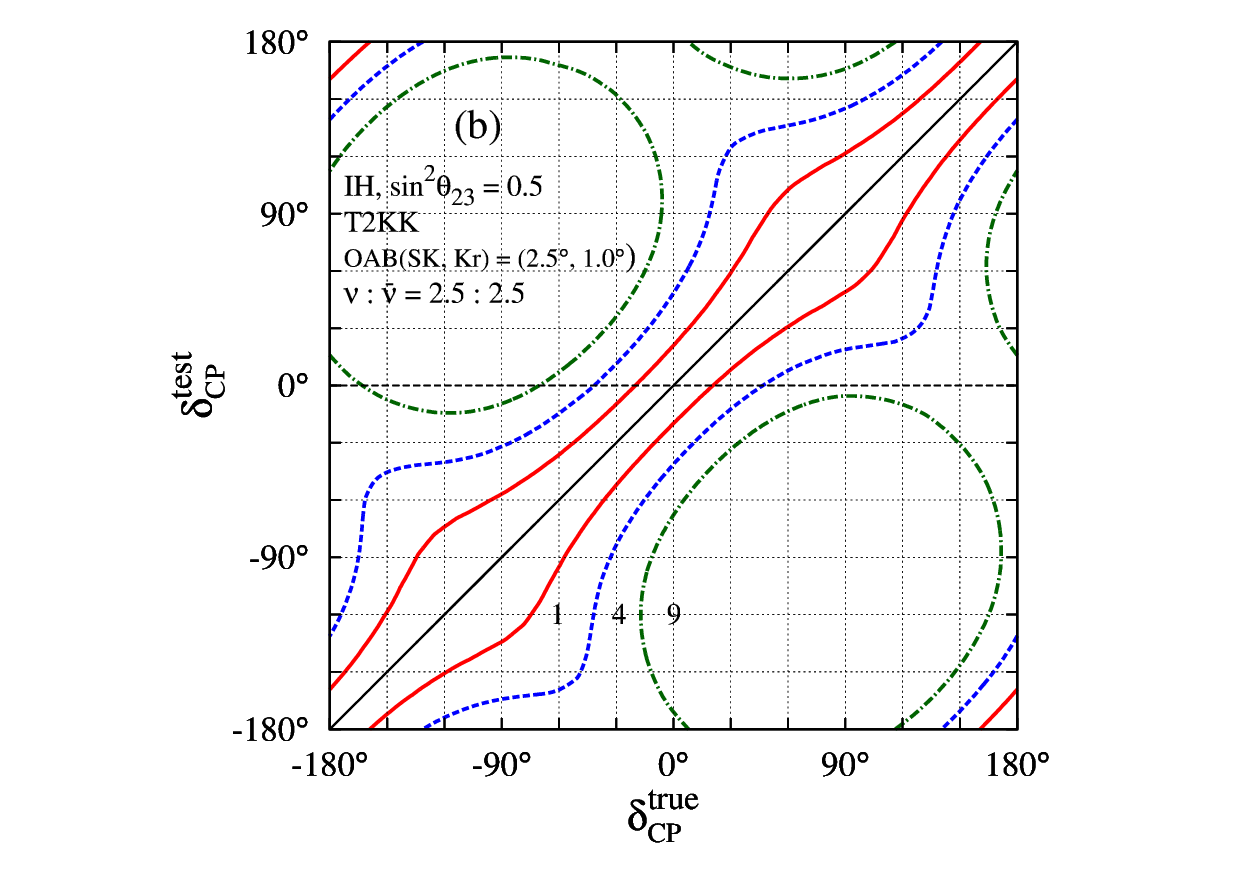} 
}
\resizebox{1.0\textwidth}{!}{
\hspace{-5em}
\includegraphics[width=0.68\textwidth]{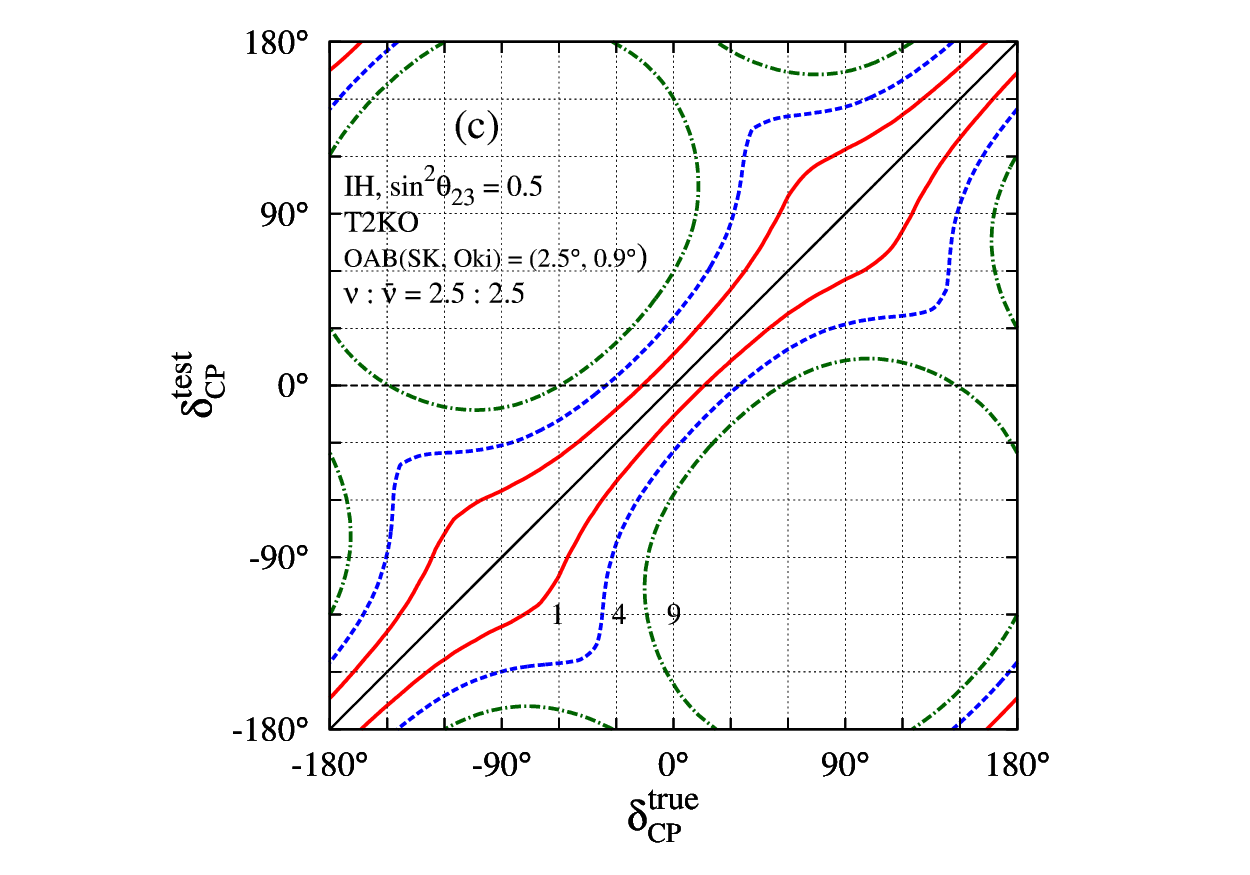} 
\hspace{-8em}
\includegraphics[width=0.68\textwidth]{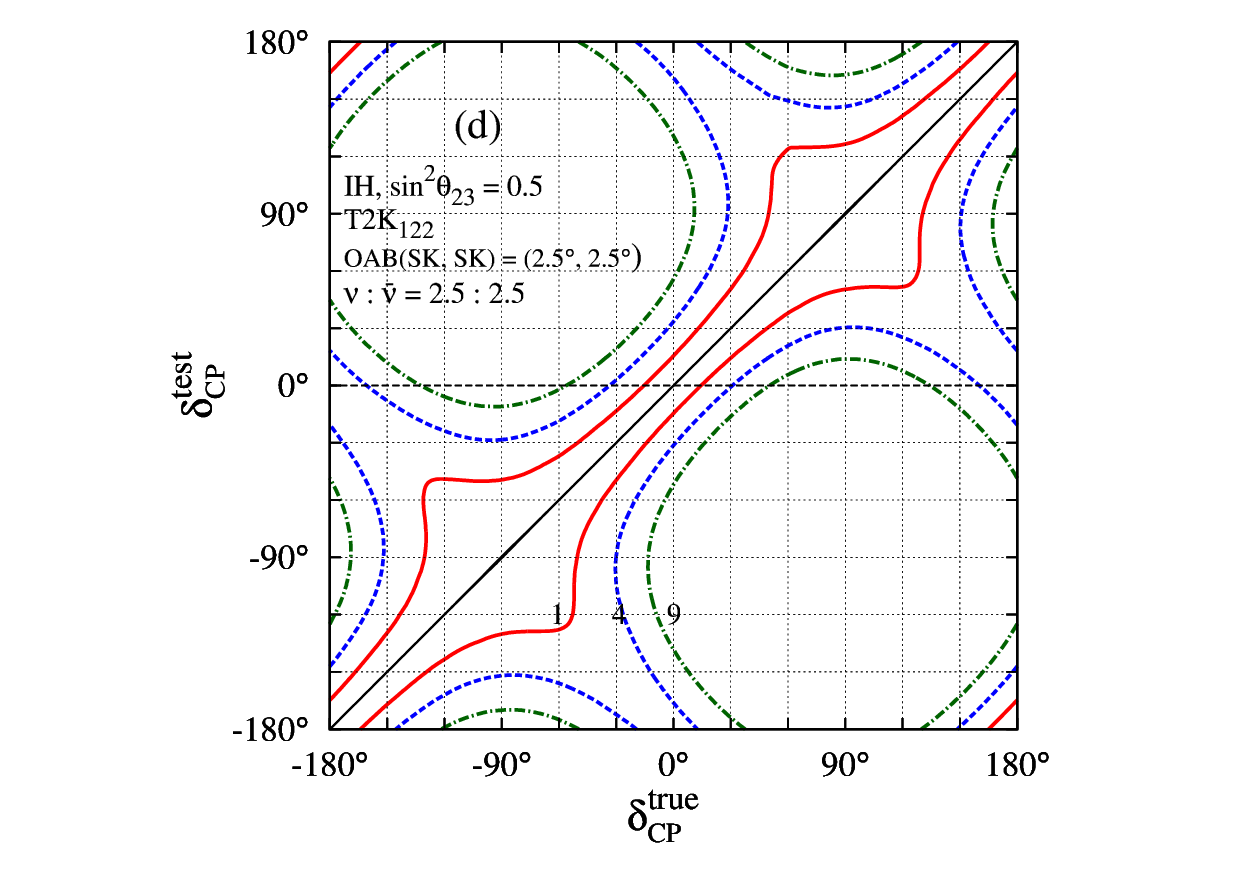} 
}
\caption{Same as Fig.~\ref{fig:3sigma_nh} but for the inverted hierarchy case.}
\label{fig:3sigma_ih}
\end{figure}

The $\ssq{}23$ dependence of the sensitivity to
the CP phase measurements is shown in Fig.~\ref{fig:3sigma_th23}.  
These are the same sensitivity plots as Fig.~\ref{fig:3sigma_nh}(a) but for $\ssq{}23 = 0.4$ (left plot) and
0.6 (right plot) for the T2KK experiment with $3.0^\circ$ OAB at the
SK and $0.5^\circ$ OAB at the Kr. It is assumed that the mass hierarchy is known to be the normal hierarchy.   
\begin{figure}[t]
\centering
\resizebox{1.0\textwidth}{!}{
\hspace{-5em}
\includegraphics[width=0.68\textwidth]{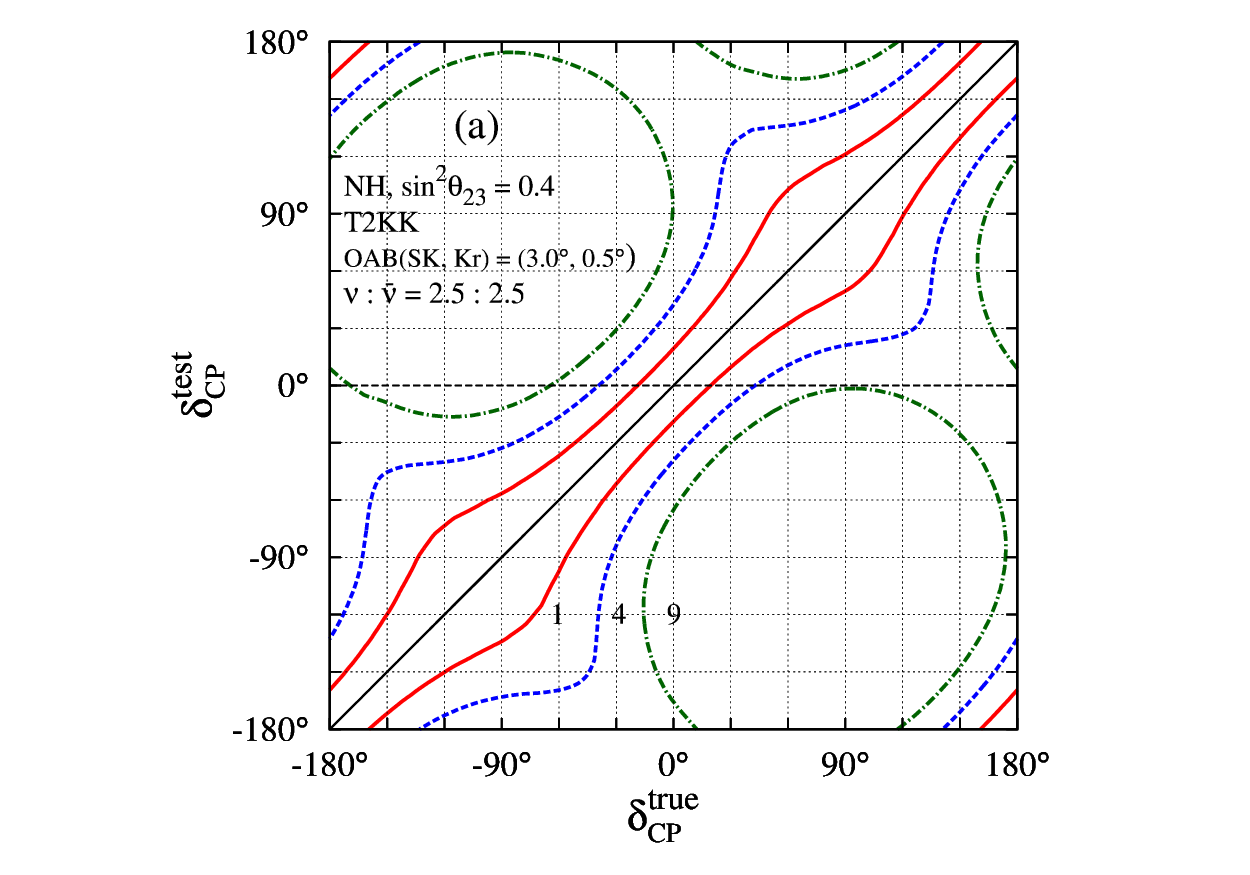} 
\hspace{-8em}
\includegraphics[width=0.68\textwidth]{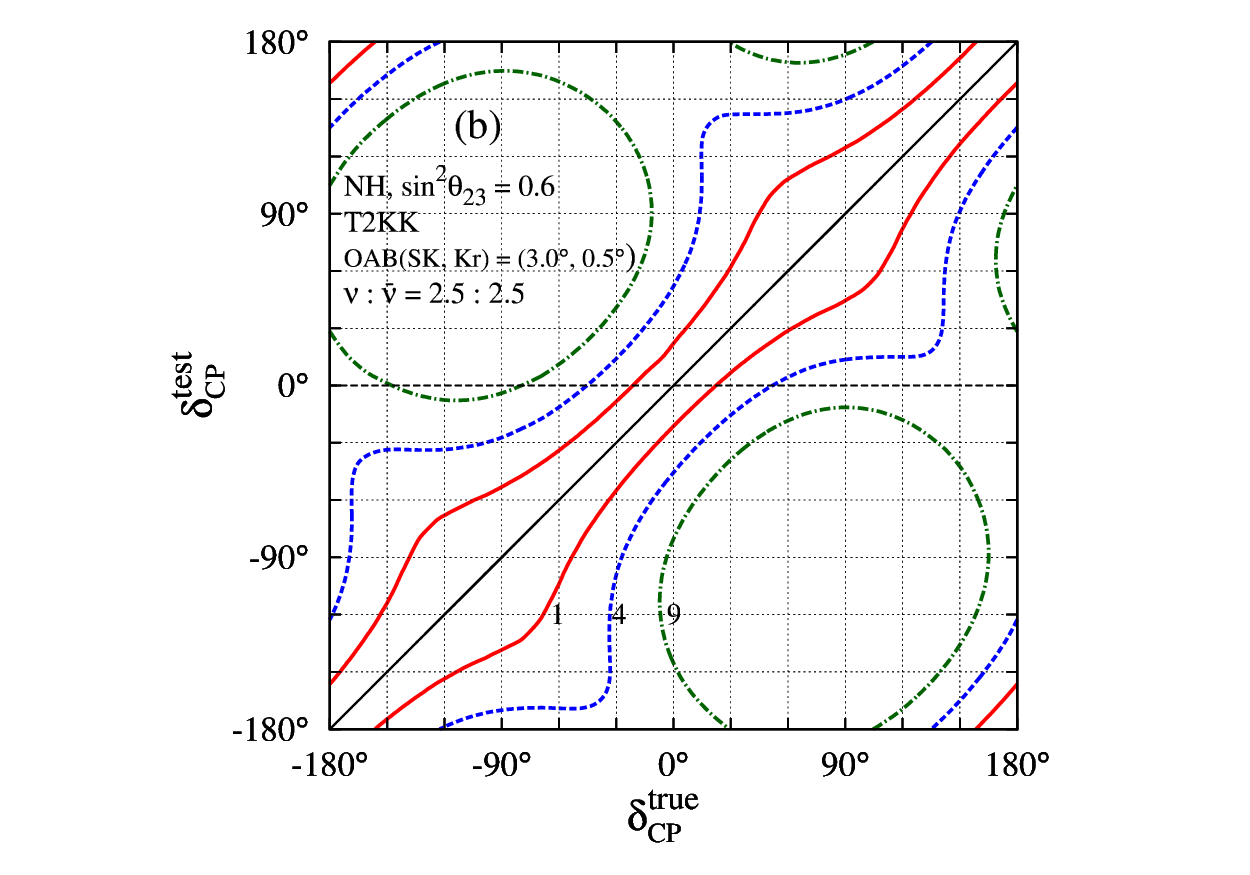} 
}
\caption{Same as Fig.~\ref{fig:3sigma_nh} (a) but for $\sin^2\theta_{23} =
 0.4$ (left) and $\sin^2\theta_{23} = 0.6$ (right).}
\label{fig:3sigma_th23}
\end{figure}
The sensitivity is better for smaller
$\ssq{}23$ as shown in the figure. 
This is because the coefficients of the $\sin\dmns$ and $\cos\dmns$ in
the $\nu_\mu \ra \nu_e$ oscillation probability, Eq.~\eqref{eq:Pmue}, is
proportional to $1/\tan\theta_{23}$. 
The percentages of the regions rejected with $\dchisqmin >
1, 4, 9$ in the test-$\dmns$
vs. true-$\dmns$ plane are
83\% (80\%) , 67\% (60\%) and 44\% (35\%), respectively, for $\ssq{}23
 = 0.4\,(0.6)$. Similar dependences are found for the inverted hierarchy
 case and other experiments. 

\section{Summary and conclusion}
\label{sec:conclusion}
In this paper, we have revisited the previous analysis of
Ref.~\cite{Hagiwara:2009bb,Hagiwara:2011kw,Hagiwara:2012mg} on the sensitivities to
the mass hierarchy determination and leptonic CP phase measurements of the
Tokai-to-Kamioka-and-Korea
(T2KK)~\cite{Hagiwara:2004iq,Ishitsuka:2005qi,Hagiwara:2005pe,Hagiwara:2006vn,Kajita:2006bt,Hagiwara:2006nn,Huber:2007em,Hagiwara:2009bb,Dufour:2010vr,Hagiwara:2011kw,Hagiwara:2012mg,Dufour:2012zr} and Tokai-to-Kamioka-and-Oki (T2KO) experiments~\cite{Badertscher:2008bp,Hagiwara:2012mg}, putting
emphasis on the $\nu_\mu$ and $\bar{\nu}_\mu$ focusing
beam ratio with dedicated estimation of
backgrounds. We place a
Super-Kamiokande (SK) type water \cerenkov\, detector of 100 kton fiducial volume in
 Korea (T2KK) or Oki island (T2KO) at 1000 km and 653 km away from the
J-PARC neutrino facility, respectively.
The
neutral current (NC)
single-$\pi^0$ background and its uncertainty are estimated by using the realistic $\pi^0$
rejection probability based on the \polfit
algorithm~\cite{Plfit_okumura}, taking into
account the coherent $\pi^0$ production processes, which is neglected in
the previous analysis~\cite{Hagiwara:2009bb}, and including the uncertainty of axial
masses in the neutrino-nucleus interaction
model~\cite{Rein:1980wg,Rein:1982pf}.
The sensitivities are then evaluated using the standard $\chi^2$
analysis.

We found that the wrong mass hierarchy is rejected 
 with $|\odchisqmh| > 10$ in the T2KK and $|\odchisqmh| > 3$ in the T2KO experiment for
 any CP phases when $\ssq{}23 = 0.5$, using the $\nu_\mu$ and $\bar{\nu}_\mu$
 focusing beam ratio between
 3\,:\,2 and 2.5\,:\,2.5 (in unit of $10^{21}$POT with the proton energy of
 40 GeV). 
It should be noted that the $|\odchisqmh|$ quoted in
 this study is regarded as the average sensitivity expected for an experiment
 because we neglect the statistical fluctuations in input data set. 
Although a rigorous interpretation of the
 $|\odchisqmh|$ needs dedicated statistical consideration, 
above
 $|\odchisqmh|$ may be roughly interpreted as the $80\%$ probabilities
of determining the mass
hierarchy with $> 2.6\,\sigma$ for T2KK
and $> 1.3\,\sigma$ for T2KO,
 respectively, assuming the Gaussian distribution for the
$\dchisqmh$ in the T2KK and T2KO experiments (See Fig.2 in
 Ref.~\cite{Blennow:2013oma} for the interpretation).

In the most sensitive
region around $\dmns \sim -90^\circ$ for the normal hierarchy case, we reject
the wrong mass hierarchy with $|\odchisqmh| \sim 32$ in the T2KK
experiment 
 ($3.0^\circ$ OAB at the SK) and with $|\odchisqmh| \sim 20$ in the T2KO
experiment, using the 3\,:\,2 - 2.5\,:\,2.5 beam
ratio. 
These $|\odchisqmh|$ correspond to the $80\%$
probabilities of the mass hierarchy determination with $> 4.9\,\sigma$
for the T2KK experiment and with $> 3.8\, \sigma$ for the T2KO
 experiment. 
On the other hand, for the inverted hierarchy case, we reject
the wrong mass hierarchy with $|\odchisqmh| \sim 30$ in the T2KK
experiment 
 ($3.0^\circ$ OAB at the SK) and with $|\odchisqmh| \sim 18$ in the T2KO
experiment around $\dmns \sim 90^\circ$, using the same
beam ratio as the normal hierarchy case.  
These $|\odchisqmh|$ correspond to the $80\%$
probabilities of the mass hierarchy determination with $> 4.7\,\sigma$
for the T2KK experiment and with $> 3.6\, \sigma$ for the T2KO
 experiment. 
These sensitivities are obtained for $\ssq{}23 = 0.5$ and enhanced
(reduced) by $30\%$ - $40\%$ in $|\odchisqmh|$ for
$\ssq{}23 = 0.6\, (0.4)$.

We also examined the sensitivity to the CP phase measurements. 
The $\nu_\mu$ and $\bar{\nu}_\mu$ focusing beam ratio between 3.5\,:\,1.5 and 1.5\,:\,3.5 give the smallest
uncertainty for most of the CP phases.
Employing the 2.5\,:\,2.5 beam ratio, the T2KK and T2KO experiments
measure the CP phase with the uncertainty of $\sim 20^\circ$ - $50^\circ$ (T2KK with $3.0^\circ$
OAB at the SK), $\sim 20^\circ$ - $45^\circ$ (T2KK with $2.5^\circ$ OAB at
the SK and T2KO)
, depending on the CP phase. 
We can measure the CP phase most
 accurately around $\dmns \sim 0^\circ$ and $\sim 180^\circ$, while the
 uncertainty is largest around $\dmns \sim \pm
60^\circ$ and $\sim \pm 120^\circ$.
A long baseline is helpful to improve the CP phase
measurements around those large uncertainty regions. 
The mass hierarchy and $\ssq{}23$ dependence of the CP phase measurements
are not so large.
The CP violation in the lepton sector is detected with $\dchisqmin >
9$ for $-100^\circ < \dmns < -60^\circ$ and $70^\circ < \dmns < 120^\circ$ by the
T2KO experiment, while the T2KK experiment detects the CP violation only
with $\dchisqmin > 4$.
In either experiments, we need larger statistics
to establish the CP violation in wide range of the CP phases. 

As discussed in this paper, the T2KK and T2KO experiments can
improve their sensitivity to both the mass hierarchy determination and
leptonic CP phase measurement using $\nu_\mu$ and $\bar{\nu}_\mu$ focusing
beams with 3\,:\,2 - 2.5\,:\,2.5 beam ratio. 
This improvement is significant especially for the mass hierarchy
determination, lifting the highest sensitivities in the T2KK (both
2.5$^\circ$ and 3.0$^\circ$ OAB at the SK) and T2KO
experiments. The lowest sensitivities 
are improved in the T2KK experiment with $2.5^\circ$ OAB at the SK,
while the improvement is not so evident in the other experiments.
The T2KK experiment allows us to determine the mass hierarchy and
measure the leptonic CP phase simultaneously. The T2KO experiment also
has the sensitivity to the CP phase
measurement, while its physics potential for the
mass hierarchy determination is not as good as that of the T2KK
experiment.

\vspace{2em}
\noindent {\bf Acknowledgment}
\vspace{1em}

\noindent
We would like to thank T. Nakaya and M. Yokoyama for useful discussions
and comments on the CP phase sensitivity study of $\T2K122$ setup. 
Y.T. wishes to thank the Korea Neutrino Research Center and KIAS, where part of this
work was done.
This work is in part supported by the Grant in Aid for Scientific Research
No.25400287 (K.H.) and No.26400254 (N.O. and Y.T.) from MEXT, Japan, by National Research Foundation of Korea (NRF)
Research Grant NRF-2015R1A2A1A05001869 (P.K.), and by the NRF grant funded by the
Korea government (MSIP) (No. 2009-0083526) through Korea Neutrino
Research Center at Seoul National University (P.K. and Y.T.).

\clearpage
\appendix
\setcounter{equation}{0}

\noindent \textbf{\Large Appendix}

\section{Signal cross sections}
\label{sec:appA0}
We parameterize the CCQE cross sections for oxygen nuclei after imposing
the selection criteria \eqref{eq:criteriaCC} as follows:  
\begin{equation}
{\hat{\sigma}_{\nu O}^{\rm CCQE}(E_\nu)}
=
{{\sigma}_{\nu}^{\rm CCQE}(E_\nu)} \times
\left\{
\begin{array}{ll}
\displaystyle\frac{\alpha^\nu}{1 +(\beta^\nu/E_\nu)^8}\,
&(0.2 \leq E_\nu < 0.8 \,{\rm GeV}), \\
c_0^{\nu} +c_1^{\nu}E_\nu +c_2^{\nu} E_\nu^2 +c_3^\nu E_\nu^3\,
&(0.8 \leq E_\nu \leq 5 \,{\rm GeV}),
\end{array}
\right.
\label{eq:Xsec_CCQE_nu}
\end{equation}
where $\nu$ denotes $\nu_\mu, \nu_e, \bar{\nu}_\mu$ and $\bar{\nu_e}$,
and ${\sigma}_{\nu}^{\rm CCQE}$ is the CCQE cross
sections for a water target without any cuts~\cite{Smith:1972xh}, and $E_\nu$ is in GeV unit.
The parameters $\alpha^\nu, \beta^\nu$ and $c^\nu$'s are summarized in Table~\ref{tb:params_CCQEO}.
\begin{table}[t]
\begin{center}
\begin{tabular}{ccccccc}
\hline\hline\addlinespace[2 pt]
CCQE-$O$ & $\alpha$ & $\beta$ & $c_0$ & $c_1$ & $c_2$ & $c_3$ \\
\hline\addlinespace[2 pt]
$\nu_\mu = \nu_e$ & 0.998 & 0.284 & 1.02 & -0.0323 & 0.00413 & 0 \\
$\bar{\nu}_\mu = \bar{\nu}_e$ & 0.733 & 0.275 & 0.718 & 0.0253 & 0.00990 & 0.00107 \\
\hline\hline
\end{tabular}
\end{center}
\caption{The parameters for the parameterization,
 Eq.~\eqref{eq:Xsec_CCQE_nu}, of the oxygen
 CCQE cross sections after imposing the CCQE selection criteria~\eqref{eq:criteriaCC}.}
\label{tb:params_CCQEO}
\end{table}

The hydrogen CCQE cross sections can be parameterized as
\begin{equation}
{\hat{\sigma}_{\bar{\nu}_{\mu/e} H}^{\rm CCQE}(E_\nu)}
=
{{\sigma}_{\bar{\nu}_{\mu/e}}^{\rm CCQE}(E_\nu)} \times
\left\{
\begin{array}{ll}
0.324  -0.116 E_\nu\,
&(0.2 \leq E_\nu < 0.8 \,{\rm GeV}), \\
0.271 -0.0580 E_\nu +0.0172 E_\nu^2 -0.00169 E_\nu^3\,
&(0.8 \leq E_\nu \leq 5 \,{\rm GeV}).
\end{array}
\right.
\label{eq:Xsec_CCQE_aH}
\end{equation}
Note that only anti-neutrinos interact with hydrogens via the CCQE interactions.

The non-CCQE signal cross sections
, $\hat{\sigma}_{\nu Z}^{\rm nonCCQE} (E_\nu)$,
are the total cross sections of all the non-CCQE events that satisfy the CCQE selection
criteria~\eqref{eq:criteriaCC} and can be parameterized for the
$\nu$\,-\,$O$ and $\nubar$\,-\,$O$ interactions as
\begin{eqnarray}
{\hat{\sigma}_{\nu O}^{\rm nonCCQE}(E_\nu)}
=
{{\sigma}_{\nu}^{\rm CCQE}(E_\nu)} \times
\left\{
\begin{array}{ll}
\displaystyle\frac{\alpha^\nu}{1 +(\beta^\nu/E_\nu)^5}\,
&(0.2 \leq E_\nu < 1.1 \,{\rm GeV}), \\
c_0^{\nu} +c_1^{\nu}E_\nu +c_2^{\nu} E_\nu^2 +c_3^\nu E_\nu^3\,
&(1.1 \leq E_\nu \leq 5 \,{\rm GeV}),
\end{array}
\right.
\label{eq:Xsec_CCRes}
\end{eqnarray}
and for the $\nu$\,-\,$H$ and $\nubar$\,-\,$H$ interactions as 
\begin{subequations} 
\begin{eqnarray}
{\hat{\sigma}_{\nu_{\mu/e} H}^{\rm nonCCQE}(E_\nu)}
=
{{\sigma}_{\nu}^{\rm CCQE}(E_\nu)} \times
\left\{
\begin{array}{ll}
\displaystyle\frac{\alpha^\nu}{1 +(\beta^\nu/E_\nu)^5}\,
&(0.2 \leq E_\nu < 1.1 \,{\rm GeV}), \\
c_0^{\nu} +c_1^{\nu}E_\nu +c_2^{\nu} E_\nu^2 +c_3^\nu E_\nu^3\,
&(1.1 \leq E_\nu \leq 5 \,{\rm GeV}),
\end{array}
\right.
\label{eq:Xsec_nonCCQE_nu}
\end{eqnarray}
\begin{eqnarray}
{\hat{\sigma}_{\bar{\nu}_{\mu/e} H}^{\rm nonCCQE}(E_\nu)}
=
{{\sigma}_{\bar{\nu}}^{\rm CCQE}(E_\nu)} \times
\left\{
\begin{array}{ll}
\displaystyle\frac{\alpha^{\bar{\nu}}}{1 +(\beta^{\bar{\nu}}/E_\nu)^{10}}\,
&(0.2 \leq E_\nu < 0.6 \,{\rm GeV}), \\
c_0^{\bar{\nu}} +c_1^{\bar{\nu}}E_\nu +c_2^{\bar{\nu}} E_\nu^2 +c_3^{\bar{\nu}} E_\nu^3\,
&(0.6 \leq E_\nu \leq 5 \,{\rm GeV}).
\end{array}
\right.
\label{eq:Xsec_nonCCQE_anu}
\end{eqnarray}
\label{eq:Xsec_CCRes-H}
\end{subequations}
The parameters $\alpha^\nu, \beta^\nu$ and $c^\nu$ are summarized in Table~\ref{tb:params_CCRes}.
\begin{table}[t]
\begin{center}
\begin{tabular}{ccccccc}
\hline\hline\addlinespace[2 pt]
 & $\alpha$ & $\beta$ & $c_0$ & $c_1$ & $c_2$ & $c_3$ \\
\hline\addlinespace[2 pt]
nonCCQE-$O$ &&&&&& \\
$\nu_\mu$ & 0.418 & 0.820 & 0.0429 & 0.355 & -0.0767 & 0.00608 \\
$\nu_e$ & 0.434 & 0.764 & 0.0318 & -0.403 & -0.0921 & 0.00753 \\
$\bar{\nu}_\mu$ & 0.232 & 0.747 & -0.0433 & 0.291 & -0.0648 & 0.00524 \\
$\bar{\nu}_e$ & 0.242 & 0.688 & -0.0289 & 0.301 & -0.0691 & 0.00571 \\
\hline
nonCCQE-$H$ &&&&&& \\
$\nu_\mu$ & 0.0396 & 0.676 & 0.0184 & 0.0230 & -0.00669 & 0.000674 \\
$\nu_e$ & 0.0422 & 0.637 & 0.0286 & 0.0127 & -0.00262 & 0.000183 \\
$\bar{\nu}_\mu$ & 0.0213 & 0.484 & 0.0259 & -0.00877 & 0.00274 & -0.000271 \\
$\bar{\nu}_e$ & 0.0238 & 0.463 & 0.0244 & -0.00538 & 0.00152 & -0.000142 \\
\hline\hline
\end{tabular}
\end{center}
\caption{Same as Table~\ref{tb:params_CCQEO} but for the non-CCQE
 neutrino-nucleus interactions, Eqs.~\eqref{eq:Xsec_CCRes} and \eqref{eq:Xsec_CCRes-H}.}
\label{tb:params_CCRes}
\end{table}
We find that those parameterizations reproduce well the results of \nuance\, for $E_\nu \leq 5$ GeV.

\section{Smearing functions $S^X_{\bar{\nu}_\alpha Z}(E_\nu, \Erec)$}
\label{sec:appA}

In this appendix,
we show our parameterizations of the smearing functions, 
$S^X_{\bar{\nu}_\alpha Z}(E_\nu, \Erec)$,
which map the incoming anti-neutrino energy, $E_\nu$, onto the
reconstructed energy, $\Erec$, for the charged-current (CC) events. Smearing
functions for neutrinos was already constructed in Ref.~\cite{Hagiwara:2009bb}.
The superscript $X$ denotes the event type,
$X=$ CCQE for the charged-current quasi-elastic (CCQE) events or
$X=$ non-CCQE for the other CC events
that pass the CCQE selection criteria of Eq.~(\ref{eq:criteriaCC}).
The subscript $\bar{\nu}_\alpha$ and $Z$ denote an incoming anti-neutrino
($\nubar_\mu$ and $\nubar_e$) and
a target nucleus, respectively.
These functions take account of 
the Fermi motion of target nucleons
inside an oxygen nucleus
and
the finite detector resolutions for muons and 
electrons in a water \cerenkov\, detector
 and valid in the region of
$0.3\mbox{{~GeV}}\leq E_\nu \leq 5.0\mbox{{~GeV}}
\mbox{{~~for~~}}\,
0.4\mbox{{~GeV}}\leq \Erec \leq 6.0\mbox{{~GeV}}$.

\subsection{CCQE events}
\label{sec:appA1}
\subsubsection{$^{16}O$ interaction}

The $\Erec$ distributions of the anti-neutrino induced CCQE events via interactions with oxygen nuclei,
which are generated by \nuance\, v.3.504 package \cite{Casper:2002sd},
can be parameterized by three Gaussians,
\begin{eqnarray}
 S^{\rm CCQE}_{\bar{\nu}_\alpha O}(E_\nu, \Erec) =
 \dfrac{1}{A^\alpha(E_\nu)}
\sum_{n=1}^{3}
r^\alpha_n(E_\nu) 
\exp
\left\{-\dfrac{(\Erec-E_\nu +{\delta}E^\alpha_n(E_\nu))^2}
{2(\sigma^\alpha_n(E_\nu))^2}
\right\}.
\label{eq:fit_ccqeA0}
\end{eqnarray}
Here the incoming anti-neutrino energy, $E_\nu$, and $\Erec$ are in MeV unit.
Each function is normalized by
\begin{eqnarray}
 A^\alpha(E_\nu)
= \sqrt{2\pi}\sum_{n=1}^{3}
  r^\alpha_n(E_\nu)
     \sigma_n^\alpha(E_\nu)\,.
\label{eq:normalzationCCQE}
\end{eqnarray}

For $\nubar_\mu$ case, the weight factors $r^\alpha_{n}$, variances $\sigma^\alpha_n$, 
and the energy shifts ${\delta}E^\alpha_n$,
are functions of the dimensionless parameter, $z \equiv
E_\nu/(1000\,{\rm MeV})$, and expressed as
\begin{align}
 r^{\mu}_1 &= 3.20 -2.16\, \,z +0.562\, \,z^2 -0.0504\, \,z^3, \nn\\
 r^{\mu}_2 &= 2.05 -1.52 \,z +0.406 \,z^2 -0.0360 \,z^3, \nn\\
 r^{\mu}_3 &= 0.110 -0.0828 \,z +0.0224 \,z^2 -0.00198 \,z^3,
\label{eq:r_ccqe_mu}
\end{align}
\begin{align}
 \sigma^{\mu}_1 &= 11.3 +27.4 \,z -1.01 \,z^2, \nn\\
 \sigma^{\mu}_2 &= 36.0 +49.0 \,z -3.31 \,z^2, \nn\\
 \sigma^{\mu}_3 &= -18.0 +241 \,z -56.8 \,z^2 +4.85 \,z^3,
\label{eq:sig_ccqe_mu}
\end{align} 
\begin{align}
 \delta E^{\mu}_1 &= 43.5 -9.64 \,z +2.86 \,z^2 -0.229 \,z^3, \nn\\
 \delta E^{\mu}_2 &= 28.1 -2.89 \,z -0.482 \,z^2, \nn\\
 \delta E^{\mu}_3 &= -4.84 -5.34 \,z -1.59 \,z^2.
\label{eq:dE_ccqe_mu}
\end{align}
The variances and the energy-shift terms $\delta E_n^\mu$ are given in units of MeV.

For $\nubar_e$ case, we find
\begin{align}
 r^{e}_1 &= 1, \nn\\
 r^{e}_2 &= 0.586 -0.258 \,z +0.122 \,z^2 -0.0137 \,z^3, \nn\\
 r^{e}_3 &= 0.0140 -0.00131 \,z +0.00502 \,z^2,
\label{eq:r_ccqe_e}
\end{align}
\begin{align}
 \sigma^{e}_1 &= 23.3 +25.8 \,z -1.54 \,z^2, \nn\\
 \sigma^{e}_2 &= 53.8 +46.5 \,z -2.79 \,z^2, \nn\\
 \sigma^{e}_3 &= 12.5 +250 \,z -66.1 \,z^2 +6.43 \,z^3,
\end{align} 
\begin{align}
 \delta E^{e}_1 &= 42.9 -11.0 \,z +4.42 \,z^2 -0.382 \,z^3, \nn\\
 \delta E^{e}_2 &= 27.3 -6.43 \,z, \nn\\
 \delta E^{e}_3 &= -176 +178 \,z -86.5 \,z^2 +15.5 \,z^3
 -1.04 \,z^4.
\end{align}

\subsubsection{proton interaction}
The $\Erec$ distributions of the $\nubar_\mu$ induced CCQE events via interactions with
protons
can be parameterized by up to two Gaussians,
\begin{align}
S^{\rm CCQE}_{\bar{\nu}_\mu H}(E_\nu, \Erec) &= \dfrac{1}{A^\alpha(E_\nu)}\left[
r^\alpha_1(E_\nu) \exp\left\{-\dfrac{(\Erec-E_\nu+\delta
 				   E_1^\alpha(E_\nu))^2}{2(\sigma_1^\alpha(E_\nu))^2}
 				  \right\}\right. \nn\\
 +& \left. r^\alpha_2(E_\nu) \exp\left\{-\dfrac{(\Erec-E_\nu+\delta
 				   E_2^\alpha(E_\nu))^2}{2(\sigma_2^\alpha(E_\nu))^2}
 				  \right\} \Theta(\,z -0.7)\right],
\label{eq:fit_ResO}
\end{align}
where $\Theta$ is a step function.
The weight factors
$r^\alpha_{n}$, variances $\sigma^\alpha_n$ (MeV)
and energy shifts ${\delta}E^\alpha_n$ (MeV) 
are expressed as
\begin{align}
 r^{\mu}_1 &= 1, \nn\\
 r^{\mu}_2 &= 0.106,
\label{eq:r_ccqe_mu}
\end{align}
\vspace{-2em}
\begin{align}
 \sigma^{\mu}_1 &= 3.20 +25.5 \,z, \nn\\
 \sigma^{\mu}_2 &= -3.73 +52.7 \,z,
\label{eq:sig_ccqe_mu}
\end{align} 
\vspace{-2em}
\begin{align}
 \delta E^{\mu}_1 &= 0.00, \nn\\
 \delta E^{\mu}_2 &= 11.6 -17.0 \,z.
\label{eq:dE_ccqe_mu}
\end{align}

For $\bar{\nu}_e$ case,
the $\Erec$ distributions
are parameterized by up to three Gaussians,
\begin{align}
S^{\rm CCQE}_{\nubar_e H}(E_\nu, \Erec) &= \dfrac{1}{A^\alpha(E_\nu)}\left[
r^\alpha_1(E_\nu) \exp\left\{-\dfrac{(\Erec-E_\nu+\delta
 				   E_1^\alpha(E_\nu))^2}{2(\sigma_1^\alpha(E_\nu))^2}
 				  \right\}\right.\\
+&r^\alpha_2(E_\nu) \exp\left\{-\dfrac{(\Erec-E_\nu+\delta
 				   E_2^\alpha(E_\nu))^2}{2(\sigma_2^\alpha(E_\nu))^2}
 				  \right\} \\
+&\left.r^\alpha_3(E_\nu) \exp\left\{-\dfrac{(\Erec-E_\nu+\delta
 				   E_3^\alpha(E_\nu))^2}{2(\sigma_3^\alpha(E_\nu))^2}
 				  \right\} \Theta(\,z -1.0)\right].
\label{eq:fit_ResO}
\end{align}
The weight factors, variances $\sigma^\alpha_n$ (MeV)
and energy shifts ${\delta}E^\alpha_n$ (MeV) 
are
\begin{align}
 r^{e}_1 &= 1, \nn\\
 r^{e}_2 &= 4.47 (e^{0.114 \,z -2.59} -e^{-0.698
 \,z -2.36}), \nn\\
 r^{e}_3 &= 0.00532 \,e^{0.620 \,z},
\label{eq:r_ccqe_e}
\end{align}
\begin{align}
 \sigma^{e}_1 &= 268\left( e^{-0.0225 \,z} -e^{-0.135 \,z
 -0.0470} \right), \nn\\
 \sigma^{e}_2 &= 297\left( 1 -e^{-0.235 \,z} \right), \nn\\
 \sigma^{e}_3 &= 4.41 \left( e^{ 0.0584 \,z +4.27} -e^{ -0.489
 \,z +4.47} \right),
\label{eq:sig_ccqe_e}
\end{align} 
\begin{align}
 \delta E^{e}_1 &= \frac{18.3}{1 +34.0 \,e^{ -1.02 \,z }}, \nn\\
 \delta E^{e}_2 &= -7.26 -5.66 \,z, \nn\\
 \delta E^{e}_3 &= 665 \left(e^{ -0.261 \,z +0.0811} -e^{
 -0.107 \,z} \right).
\label{eq:dE_ccqe_e}
\end{align}

\subsection{non-CCQE Events}
\label{sec:appA2}
\subsubsection{$^{16}O$ interaction}
The $\Erec$ distributions of the anti-neutrino induced non-CCQE events via
interactions with oxygen nuclei can be parameterized by up to four Gaussians,
\begin{align}
S^{\rm nonCCQE}_{\bar{\nu}_\alpha O}(E_\nu, \Erec) &= \dfrac{1}{A^\alpha(E_\nu)}\left[
r^\alpha_1(E_\nu) \exp\left\{-\dfrac{(\Erec-E_\nu+\delta
 				   E_1^\alpha(E_\nu))^2}{2(\sigma_1^\alpha(E_\nu))^2}
 				  \right\}\right. \nn\\
+&r^\alpha_2(E_\nu) \exp\left\{-\dfrac{(\Erec-E_\nu+\delta
 				   E_2^\alpha(E_\nu))^2}{2(\sigma_2^\alpha(E_\nu))^2}
 				  \right\} \Theta(\,z -0.7) \nn\\
+&r^\alpha_3(E_\nu) \exp\left\{-\dfrac{(\Erec-E_\nu+\delta
 				   E_3^\alpha(E_\nu))^2}{2(\sigma_3^\alpha(E_\nu))^2}
 				  \right\} \Theta(\,z -1.0) \nn\\
+&\left.r^\alpha_4(E_\nu) \exp\left\{-\dfrac{(\Erec-E_\nu+\delta
 				   E_4^\alpha(E_\nu))^2}{2(\sigma_4^\alpha(E_\nu))^2}
 				  \right\} \Theta(1.5 - z)\right].
\label{eq:fit_ResO}
\end{align}
Each function is normalized as in Eq.~(\ref{eq:normalzationCCQE}).
The fourth Gaussian takes into account the effects of the event selection
cut.

For $\nubar_\mu$ case, the weight factors
$r^\alpha_{n}$, variances $\sigma^\alpha_n$ (MeV)
and energy shifts ${\delta}E^\alpha_n$ (MeV) 
are
\begin{align}
r^\mu_1 =& 1, \nn\\
r^\mu_2 =& 0.325 -0.0652 \,z +0.0142 \,z^2, \nn\\
r^\mu_3 =& 0.0804 -0.00176 \,z +0.00194 \,z^2, \nn\\
r^\mu_4 =& -0.326 +0.229 \,z,
\label{eq:sig_ResOmu} 
\end{align} 
\begin{align}
\sigma^\mu_1 =& 70.3 +13.9 \,z, \nn\\
\sigma^\mu_2 =& 114 +40.1 \,z, \nn\\
\sigma^\mu_3 =& 452 \left( e^{0.0105 \,z} -e^{-1.67\,z +1.98}
 \right), \nn\\
\sigma^\mu_4 =& 111 \left( e^{-1.67 \,z} -e^{-323 \,z} \right),
\label{eq:dE_ResOmu} 
\end{align}
\begin{align}
\delta E^\mu_1 =& 355, \nn\\
\delta E^\mu_2 =& \frac{539}{1 +1.53 \,e^{-2.76 \,z}}, \nn\\
\delta E^\mu_3 =& 850 +32 \,z, \nn\\
\delta E^\mu_4 =& -214 +998 \,z.
\label{eq:r_ResOmu} 
\end{align} 

For the $\nubar_e$ case, we find
\begin{align}
r^e_1 =&\, 1, \nn\\
r^e_2 =&\, 0.143 +0.0605 \,z, \nn\\
r^e_3 =&\, -0.0459 +0.105 \,z -0.0134 \,z^2, \nn\\
r^e_4 =&\, 0,
\label{eq:r_ResOe} 
\end{align} 
\begin{align}
\sigma^e_1 =&\, 72.9 +12.8 \,z, \nn\\
\sigma^e_2 =&\, 105 +40.2 \,z, \nn\\
\sigma^e_3 =&\, 147 +92.7 \,z,
\label{eq:sig_ResOe} 
\end{align} 
\begin{align}
\delta E^e_1 =&\, 348 +0.352 \,z, \nn\\
\delta E^e_2 =&\, \frac{443}{1 -0.434 \,e^{-0.902 \,z}}, \nn\\
\delta E^e_3 =&\, 753 +17.6 \,z.
\label{eq:dE_ResOe} 
\end{align}

\subsubsection{proton interaction}
The $\Erec$ distributions of the $\nubar_\mu$ induced non-CCQE events via interactions with protons
can be parameterized by up to four Gaussians,
\begin{align}
S^{\rm nonCCQE}_{\nubar_\mu H}(E_\nu, \Erec) &= \dfrac{1}{A^\alpha(E_\nu)}\left[
r^\alpha_1(E_\nu) \exp\left\{-\dfrac{(\Erec-E_\nu+\delta
 				   E_1^\alpha(E_\nu))^2}{2(\sigma_1^\alpha(E_\nu))^2}
 				  \right\}\right. \nn\\
+&r^\alpha_2(E_\nu) \exp\left\{-\dfrac{(\Erec-E_\nu+\delta
 				   E_2^\alpha(E_\nu))^2}{2(\sigma_2^\alpha(E_\nu))^2}
 				  \right\} \nn\\
+&r^\alpha_3(E_\nu) \exp\left\{-\dfrac{(\Erec-E_\nu+\delta
 				   E_3^\alpha(E_\nu))^2}{2(\sigma_3^\alpha(E_\nu))^2}
 				  \right\} \nn\\
+&\left.r^\alpha_4(E_\nu) \exp\left\{-\dfrac{(\Erec-E_\nu+\delta
 				   E_4^\alpha(E_\nu))^2}{2(\sigma_4^\alpha(E_\nu))^2}
 				  \right\} \Theta(z -1.0)\right].
\label{eq:fit_ResH}
\end{align}
The function is normalized as in Eq.~(\ref{eq:normalzationCCQE}).
The fourth Gaussian parameterizes the long low-energy tail of the distribution.
The weight factors
$r^\alpha_{n}$, variances $\sigma^\alpha_n$ (MeV)
and energy shifts ${\delta}E^\alpha_n$ (MeV) 
are
\begin{align}
r^\mu_1 =& e^{-0.442 z} +7.84 e^{-2.90 z}, \nn\\
r^\mu_2 =& 1.7 e^{-0.912 z}, \nn\\
r^\mu_3 =& 0.893 e^{-1.53 z}, \nn\\
r^\mu_4 =& 0.0624 \left( e^{-0.352 z} -e^{-0.434 z +0.0365} \right),
\label{eq:r_ResHmu} 
\end{align} 
\begin{align}
\sigma^\mu_1 =& 18.0 +22.8 \,z, \nn\\
\sigma^\mu_2 =& 8.61 +13.5 \,z, \nn\\
\sigma^\mu_3 =& 31.5 +46.6 \,z -6.27 \,z^2, \nn\\
\sigma^\mu_4 =& -498 +763 \,z -203 \,z^2 +18.7 \,z^3,
\label{eq:sig_ResHmu} 
\end{align} 
\begin{align}
\delta E^\mu_1 =& 212 +3.81 \,z -0.536 \,z^2, \nn\\
\delta E^\mu_2 =& 202 -1.63 \,z, \nn\\
\delta E^\mu_3 =& 284 -6.23 \,z +16.1 \,z^2, \nn\\
\delta E^\mu_4 =& 655 +35 \,z.
\label{eq:dE_ResHmu} 
\end{align}

For $\nubar_e$ case,
$\Erec$ distributions are also parameterized by up to four Gaussians,
\begin{align}
S^{\rm nonCCQE}_{\nubar_e H}(E_\nu, \Erec) &= \dfrac{1}{A^\alpha(E_\nu)}\left[
r^\alpha_1(E_\nu) \exp\left\{-\dfrac{(\Erec-E_\nu+\delta
 				   E_1^\alpha(E_\nu))^2}{2(\sigma_1^\alpha(E_\nu))^2}
 				  \right\}\right. \nn\\
+&r^\alpha_2(E_\nu) \exp\left\{-\dfrac{(\Erec-E_\nu+\delta
 				   E_2^\alpha(E_\nu))^2}{2(\sigma_2^\alpha(E_\nu))^2}
 				  \right\} \nn\\
+&r^\alpha_3(E_\nu) \exp\left\{-\dfrac{(\Erec-E_\nu+\delta
 				   E_3^\alpha(E_\nu))^2}{2(\sigma_3^\alpha(E_\nu))^2}
 				  \right\} \nn\\
+&\left.r^\alpha_4(E_\nu) \exp\left\{-\dfrac{(\Erec-E_\nu+\delta
 				   E_4^\alpha(E_\nu))^2}{2(\sigma_4^\alpha(E_\nu))^2}
 				  \right\} \Theta(z -1.5)\right].
\label{eq:fit_ResHe}
\end{align}
The weight factors, variances $\sigma^\alpha_n$ (MeV)
and energy shifts ${\delta}E^\alpha_n$ (MeV) 
are
\begin{align}
r^e_1 =&\, 1 +6.60 \,e^{-1.51 \,z}, \nn\\
r^e_2 =&\, 2.27 \,e^{-2.76 \,z} +0.507 \,e^{-0.247 \,z}, \nn\\
r^e_3 =&\, 0.0117 \,e^{0.196 \,z} +0.375 \,e^{-3.88 \,z}, \nn\\
r^e_4 =&\, 0.0122 \,e^{0.003 \,z},
\label{eq:r_ResOe} 
\end{align} 
\vspace{-1em}
\begin{align}
\sigma^e_1 =&\, 17.0 +23.9 \,z -1.16 \,z^2, \nn\\
\sigma^e_2 =&\, 29.5 +46.2 \,z -1.70 \,z^2, \nn\\
\sigma^e_3 =&\, -162 +345 \,z -42.6 \,z^2, \nn\\ 
\sigma^e_4 =&\, 62.6 \,e^{0.651 \,z},
\label{eq:sig_ResOe} 
\end{align} 
\begin{align}
\delta E^e_1 =&\, 206, \nn\\
\delta E^e_2 =&\, 272 -3.13 \,e^{0.682 \,z}, \nn\\
\delta E^e_3 =&\, 210 +387 \,z -60 \,z^2, \nn\\
\delta E^e_4 =&\, -1.43\times 10^3 +1.66\times 10^3 \,z -118 \,z^2.
\label{eq:dE_ResOe} 
\end{align}

\bibliographystyle{biblio/general}
\bibliography{biblio/reference}

\begin{thebibliography}{55}%
\makeatletter
\providecommand \@ifxundefined [1]{%
 \@ifx{#1\undefined}
}%
\providecommand \@ifnum [1]{%
 \ifnum #1\expandafter \@firstoftwo
 \else \expandafter \@secondoftwo
 \fi
}%
\providecommand \@ifx [1]{%
 \ifx #1\expandafter \@firstoftwo
 \else \expandafter \@secondoftwo
 \fi
}%
\providecommand \natexlab [1]{#1}%
\providecommand \enquote  [1]{``#1''}%
\providecommand \bibnamefont  [1]{#1}%
\providecommand \bibfnamefont [1]{#1}%
\providecommand \citenamefont [1]{#1}%
\providecommand \href@noop [0]{\@secondoftwo}%
\providecommand \href [0]{\begingroup \@sanitize@url \@href}%
\providecommand \@href[1]{\@@startlink{#1}\@@href}%
\providecommand \@@href[1]{\endgroup#1\@@endlink}%
\providecommand \@sanitize@url [0]{\catcode `\\12\catcode `\$12\catcode
  `\&12\catcode `\#12\catcode `\^12\catcode `\_12\catcode `\%12\relax}%
\providecommand \@@startlink[1]{}%
\providecommand \@@endlink[0]{}%
\providecommand \url  [0]{\begingroup\@sanitize@url \@url }%
\providecommand \@url [1]{\endgroup\@href {#1}{\urlprefix }}%
\providecommand \urlprefix  [0]{URL }%
\providecommand \Eprint [0]{\href }%
\providecommand \doibase [0]{http://dx.doi.org/}%
\providecommand \selectlanguage [0]{\@gobble}%
\providecommand \bibinfo  [0]{\@secondoftwo}%
\providecommand \bibfield  [0]{\@secondoftwo}%
\providecommand \translation [1]{[#1]}%
\providecommand \BibitemOpen [0]{}%
\providecommand \bibitemStop [0]{}%
\providecommand \bibitemNoStop [0]{.\EOS\space}%
\providecommand \EOS [0]{\spacefactor3000\relax}%
\providecommand \BibitemShut  [1]{\csname bibitem#1\endcsname}%
\let\auto@bib@innerbib\@empty
\bibitem [{\citenamefont {An}\ \emph {et~al.}(2012)\citenamefont {An} \emph
  {et~al.}}]{An:2012eh}%
  \BibitemOpen
  \bibfield  {author} {\bibinfo {author} {\bibfnamefont {F.~P.}\ \bibnamefont
  {An}} \emph {et~al.} (\bibinfo {collaboration} {Daya Bay Collaboration}),\
  }\bibfield  {booktitle} {\href {\doibase 10.1103/PhysRevLett.108.171803}
  {}\bibfield  {journal} {\bibinfo  {journal} {Phys.Rev.Lett.}\ }\textbf
  {\bibinfo {volume} {108}\ }(\bibinfo {year} {2012})\ \bibinfo {pages}
  {171803} [\Eprint {http://arxiv.org/abs/1203.1669}
  {arXiv:1203.1669}]}\BibitemShut {NoStop}%
\bibitem [{\citenamefont {An}\ \emph {et~al.}(2013)\citenamefont {An} \emph
  {et~al.}}]{An:2013uza}%
  \BibitemOpen
  \bibfield  {author} {\bibinfo {author} {\bibfnamefont {F.}~\bibnamefont {An}}
  \emph {et~al.} (\bibinfo {collaboration} {Daya Bay Collaboration}),\
  }\bibfield  {booktitle} {\href {\doibase 10.1088/1674-1137/37/1/011001}
  {}\bibfield  {journal} {\bibinfo  {journal} {Chin.Phys.}\ }\textbf {\bibinfo
  {volume} {C37}\ }(\bibinfo {year} {2013})\ \bibinfo {pages} {011001} [\Eprint
  {http://arxiv.org/abs/1210.6327} {arXiv:1210.6327}]}\BibitemShut {NoStop}%
\bibitem [{\citenamefont {An}\ \emph {et~al.}(2014{\natexlab{a}})\citenamefont
  {An} \emph {et~al.}}]{An:2013zwz}%
  \BibitemOpen
  \bibfield  {author} {\bibinfo {author} {\bibfnamefont {F.}~\bibnamefont {An}}
  \emph {et~al.} (\bibinfo {collaboration} {Daya Bay Collaboration}),\
  }\bibfield  {booktitle} {\href {\doibase 10.1103/PhysRevLett.112.061801}
  {}\bibfield  {journal} {\bibinfo  {journal} {Phys.Rev.Lett.}\ }\textbf
  {\bibinfo {volume} {112}\ }(\bibinfo {year} {2014}{\natexlab{a}})\ \bibinfo
  {pages} {061801} [\Eprint {http://arxiv.org/abs/1310.6732}
  {arXiv:1310.6732}]}\BibitemShut {NoStop}%
\bibitem [{\citenamefont {An}\ \emph {et~al.}(2014{\natexlab{b}})\citenamefont
  {An} \emph {et~al.}}]{An:2014ehw}%
  \BibitemOpen
  \bibfield  {author} {\bibinfo {author} {\bibfnamefont {F.~P.}\ \bibnamefont
  {An}} \emph {et~al.} (\bibinfo {collaboration} {Daya Bay Collaboration}),\
  }\bibfield  {booktitle} {\href {\doibase 10.1103/PhysRevD.90.071101}
  {}\bibfield  {journal} {\bibinfo  {journal} {Phys. Rev.}\ }\textbf {\bibinfo
  {volume} {D90}\ }(\bibinfo {year} {2014}{\natexlab{b}})\ \bibinfo {pages}
  {071101} [\Eprint {http://arxiv.org/abs/1406.6468}
  {arXiv:1406.6468}]}\BibitemShut {NoStop}%
\bibitem [{\citenamefont {An}\ \emph {et~al.}(2015)\citenamefont {An} \emph
  {et~al.}}]{An:2015rpe}%
  \BibitemOpen
  \bibfield  {author} {\bibinfo {author} {\bibfnamefont {F.~P.}\ \bibnamefont
  {An}} \emph {et~al.} (\bibinfo {collaboration} {Daya Bay Collaboration}),\
  }\bibfield  {booktitle} {\href {\doibase 10.1103/PhysRevLett.115.111802}
  {}\bibfield  {journal} {\bibinfo  {journal} {Phys. Rev. Lett.}\ }\textbf
  {\bibinfo {volume} {115}\ }(\bibinfo {year} {2015})\ \bibinfo {pages}
  {111802} [\Eprint {http://arxiv.org/abs/1505.03456}
  {arXiv:1505.03456}]}\BibitemShut {NoStop}%
\bibitem [{\citenamefont {An}\ \emph {et~al.}(2016)\citenamefont {An} \emph
  {et~al.}}]{An:2016bvr}%
  \BibitemOpen
  \bibfield  {author} {\bibinfo {author} {\bibfnamefont {F.~P.}\ \bibnamefont
  {An}} \emph {et~al.} (\bibinfo {collaboration} {Daya Bay Collaboration}),\
  }\bibfield  {booktitle} {\href {\doibase 10.1103/PhysRevD.93.072011}
  {}\bibfield  {journal} {\bibinfo  {journal} {Phys. Rev.}\ }\textbf {\bibinfo
  {volume} {D93}\ }(\bibinfo {year} {2016})\ \bibinfo {pages} {072011} [\Eprint
  {http://arxiv.org/abs/1603.03549} {arXiv:1603.03549}]}\BibitemShut {NoStop}%
\bibitem [{\citenamefont {Ahn}\ \emph {et~al.}(2012)\citenamefont {Ahn} \emph
  {et~al.}}]{Ahn:2012nd}%
  \BibitemOpen
  \bibfield  {author} {\bibinfo {author} {\bibfnamefont {J.~K.}\ \bibnamefont
  {Ahn}} \emph {et~al.} (\bibinfo {collaboration} {RENO Collaboration}),\
  }\bibfield  {booktitle} {\href {\doibase 10.1103/PhysRevLett.108.191802}
  {}\bibfield  {journal} {\bibinfo  {journal} {Phys.Rev.Lett.}\ }\textbf
  {\bibinfo {volume} {108}\ }(\bibinfo {year} {2012})\ \bibinfo {pages}
  {191802} [\Eprint {http://arxiv.org/abs/1204.0626}
  {arXiv:1204.0626}]}\BibitemShut {NoStop}%
\bibitem [{\citenamefont {Choi}\ \emph {et~al.}(2016)\citenamefont {Choi} \emph
  {et~al.}}]{RENO:2015ksa}%
  \BibitemOpen
  \bibfield  {author} {\bibinfo {author} {\bibfnamefont {J.~H.}\ \bibnamefont
  {Choi}} \emph {et~al.} (\bibinfo {collaboration} {RENO Collaboration}),\
  }\bibfield  {booktitle} {\href {\doibase 10.1103/PhysRevLett.116.211801}
  {}\bibfield  {journal} {\bibinfo  {journal} {Phys. Rev. Lett.}\ }\textbf
  {\bibinfo {volume} {116}\ }(\bibinfo {year} {2016})\ \bibinfo {pages}
  {211801} [\Eprint {http://arxiv.org/abs/1511.05849}
  {arXiv:1511.05849}]}\BibitemShut {NoStop}%
\bibitem [{\citenamefont {Abe}\ \emph {et~al.}(2012)\citenamefont {Abe} \emph
  {et~al.}}]{Abe:2012tg}%
  \BibitemOpen
  \bibfield  {author} {\bibinfo {author} {\bibfnamefont {Y.}~\bibnamefont
  {Abe}} \emph {et~al.} (\bibinfo {collaboration} {Double Chooz
  Collaboration}),\ }\bibfield  {booktitle} {\href {\doibase
  10.1103/PhysRevD.86.052008} {}\bibfield  {journal} {\bibinfo  {journal}
  {Phys.Rev.}\ }\textbf {\bibinfo {volume} {D86}\ }(\bibinfo {year} {2012})\
  \bibinfo {pages} {052008} [\Eprint {http://arxiv.org/abs/1207.6632}
  {arXiv:1207.6632}]}\BibitemShut {NoStop}%
\bibitem [{\citenamefont {Abe}\ \emph {et~al.}(2013)\citenamefont {Abe} \emph
  {et~al.}}]{Abe:2013sxa}%
  \BibitemOpen
  \bibfield  {author} {\bibinfo {author} {\bibfnamefont {Y.}~\bibnamefont
  {Abe}} \emph {et~al.} (\bibinfo {collaboration} {Double Chooz
  Collaboration}),\ }\bibfield  {booktitle} {\href {\doibase
  10.1016/j.physletb.2013.04.050} {}\bibfield  {journal} {\bibinfo  {journal}
  {Phys.Lett.}\ }\textbf {\bibinfo {volume} {B723}\ }(\bibinfo {year} {2013})\
  \bibinfo {pages} {66} [\Eprint {http://arxiv.org/abs/1301.2948}
  {arXiv:1301.2948}]}\BibitemShut {NoStop}%
\bibitem [{\citenamefont {Abe}\ \emph {et~al.}(2014{\natexlab{a}})\citenamefont
  {Abe} \emph {et~al.}}]{Abe:2014lus}%
  \BibitemOpen
  \bibfield  {author} {\bibinfo {author} {\bibfnamefont {Y.}~\bibnamefont
  {Abe}} \emph {et~al.} (\bibinfo {collaboration} {Double Chooz
  Collaboration}),\ }\bibfield  {booktitle} {\href {\doibase
  10.1016/j.physletb.2014.04.045} {}\bibfield  {journal} {\bibinfo  {journal}
  {Phys.Lett.}\ }\textbf {\bibinfo {volume} {B735}\ }(\bibinfo {year}
  {2014}{\natexlab{a}})\ \bibinfo {pages} {51} [\Eprint
  {http://arxiv.org/abs/1401.5981} {arXiv:1401.5981}]}\BibitemShut {NoStop}%
\bibitem [{\citenamefont {Abe}\ \emph {et~al.}(2014{\natexlab{b}})\citenamefont
  {Abe} \emph {et~al.}}]{Abe:2014bwa}%
  \BibitemOpen
  \bibfield  {author} {\bibinfo {author} {\bibfnamefont {Y.}~\bibnamefont
  {Abe}} \emph {et~al.} (\bibinfo {collaboration} {Double Chooz
  Collaboration}),\ }\bibfield  {booktitle} {\href {\doibase
  10.1007/JHEP02(2015)074, 10.1007/JHEP10(2014)086} {}\bibfield  {journal}
  {\bibinfo  {journal} {JHEP}\ }\textbf {\bibinfo {volume} {1410}\ }(\bibinfo
  {year} {2014}{\natexlab{b}})\ \bibinfo {pages} {086} [\Eprint
  {http://arxiv.org/abs/1406.7763} {arXiv:1406.7763}]}\BibitemShut {NoStop}%
\bibitem [{\citenamefont {Abe}\ \emph {et~al.}(2016)\citenamefont {Abe} \emph
  {et~al.}}]{Abe:2015rcp}%
  \BibitemOpen
  \bibfield  {author} {\bibinfo {author} {\bibfnamefont {Y.}~\bibnamefont
  {Abe}} \emph {et~al.} (\bibinfo {collaboration} {Double Chooz
  Collaboration}),\ }\bibfield  {booktitle} {\href {\doibase
  10.1007/JHEP01(2016)163} {}\bibfield  {journal} {\bibinfo  {journal} {JHEP}\
  }\textbf {\bibinfo {volume} {01}\ }(\bibinfo {year} {2016})\ \bibinfo {pages}
  {163} [\Eprint {http://arxiv.org/abs/1510.08937}
  {arXiv:1510.08937}]}\BibitemShut {NoStop}%
\bibitem [{\citenamefont {Maki}\ \emph {et~al.}(1962)\citenamefont {Maki},
  \citenamefont {Nakagawa},\ and\ \citenamefont {Sakata}}]{Maki:1962mu}%
  \BibitemOpen
  \bibfield  {author} {\bibinfo {author} {\bibfnamefont {Z.}~\bibnamefont
  {Maki}}, \bibinfo {author} {\bibfnamefont {M.}~\bibnamefont {Nakagawa}}\ and\
  \bibinfo {author} {\bibfnamefont {S.}~\bibnamefont {Sakata}},\ }\bibfield
  {booktitle} {\href {\doibase 10.1143/PTP.28.870} {}\bibfield  {journal}
  {\bibinfo  {journal} {Prog.Theor.Phys.}\ }\textbf {\bibinfo {volume} {28}\
  }(\bibinfo {year} {1962})\ \bibinfo {pages} {870}}\BibitemShut {NoStop}%
\bibitem [{\citenamefont {Hagiwara}(2004)}]{Hagiwara:2004iq}%
  \BibitemOpen
  \bibfield  {author} {\bibinfo {author} {\bibfnamefont {K.}~\bibnamefont
  {Hagiwara}},\ }\bibfield  {booktitle} {\href {\doibase
  10.1016/j.nuclphysbps.2004.10.054} {}\bibfield  {journal} {\bibinfo
  {journal} {Nucl.Phys.Proc.Suppl.}\ }\textbf {\bibinfo {volume} {137}\
  }(\bibinfo {year} {2004})\ \bibinfo {pages} {84} [\Eprint
  {http://arxiv.org/abs/hep-ph/0410229} {arXiv:hep-ph/0410229}]}\BibitemShut
  {NoStop}%
\bibitem [{\citenamefont {Ishitsuka}\ \emph {et~al.}(2005)\citenamefont
  {Ishitsuka}, \citenamefont {Kajita}, \citenamefont {Minakata},\ and\
  \citenamefont {Nunokawa}}]{Ishitsuka:2005qi}%
  \BibitemOpen
  \bibfield  {author} {\bibinfo {author} {\bibfnamefont {M.}~\bibnamefont
  {Ishitsuka}}, \bibinfo {author} {\bibfnamefont {T.}~\bibnamefont {Kajita}},
  \bibinfo {author} {\bibfnamefont {H.}~\bibnamefont {Minakata}}\ and\ \bibinfo
  {author} {\bibfnamefont {H.}~\bibnamefont {Nunokawa}},\ }\bibfield
  {booktitle} {\href {\doibase 10.1103/PhysRevD.72.033003} {}\bibfield
  {journal} {\bibinfo  {journal} {Phys.Rev.}\ }\textbf {\bibinfo {volume}
  {D72}\ }(\bibinfo {year} {2005})\ \bibinfo {pages} {033003} [\Eprint
  {http://arxiv.org/abs/hep-ph/0504026} {arXiv:hep-ph/0504026}]}\BibitemShut
  {NoStop}%
\bibitem [{\citenamefont {Hagiwara}\ \emph {et~al.}(2006)\citenamefont
  {Hagiwara}, \citenamefont {Okamura},\ and\ \citenamefont
  {Senda}}]{Hagiwara:2005pe}%
  \BibitemOpen
  \bibfield  {author} {\bibinfo {author} {\bibfnamefont {K.}~\bibnamefont
  {Hagiwara}}, \bibinfo {author} {\bibfnamefont {N.}~\bibnamefont {Okamura}}\
  and\ \bibinfo {author} {\bibfnamefont {K.-i.}\ \bibnamefont {Senda}},\
  }\bibfield  {booktitle} {\href {\doibase 10.1016/j.physletb.2006.09.003,
  10.1016/j.physletb.2006.04.041} {}\bibfield  {journal} {\bibinfo  {journal}
  {Phys.Lett.}\ }\textbf {\bibinfo {volume} {B637}\ }(\bibinfo {year} {2006})\
  \bibinfo {pages} {266} [\Eprint {http://arxiv.org/abs/hep-ph/0504061}
  {arXiv:hep-ph/0504061}]}\BibitemShut {NoStop}%
\bibitem [{\citenamefont {Hagiwara}\ \emph {et~al.}(2007)\citenamefont
  {Hagiwara}, \citenamefont {Okamura},\ and\ \citenamefont
  {Senda}}]{Hagiwara:2006vn}%
  \BibitemOpen
  \bibfield  {author} {\bibinfo {author} {\bibfnamefont {K.}~\bibnamefont
  {Hagiwara}}, \bibinfo {author} {\bibfnamefont {N.}~\bibnamefont {Okamura}}\
  and\ \bibinfo {author} {\bibfnamefont {K.-i.}\ \bibnamefont {Senda}},\
  }\bibfield  {booktitle} {\href {\doibase 10.1103/PhysRevD.76.093002}
  {}\bibfield  {journal} {\bibinfo  {journal} {Phys.Rev.}\ }\textbf {\bibinfo
  {volume} {D76}\ }(\bibinfo {year} {2007})\ \bibinfo {pages} {093002} [\Eprint
  {http://arxiv.org/abs/hep-ph/0607255} {arXiv:hep-ph/0607255}]}\BibitemShut
  {NoStop}%
\bibitem [{\citenamefont {Kajita}\ \emph {et~al.}(2007)\citenamefont {Kajita},
  \citenamefont {Minakata}, \citenamefont {Nakayama},\ and\ \citenamefont
  {Nunokawa}}]{Kajita:2006bt}%
  \BibitemOpen
  \bibfield  {author} {\bibinfo {author} {\bibfnamefont {T.}~\bibnamefont
  {Kajita}}, \bibinfo {author} {\bibfnamefont {H.}~\bibnamefont {Minakata}},
  \bibinfo {author} {\bibfnamefont {S.}~\bibnamefont {Nakayama}}\ and\ \bibinfo
  {author} {\bibfnamefont {H.}~\bibnamefont {Nunokawa}},\ }\bibfield
  {booktitle} {\href {\doibase 10.1103/PhysRevD.75.013006} {}\bibfield
  {journal} {\bibinfo  {journal} {Phys.Rev.}\ }\textbf {\bibinfo {volume}
  {D75}\ }(\bibinfo {year} {2007})\ \bibinfo {pages} {013006} [\Eprint
  {http://arxiv.org/abs/hep-ph/0609286} {arXiv:hep-ph/0609286}]}\BibitemShut
  {NoStop}%
\bibitem [{\citenamefont {Hagiwara}\ and\ \citenamefont
  {Okamura}(2008)}]{Hagiwara:2006nn}%
  \BibitemOpen
  \bibfield  {author} {\bibinfo {author} {\bibfnamefont {K.}~\bibnamefont
  {Hagiwara}}\ and\ \bibinfo {author} {\bibfnamefont {N.}~\bibnamefont
  {Okamura}},\ }\bibfield  {booktitle} {\href {\doibase
  10.1088/1126-6708/2008/01/022} {}\bibfield  {journal} {\bibinfo  {journal}
  {JHEP}\ }\textbf {\bibinfo {volume} {0801}\ }(\bibinfo {year} {2008})\
  \bibinfo {pages} {022} [\Eprint {http://arxiv.org/abs/hep-ph/0611058}
  {arXiv:hep-ph/0611058}]}\BibitemShut {NoStop}%
\bibitem [{\citenamefont {Huber}\ \emph {et~al.}(2008)\citenamefont {Huber},
  \citenamefont {Mezzetto},\ and\ \citenamefont {Schwetz}}]{Huber:2007em}%
  \BibitemOpen
  \bibfield  {author} {\bibinfo {author} {\bibfnamefont {P.}~\bibnamefont
  {Huber}}, \bibinfo {author} {\bibfnamefont {M.}~\bibnamefont {Mezzetto}}\
  and\ \bibinfo {author} {\bibfnamefont {T.}~\bibnamefont {Schwetz}},\
  }\bibfield  {booktitle} {\href {\doibase 10.1088/1126-6708/2008/03/021}
  {}\bibfield  {journal} {\bibinfo  {journal} {JHEP}\ }\textbf {\bibinfo
  {volume} {0803}\ }(\bibinfo {year} {2008})\ \bibinfo {pages} {021} [\Eprint
  {http://arxiv.org/abs/0711.2950} {arXiv:0711.2950}]}\BibitemShut {NoStop}%
\bibitem [{\citenamefont {Hagiwara}\ and\ \citenamefont
  {Okamura}(2009)}]{Hagiwara:2009bb}%
  \BibitemOpen
  \bibfield  {author} {\bibinfo {author} {\bibfnamefont {K.}~\bibnamefont
  {Hagiwara}}\ and\ \bibinfo {author} {\bibfnamefont {N.}~\bibnamefont
  {Okamura}},\ }\bibfield  {booktitle} {\href {\doibase
  10.1088/1126-6708/2009/07/031} {}\bibfield  {journal} {\bibinfo  {journal}
  {JHEP}\ }\textbf {\bibinfo {volume} {0907}\ }(\bibinfo {year} {2009})\
  \bibinfo {pages} {031} [\Eprint {http://arxiv.org/abs/0901.1517}
  {arXiv:0901.1517}]}\BibitemShut {NoStop}%
\bibitem [{\citenamefont {Dufour}\ \emph {et~al.}(2010)\citenamefont {Dufour},
  \citenamefont {Kajita}, \citenamefont {Kearns},\ and\ \citenamefont
  {Okumura}}]{Dufour:2010vr}%
  \BibitemOpen
  \bibfield  {author} {\bibinfo {author} {\bibfnamefont {F.}~\bibnamefont
  {Dufour}}, \bibinfo {author} {\bibfnamefont {T.}~\bibnamefont {Kajita}},
  \bibinfo {author} {\bibfnamefont {E.}~\bibnamefont {Kearns}}\ and\ \bibinfo
  {author} {\bibfnamefont {K.}~\bibnamefont {Okumura}},\ }\bibfield
  {booktitle} {\href {\doibase 10.1103/PhysRevD.81.093001} {}\bibfield
  {journal} {\bibinfo  {journal} {Phys.Rev.}\ }\textbf {\bibinfo {volume}
  {D81}\ }(\bibinfo {year} {2010})\ \bibinfo {pages} {093001} [\Eprint
  {http://arxiv.org/abs/1001.5165} {arXiv:1001.5165}]}\BibitemShut {NoStop}%
\bibitem [{\citenamefont {Hagiwara}\ \emph {et~al.}(2011)\citenamefont
  {Hagiwara}, \citenamefont {Okamura},\ and\ \citenamefont
  {Senda}}]{Hagiwara:2011kw}%
  \BibitemOpen
  \bibfield  {author} {\bibinfo {author} {\bibfnamefont {K.}~\bibnamefont
  {Hagiwara}}, \bibinfo {author} {\bibfnamefont {N.}~\bibnamefont {Okamura}}\
  and\ \bibinfo {author} {\bibfnamefont {K.}~\bibnamefont {Senda}},\ }\bibfield
   {booktitle} {\href {\doibase 10.1007/JHEP09(2011)082} {}\bibfield  {journal}
  {\bibinfo  {journal} {JHEP}\ }\textbf {\bibinfo {volume} {1109}\ }(\bibinfo
  {year} {2011})\ \bibinfo {pages} {082} [\Eprint
  {http://arxiv.org/abs/1107.5857} {arXiv:1107.5857}]}\BibitemShut {NoStop}%
\bibitem [{\citenamefont {Hagiwara}\ \emph {et~al.}(2013)\citenamefont
  {Hagiwara}, \citenamefont {Kiwanami}, \citenamefont {Okamura},\ and\
  \citenamefont {Senda}}]{Hagiwara:2012mg}%
  \BibitemOpen
  \bibfield  {author} {\bibinfo {author} {\bibfnamefont {K.}~\bibnamefont
  {Hagiwara}}, \bibinfo {author} {\bibfnamefont {T.}~\bibnamefont {Kiwanami}},
  \bibinfo {author} {\bibfnamefont {N.}~\bibnamefont {Okamura}}\ and\ \bibinfo
  {author} {\bibfnamefont {K.-i.}\ \bibnamefont {Senda}},\ }\bibfield
  {booktitle} {\href {\doibase 10.1007/JHEP06(2013)036} {}\bibfield  {journal}
  {\bibinfo  {journal} {JHEP}\ }\textbf {\bibinfo {volume} {06}\ }(\bibinfo
  {year} {2013})\ \bibinfo {pages} {036} [\Eprint
  {http://arxiv.org/abs/1209.2763} {arXiv:1209.2763}]}\BibitemShut {NoStop}%
\bibitem [{\citenamefont {Dufour}(2012)}]{Dufour:2012zr}%
  \BibitemOpen
  \bibfield  {author} {\bibinfo {author} {\bibfnamefont {F.}~\bibnamefont
  {Dufour}},\ }\bibfield  {booktitle} {\href@noop {} {}\bibfield  {journal}
  {\bibinfo  {journal} {\Eprint {http://arxiv.org/abs/1211.3884}
  {arXiv:1211.3884}}\ }(\bibinfo {year} {2012})}\BibitemShut {NoStop}%
\bibitem [{\citenamefont {Badertscher}\ \emph {et~al.}(2008)\citenamefont
  {Badertscher}, \citenamefont {Hasegawa}, \citenamefont {Kobayashi},
  \citenamefont {Marchionni}, \citenamefont {Meregaglia} \emph
  {et~al.}}]{Badertscher:2008bp}%
  \BibitemOpen
  \bibfield  {author} {\bibinfo {author} {\bibfnamefont {A.}~\bibnamefont
  {Badertscher}}, \bibinfo {author} {\bibfnamefont {T.}~\bibnamefont
  {Hasegawa}}, \bibinfo {author} {\bibfnamefont {T.}~\bibnamefont {Kobayashi}},
  \bibinfo {author} {\bibfnamefont {A.}~\bibnamefont {Marchionni}}, \bibinfo
  {author} {\bibfnamefont {A.}~\bibnamefont {Meregaglia}} \emph {et~al.},\
  }\bibfield  {booktitle} {\href@noop {} {}\bibfield  {journal} {\bibinfo
  {journal} {\Eprint {http://arxiv.org/abs/0804.2111} {arXiv:0804.2111}}\
  }(\bibinfo {year} {2008})}\BibitemShut {NoStop}%
\bibitem [{\citenamefont {Ciuffoli}\ \emph {et~al.}(2014)\citenamefont
  {Ciuffoli}, \citenamefont {Evslin},\ and\ \citenamefont
  {Zhang}}]{Ciuffoli:2014ika}%
  \BibitemOpen
  \bibfield  {author} {\bibinfo {author} {\bibfnamefont {E.}~\bibnamefont
  {Ciuffoli}}, \bibinfo {author} {\bibfnamefont {J.}~\bibnamefont {Evslin}}\
  and\ \bibinfo {author} {\bibfnamefont {X.}~\bibnamefont {Zhang}},\ }\bibfield
   {booktitle} {\href {\doibase 10.1007/JHEP12(2014)051} {}\bibfield  {journal}
  {\bibinfo  {journal} {JHEP}\ }\textbf {\bibinfo {volume} {12}\ }(\bibinfo
  {year} {2014})\ \bibinfo {pages} {051} [\Eprint
  {http://arxiv.org/abs/1401.3977} {arXiv:1401.3977}]}\BibitemShut {NoStop}%
\bibitem [{\citenamefont {Evslin}\ \emph {et~al.}(2016)\citenamefont {Evslin},
  \citenamefont {Ge},\ and\ \citenamefont {Hagiwara}}]{Evslin:2015pya}%
  \BibitemOpen
  \bibfield  {author} {\bibinfo {author} {\bibfnamefont {J.}~\bibnamefont
  {Evslin}}, \bibinfo {author} {\bibfnamefont {S.-F.}\ \bibnamefont {Ge}}\ and\
  \bibinfo {author} {\bibfnamefont {K.}~\bibnamefont {Hagiwara}},\ }\bibfield
  {booktitle} {\href {\doibase 10.1007/JHEP02(2016)137} {}\bibfield  {journal}
  {\bibinfo  {journal} {JHEP}\ }\textbf {\bibinfo {volume} {02}\ }(\bibinfo
  {year} {2016})\ \bibinfo {pages} {137} [\Eprint
  {http://arxiv.org/abs/1506.05023} {arXiv:1506.05023}]}\BibitemShut {NoStop}%
\bibitem [{\citenamefont {Ge}\ \emph {et~al.}(2016)\citenamefont {Ge},
  \citenamefont {Pasquini}, \citenamefont {Tortola},\ and\ \citenamefont
  {Valle}}]{Ge:2016xya}%
  \BibitemOpen
  \bibfield  {author} {\bibinfo {author} {\bibfnamefont {S.-F.}\ \bibnamefont
  {Ge}}, \bibinfo {author} {\bibfnamefont {P.}~\bibnamefont {Pasquini}},
  \bibinfo {author} {\bibfnamefont {M.}~\bibnamefont {Tortola}}\ and\ \bibinfo
  {author} {\bibfnamefont {J.~W.~F.}\ \bibnamefont {Valle}},\ }\bibfield
  {booktitle} {\href@noop {} {}\bibfield  {journal} {\bibinfo  {journal}
  {\Eprint {http://arxiv.org/abs/1605.01670} {arXiv:1605.01670}}\ }(\bibinfo
  {year} {2016})}\BibitemShut {NoStop}%
\bibitem [{Bar()}]{Barszczak:2005sf}%
  \BibitemOpen
  \bibfield  {author} {\href@noop {} {}}\bibinfo {note} {T. Barszczak, Ph.D.
  thesis, University of California, Irvine (2005).}\BibitemShut {Stop}%
\bibitem [{\citenamefont {Rein}\ and\ \citenamefont
  {Sehgal}(1981)}]{Rein:1980wg}%
  \BibitemOpen
  \bibfield  {author} {\bibinfo {author} {\bibfnamefont {D.}~\bibnamefont
  {Rein}}\ and\ \bibinfo {author} {\bibfnamefont {L.~M.}\ \bibnamefont
  {Sehgal}},\ }\bibfield  {booktitle} {\href {\doibase
  10.1016/0003-4916(81)90242-6} {}\bibfield  {journal} {\bibinfo  {journal}
  {Annals Phys.}\ }\textbf {\bibinfo {volume} {133}\ }(\bibinfo {year} {1981})\
  \bibinfo {pages} {79}}\BibitemShut {NoStop}%
\bibitem [{\citenamefont {Rein}\ and\ \citenamefont
  {Sehgal}(1983)}]{Rein:1982pf}%
  \BibitemOpen
  \bibfield  {author} {\bibinfo {author} {\bibfnamefont {D.}~\bibnamefont
  {Rein}}\ and\ \bibinfo {author} {\bibfnamefont {L.~M.}\ \bibnamefont
  {Sehgal}},\ }\bibfield  {booktitle} {\href {\doibase
  10.1016/0550-3213(83)90090-1} {}\bibfield  {journal} {\bibinfo  {journal}
  {Nucl.Phys.}\ }\textbf {\bibinfo {volume} {B223}\ }(\bibinfo {year} {1983})\
  \bibinfo {pages} {29}}\BibitemShut {NoStop}%
\bibitem [{\citenamefont {Beringer}\ \emph {et~al.}(2012)\citenamefont
  {Beringer} \emph {et~al.}}]{Beringer:1900zz}%
  \BibitemOpen
  \bibfield  {author} {\bibinfo {author} {\bibfnamefont {J.}~\bibnamefont
  {Beringer}} \emph {et~al.} (\bibinfo {collaboration} {Particle Data Group}),\
  }\bibfield  {booktitle} {\href {\doibase 10.1103/PhysRevD.86.010001}
  {}\bibfield  {journal} {\bibinfo  {journal} {Phys.Rev.}\ }\textbf {\bibinfo
  {volume} {D86}\ }(\bibinfo {year} {2012})\ \bibinfo {pages}
  {010001}}\BibitemShut {NoStop}%
\bibitem [{\citenamefont {Itow}\ \emph {et~al.}(2001)\citenamefont {Itow} \emph
  {et~al.}}]{Itow:2001ee}%
  \BibitemOpen
  \bibfield  {author} {\bibinfo {author} {\bibfnamefont {Y.}~\bibnamefont
  {Itow}} \emph {et~al.} (\bibinfo {collaboration} {T2K Collaboration}),\
  }\bibfield  {booktitle} {\href@noop {} {}\bibfield  {journal} {\bibinfo
  {journal} {\Eprint {http://arxiv.org/abs/hep-ex/0106019}
  {arXiv:hep-ex/0106019}}\ }(\bibinfo {year} {2001})\ \bibinfo {pages}
  {239}}\BibitemShut {NoStop}%
\bibitem [{\citenamefont {Arafune}\ \emph {et~al.}(1997)\citenamefont
  {Arafune}, \citenamefont {Koike},\ and\ \citenamefont
  {Sato}}]{Arafune:1997hd}%
  \BibitemOpen
  \bibfield  {author} {\bibinfo {author} {\bibfnamefont {J.}~\bibnamefont
  {Arafune}}, \bibinfo {author} {\bibfnamefont {M.}~\bibnamefont {Koike}}\ and\
  \bibinfo {author} {\bibfnamefont {J.}~\bibnamefont {Sato}},\ }\bibfield
  {booktitle} {\href {\doibase 10.1103/PhysRevD.60.119905,
  10.1103/PhysRevD.56.3093} {}\bibfield  {journal} {\bibinfo  {journal}
  {Phys.Rev.}\ }\textbf {\bibinfo {volume} {D56}\ }(\bibinfo {year} {1997})\
  \bibinfo {pages} {3093} [\Eprint {http://arxiv.org/abs/hep-ph/9703351}
  {arXiv:hep-ph/9703351}]}\BibitemShut {NoStop}%
\bibitem [{T2K()}]{T2Kflux}%
  \BibitemOpen
  \bibfield  {author} {\href@noop {} {}}\bibinfo {note} {A.K. Ichikawa, private
  communication; the flux data for various off-axis angles are available from
  the web page: http://www2.yukawa.kyoto-u.ac.jp/\~{}okamura/T2KK/. Some beam
  profiles are obtained with interpolations by ourselves.}\BibitemShut {Stop}%
\bibitem [{\citenamefont {Abe}\ \emph {et~al.}(2011)\citenamefont {Abe} \emph
  {et~al.}}]{Abe:2011ks}%
  \BibitemOpen
  \bibfield  {author} {\bibinfo {author} {\bibfnamefont {K.}~\bibnamefont
  {Abe}} \emph {et~al.} (\bibinfo {collaboration} {T2K Collaboration}),\
  }\bibfield  {booktitle} {\href {\doibase 10.1016/j.nima.2011.06.067}
  {}\bibfield  {journal} {\bibinfo  {journal} {Nucl.Instrum.Meth.}\ }\textbf
  {\bibinfo {volume} {A659}\ }(\bibinfo {year} {2011})\ \bibinfo {pages} {106}
  [\Eprint {http://arxiv.org/abs/1106.1238} {arXiv:1106.1238}]}\BibitemShut
  {NoStop}%
\bibitem [{\citenamefont {Abe}\ \emph {et~al.}(2014{\natexlab{c}})\citenamefont
  {Abe} \emph {et~al.}}]{Abe:2013hdq}%
  \BibitemOpen
  \bibfield  {author} {\bibinfo {author} {\bibfnamefont {K.}~\bibnamefont
  {Abe}} \emph {et~al.} (\bibinfo {collaboration} {T2K}),\ }\bibfield
  {booktitle} {\href {\doibase 10.1103/PhysRevLett.112.061802} {}\bibfield
  {journal} {\bibinfo  {journal} {Phys. Rev. Lett.}\ }\textbf {\bibinfo
  {volume} {112}\ }(\bibinfo {year} {2014}{\natexlab{c}})\ \bibinfo {pages}
  {061802} [\Eprint {http://arxiv.org/abs/1311.4750}
  {arXiv:1311.4750}]}\BibitemShut {NoStop}%
\bibitem [{\citenamefont {Casper}(2002)}]{Casper:2002sd}%
  \BibitemOpen
  \bibfield  {author} {\bibinfo {author} {\bibfnamefont {D.}~\bibnamefont
  {Casper}},\ }\bibfield  {booktitle} {\href {\doibase
  10.1016/S0920-5632(02)01756-5} {}\bibfield  {journal} {\bibinfo  {journal}
  {Nucl.Phys.Proc.Suppl.}\ }\textbf {\bibinfo {volume} {112}\ }(\bibinfo {year}
  {2002})\ \bibinfo {pages} {161} [\Eprint
  {http://arxiv.org/abs/hep-ph/0208030} {arXiv:hep-ph/0208030}]}\BibitemShut
  {NoStop}%
\bibitem [{\citenamefont {Ashie}\ \emph {et~al.}(2005)\citenamefont {Ashie}
  \emph {et~al.}}]{Ashie:2005ik}%
  \BibitemOpen
  \bibfield  {author} {\bibinfo {author} {\bibfnamefont {Y.}~\bibnamefont
  {Ashie}} \emph {et~al.} (\bibinfo {collaboration} {Super-Kamiokande
  Collaboration}),\ }\bibfield  {booktitle} {\href {\doibase
  10.1103/PhysRevD.71.112005} {}\bibfield  {journal} {\bibinfo  {journal}
  {Phys.Rev.}\ }\textbf {\bibinfo {volume} {D71}\ }(\bibinfo {year} {2005})\
  \bibinfo {pages} {112005} [\Eprint {http://arxiv.org/abs/hep-ex/0501064}
  {arXiv:hep-ex/0501064}]}\BibitemShut {NoStop}%
\bibitem [{Plf()}]{Plfit_okumura}%
  \BibitemOpen
  \bibfield  {author} {\href@noop {} {}}\bibinfo {note} {K. Okumura,'$\pi^0$
  rejection with POLfit in SK', talk at ANT11 in Philadelpia USA
  (2011).}\BibitemShut {Stop}%
\bibitem [{\citenamefont {Abe}\ \emph {et~al.}(2015)\citenamefont {Abe} \emph
  {et~al.}}]{Abe:2015awa}%
  \BibitemOpen
  \bibfield  {author} {\bibinfo {author} {\bibfnamefont {K.}~\bibnamefont
  {Abe}} \emph {et~al.} (\bibinfo {collaboration} {T2K}),\ }\bibfield
  {booktitle} {\href {\doibase 10.1103/PhysRevD.91.072010} {}\bibfield
  {journal} {\bibinfo  {journal} {Phys. Rev.}\ }\textbf {\bibinfo {volume}
  {D91}\ }(\bibinfo {year} {2015})\ \bibinfo {pages} {072010} [\Eprint
  {http://arxiv.org/abs/1502.01550} {arXiv:1502.01550}]}\BibitemShut {NoStop}%
\bibitem [{\citenamefont {Kaboth}(2013)}]{A.KabothfortheT2K:2013fva}%
  \BibitemOpen
  \bibfield  {author} {\bibinfo {author} {\bibfnamefont {A.}~\bibnamefont
  {Kaboth}} (\bibinfo {collaboration} {Collaboration A. Kaboth for the T2K}),\
  }\bibfield  {booktitle} {\href@noop {} {}\bibfield  {journal} {\bibinfo
  {journal} {\Eprint {http://arxiv.org/abs/1310.6544} {arXiv:1310.6544}}\
  }(\bibinfo {year} {2013})}\BibitemShut {NoStop}%
\bibitem [{\citenamefont {Aguilar-Arevalo}\ \emph {et~al.}(2010)\citenamefont
  {Aguilar-Arevalo} \emph {et~al.}}]{AguilarArevalo:2010zc}%
  \BibitemOpen
  \bibfield  {author} {\bibinfo {author} {\bibfnamefont {A.}~\bibnamefont
  {Aguilar-Arevalo}} \emph {et~al.} (\bibinfo {collaboration} {MiniBooNE
  Collaboration}),\ }\bibfield  {booktitle} {\href {\doibase
  10.1103/PhysRevD.81.092005} {}\bibfield  {journal} {\bibinfo  {journal}
  {Phys.Rev.}\ }\textbf {\bibinfo {volume} {D81}\ }(\bibinfo {year} {2010})\
  \bibinfo {pages} {092005} [\Eprint {http://arxiv.org/abs/1002.2680}
  {arXiv:1002.2680}]}\BibitemShut {NoStop}%
\bibitem [{\citenamefont {Zhang}(2015)}]{Zhang:2015fya}%
  \BibitemOpen
  \bibfield  {author} {\bibinfo {author} {\bibfnamefont {C.}~\bibnamefont
  {Zhang}} (\bibinfo {collaboration} {Daya Bay Collaboration}),\ }\bibfield
  {booktitle} {\href@noop {} {}\bibfield  {journal} {\bibinfo  {journal}
  {\Eprint {http://arxiv.org/abs/1501.04991} {arXiv:1501.04991}}\ }(\bibinfo
  {year} {2015})}\BibitemShut {NoStop}%
\bibitem [{\citenamefont {Ge}\ \emph {et~al.}(2013)\citenamefont {Ge},
  \citenamefont {Hagiwara}, \citenamefont {Okamura},\ and\ \citenamefont
  {Takaesu}}]{Ge:2012wj}%
  \BibitemOpen
  \bibfield  {author} {\bibinfo {author} {\bibfnamefont {S.-F.}\ \bibnamefont
  {Ge}}, \bibinfo {author} {\bibfnamefont {K.}~\bibnamefont {Hagiwara}},
  \bibinfo {author} {\bibfnamefont {N.}~\bibnamefont {Okamura}}\ and\ \bibinfo
  {author} {\bibfnamefont {Y.}~\bibnamefont {Takaesu}},\ }\bibfield
  {booktitle} {\href {\doibase 10.1007/JHEP05(2013)131} {}\bibfield  {journal}
  {\bibinfo  {journal} {JHEP}\ }\textbf {\bibinfo {volume} {1305}\ }(\bibinfo
  {year} {2013})\ \bibinfo {pages} {131} [\Eprint
  {http://arxiv.org/abs/1210.8141} {arXiv:1210.8141}]}\BibitemShut {NoStop}%
\bibitem [{\citenamefont {Qian}\ \emph {et~al.}(2012)\citenamefont {Qian},
  \citenamefont {Tan}, \citenamefont {Wang}, \citenamefont {Ling},
  \citenamefont {McKeown},\ and\ \citenamefont {Zhang}}]{Qian:2012zn}%
  \BibitemOpen
  \bibfield  {author} {\bibinfo {author} {\bibfnamefont {X.}~\bibnamefont
  {Qian}}, \bibinfo {author} {\bibfnamefont {A.}~\bibnamefont {Tan}}, \bibinfo
  {author} {\bibfnamefont {W.}~\bibnamefont {Wang}}, \bibinfo {author}
  {\bibfnamefont {J.~J.}\ \bibnamefont {Ling}}, \bibinfo {author}
  {\bibfnamefont {R.~D.}\ \bibnamefont {McKeown}}\ and\ \bibinfo {author}
  {\bibfnamefont {C.}~\bibnamefont {Zhang}},\ }\bibfield  {booktitle} {\href
  {\doibase 10.1103/PhysRevD.86.113011} {}\bibfield  {journal} {\bibinfo
  {journal} {Phys. Rev.}\ }\textbf {\bibinfo {volume} {D86}\ }(\bibinfo {year}
  {2012})\ \bibinfo {pages} {113011} [\Eprint {http://arxiv.org/abs/1210.3651}
  {arXiv:1210.3651}]}\BibitemShut {NoStop}%
\bibitem [{\citenamefont {Blennow}\ \emph {et~al.}(2014)\citenamefont
  {Blennow}, \citenamefont {Coloma}, \citenamefont {Huber},\ and\ \citenamefont
  {Schwetz}}]{Blennow:2013oma}%
  \BibitemOpen
  \bibfield  {author} {\bibinfo {author} {\bibfnamefont {M.}~\bibnamefont
  {Blennow}}, \bibinfo {author} {\bibfnamefont {P.}~\bibnamefont {Coloma}},
  \bibinfo {author} {\bibfnamefont {P.}~\bibnamefont {Huber}}\ and\ \bibinfo
  {author} {\bibfnamefont {T.}~\bibnamefont {Schwetz}},\ }\bibfield
  {booktitle} {\href {\doibase 10.1007/JHEP03(2014)028} {}\bibfield  {journal}
  {\bibinfo  {journal} {JHEP}\ }\textbf {\bibinfo {volume} {1403}\ }(\bibinfo
  {year} {2014})\ \bibinfo {pages} {028} [\Eprint
  {http://arxiv.org/abs/1311.1822} {arXiv:1311.1822}]}\BibitemShut {NoStop}%
\bibitem [{\citenamefont {Cowan}\ \emph {et~al.}(2011)\citenamefont {Cowan},
  \citenamefont {Cranmer}, \citenamefont {Gross},\ and\ \citenamefont
  {Vitells}}]{Cowan:2010js}%
  \BibitemOpen
  \bibfield  {author} {\bibinfo {author} {\bibfnamefont {G.}~\bibnamefont
  {Cowan}}, \bibinfo {author} {\bibfnamefont {K.}~\bibnamefont {Cranmer}},
  \bibinfo {author} {\bibfnamefont {E.}~\bibnamefont {Gross}}\ and\ \bibinfo
  {author} {\bibfnamefont {O.}~\bibnamefont {Vitells}},\ }\bibfield
  {booktitle} {\href {\doibase 10.1140/epjc/s10052-011-1554-0,
  10.1140/epjc/s10052-013-2501-z} {}\bibfield  {journal} {\bibinfo  {journal}
  {Eur. Phys. J.}\ }\textbf {\bibinfo {volume} {C71}\ }(\bibinfo {year}
  {2011})\ \bibinfo {pages} {1554} [\Eprint {http://arxiv.org/abs/1007.1727}
  {arXiv:1007.1727}]},\ \bibinfo {note} {[Erratum: Eur. Phys.
  J.C73,2501(2013)]}\BibitemShut {NoStop}%
\bibitem [{\citenamefont {Wilks}(1938)}]{wilks1938}%
  \BibitemOpen
  \bibfield  {author} {\bibinfo {author} {\bibfnamefont {S.~S.}\ \bibnamefont
  {Wilks}},\ }\bibfield  {booktitle} {\href {\doibase 10.1214/aoms/1177732360}
  {}\bibfield  {journal} {\bibinfo  {journal} {Ann. Math. Statist.}\ }\textbf
  {\bibinfo {volume} {9}\ }(\bibinfo {year} {1938})\ \bibinfo {pages}
  {60}}\BibitemShut {NoStop}%
\bibitem [{\citenamefont {Blennow}\ \emph {et~al.}(2015)\citenamefont
  {Blennow}, \citenamefont {Coloma},\ and\ \citenamefont
  {Fernandez-Martinez}}]{Blennow:2014sja}%
  \BibitemOpen
  \bibfield  {author} {\bibinfo {author} {\bibfnamefont {M.}~\bibnamefont
  {Blennow}}, \bibinfo {author} {\bibfnamefont {P.}~\bibnamefont {Coloma}}\
  and\ \bibinfo {author} {\bibfnamefont {E.}~\bibnamefont
  {Fernandez-Martinez}},\ }\bibfield  {booktitle} {\href {\doibase
  10.1007/JHEP03(2015)005} {}\bibfield  {journal} {\bibinfo  {journal} {JHEP}\
  }\textbf {\bibinfo {volume} {03}\ }(\bibinfo {year} {2015})\ \bibinfo {pages}
  {005} [\Eprint {http://arxiv.org/abs/1407.3274}
  {arXiv:1407.3274}]}\BibitemShut {NoStop}%
\bibitem [{\citenamefont {Elevant}\ and\ \citenamefont
  {Schwetz}(2015)}]{Elevant:2015ska}%
  \BibitemOpen
  \bibfield  {author} {\bibinfo {author} {\bibfnamefont {J.}~\bibnamefont
  {Elevant}}\ and\ \bibinfo {author} {\bibfnamefont {T.}~\bibnamefont
  {Schwetz}},\ }\bibfield  {booktitle} {\href {\doibase
  10.1007/JHEP09(2015)016} {}\bibfield  {journal} {\bibinfo  {journal} {JHEP}\
  }\textbf {\bibinfo {volume} {09}\ }(\bibinfo {year} {2015})\ \bibinfo {pages}
  {016} [\Eprint {http://arxiv.org/abs/1506.07685}
  {arXiv:1506.07685}]}\BibitemShut {NoStop}%
\bibitem [{\citenamefont {Coloma}\ \emph {et~al.}(2012)\citenamefont {Coloma},
  \citenamefont {Donini}, \citenamefont {Fernandez-Martinez},\ and\
  \citenamefont {Hernandez}}]{Coloma:2012wq}%
  \BibitemOpen
  \bibfield  {author} {\bibinfo {author} {\bibfnamefont {P.}~\bibnamefont
  {Coloma}}, \bibinfo {author} {\bibfnamefont {A.}~\bibnamefont {Donini}},
  \bibinfo {author} {\bibfnamefont {E.}~\bibnamefont {Fernandez-Martinez}}\
  and\ \bibinfo {author} {\bibfnamefont {P.}~\bibnamefont {Hernandez}},\
  }\bibfield  {booktitle} {\href {\doibase 10.1007/JHEP06(2012)073} {}\bibfield
   {journal} {\bibinfo  {journal} {JHEP}\ }\textbf {\bibinfo {volume} {1206}\
  }(\bibinfo {year} {2012})\ \bibinfo {pages} {073} [\Eprint
  {http://arxiv.org/abs/1203.5651} {arXiv:1203.5651}]}\BibitemShut {NoStop}%
\bibitem [{\citenamefont {Smith}\ and\ \citenamefont
  {Moniz}(1972)}]{Smith:1972xh}%
  \BibitemOpen
  \bibfield  {author} {\bibinfo {author} {\bibfnamefont {R.~A.}\ \bibnamefont
  {Smith}}\ and\ \bibinfo {author} {\bibfnamefont {E.~J.}\ \bibnamefont
  {Moniz}},\ }\bibfield  {booktitle} {\href {\doibase
  10.1016/0550-3213(72)90040-5} {}\bibfield  {journal} {\bibinfo  {journal}
  {Nucl. Phys.}\ }\textbf {\bibinfo {volume} {B43}\ }(\bibinfo {year} {1972})\
  \bibinfo {pages} {605}},\ \bibinfo {note} {[Erratum: Nucl.
  Phys.B101,547(1975)]}\BibitemShut {NoStop}%
\end{thebibliography}%

\end{document}